\documentclass[aps,pra,twocolumn,showkeys]{revtex4}
\usepackage{amsmath}
\usepackage{amssymb}
\usepackage{graphicx}
\usepackage{bm}

\newcommand{\beq}{\begin{equation}}
\newcommand{\eeq}{\end{equation}}
\newcommand{\req}[1]{Eq.~(\ref{#1})}
\newcommand{\bea}{\begin{eqnarray}}
\newcommand{\eea}{\end{eqnarray}}
\newcommand{\alphaf}{\alpha_\mathrm{f}}
\newcommand{\amZ}[1]{a_{\mathrm{m},#1}}
\newcommand{\am}{a_{\mathrm{m}}}
\newcommand{\aB}{a_\mathrm{B}}
\newcommand{\aBred}{a_*}
\newcommand{\dd}{\mathrm{d}}
\newcommand{\ii}{\mathrm{i}}
\newcommand{\efqm}{{\tilde{\nu}}}
\newcommand{\erm}[1]{\mathrm{e}^{#1}}
\newcommand{\magmom}{\bm{\mu}}
\newcommand{\mel}{m_e}
\newcommand{\mred}{m_*}
\newcommand{\nupar}{\nu_\|}
\newcommand{\omc}[1]{\omega_{\mathrm{c}#1}}
\newcommand{\Omc}{\omc{1}}
\newcommand{\omcmu}{\omc{2}}
\newcommand{\zmax}{z_\mathrm{max}}

\begin{document}

\title{One-electron ion in a quantizing magnetic field}\thanks{This
article is based on I.~V.~Demidov's master thesis \cite{Demidov18}.}

\author{Ivan V. Demidov}
 \email{dvsmallville@gmail.com}
 \affiliation{Res.\ \& Engn.\ Corp.
 ``Mekhanobr-Tekhnika'',
 22th Line of Vasilievsky Island, 3, Saint
 Petersburg 199106, Russia}
\author{Alexander Y. Potekhin}
\email{palex-spb@yandex.ru}
\affiliation{Ioffe Institute, Politekhnicheskaya 26, Saint Petersburg
194021, Russia}

\begin{abstract}

A charged particle in a magnetic field possesses discrete energy levels
associated with particle rotation around the field lines.  A bound
complex of particles with a nonzero net charge possesses an analogous
levels associated with its center-of-mass motion and, in addition, the
levels associated with internal degrees of freedom, that is with
relative motions of its constituent particles. The center-of-mass and
internal motions are mutually dependent, which complicates theoretical
studies of the binding energies, radiative transitions and other
properties of the complex ions moving in quantizing magnetic fields. In
this work, we present a detailed derivation of practical
expressions for the numerical treatment of such properties of the
hydrogenlike ions moving in strong quantizing magnetic fields, which
follows and supplements the previous works of Bezchastnov et al.
Second, we derive asymptotic analytic expressions for the binding
energies, oscillator strengths, and photoionization cross sections of
the moving hydrogenlike ions in the limit of an ultra-strong magnetic
field.

\end{abstract}



\maketitle

\tableofcontents

\section{Introduction}

Properties of atomic
and molecular systems in external magnetic fields have been
intensively studied for several decades (see, e.g., extensive reviews 
\cite{JHY83,Ruder,Lai01}). The majority of the studies
considered them to
   be at rest and assumed the atomic nuclei to be infinitely
   massive (fixed in space). 
The model of infinitely massive nuclei can serve as 
a convenient first approximation, but it is a gross
   simplification for astrophysical
   simulations, because
   thermal motion of atoms and ions across magnetic
   field lines breaks the axial symmetry.

The theory of motion of a system of point charges in a constant magnetic
field is reviewed in \cite{JHY83,BayeVincke90}. A comprehensive
calculation of hydrogen-atom energy spectra, taking account of the
effects of motion across the strong magnetic fields, was carried out in
Refs.~\cite{VDB92,P94}. Calculations of the rates of different types of
radiative transitions and the absorption coefficients in neutron star 
atmospheres were performed in a series of studies (e.g.,
Refs.~\cite{PC03,PCH14}, and references therein). Based on these data, a
model of the hydrogen atmosphere of a neutron star with a strong
magnetic field was elaborated \cite{KK,SPW09}. The database for
astrophysical calculations was created using this model in
Refs.~\cite{HoPC08,Ho14} (see also the review \cite{P14}).

Quantum-mechanical calculations of the characteristics of the
one-electron ion (e.g., He$^+$)  that moves in a strong magnetic field
were performed in Refs.~\cite{BPV97,PB05,BP17}, based on  the formalism
suggested by Bezchastnov \cite{Bez95}. The basic difference from the
case of a neutral atom is that the ion motion is restricted by the field
in the transverse plane, therefore it is quantized
\cite{JHY83,BayeVincke90}. The derivation of the practical formulas for
such calculations was described rather sketchy in the above-cited
papers. The primary goal of the present text is to expose a detailed
derivation of such formulas. Our second aim is to supplement the
consideration of the  bound-bound radiative transitions of a moving
one-electron ion, which were studied in \cite{PB05,BP17}, by the
bound-free transitions. Third, we derive asymptotic analytic
expressions for the  binding energies and radiative transition rates in
the limit of  ultra-strong magnetic field. For the last purpose, we
extend the method previously developed by \citet{HasegawaHoward61} for a
hydrogen atom with infinitely heavy nucleus to the present case of a
moving one-electron ion.

In Sect.~\ref{sect:gen} we present the general formalism for treating
one and two charged particles in a uniform magnetic field and introduce
basic notations. In Sect.~\ref{sect:basis} we present a detailed
derivation of the convenient basis of orbitals \cite{Bez95} for treating
the problem of two charged particles in a strong magnetic field (as a
by-product, we notice some corrections to Ref.~\cite{Bez95} near the end
of Sect.~\ref{sect:states}). Section~\ref{sect:solution} is devoted to
the solution of the Schr\"odinger equation and calculation of the main
properties of the two-particle system using the basis constructed in
Sect.~\ref{sect:basis}. In Sect.~\ref{sect:rad} we present general
formulas for treating interaction of such system of particles with
radiation, based on the solution described in Sect.~\ref{sect:solution}.
In Sect.~\ref{sect:approx}, following and extending the method of
Hasegawa \& Howard \cite{HasegawaHoward61} and using the theory
described in the preceding sections, we derive analytic approximations
for wave functions, binding energies, the transverse geometric size,
overlap integrals and oscillator strengths between different quantum
states, and photoionization cross sections of a one-electron ion.
Section~\ref{sect:concl} presents the conclusions. In Appendices we
derive some useful supplementary relations and prove some statements
from the main text.

\section{Generalities}
\label{sect:gen}

\subsection{Charged particle in uniform magnetic field}
\label{sect:1part}

The quantum-mechanical problem of motion of a charged particle in a
uniform and constant magnetic field was first solved by Rabi
\cite{Rabi28} and Landau \cite{Landau30}. In this subsection we expose
this solution for completeness and introduce some basic notations.

Let us consider a motion of a free particle with a
positive or negative charge
$Ze$, where $e$ is the elementary charge, in a uniform magnetic
field $\bm{B}$.
The Hamiltonian equals the kinetic
energy operator 
\beq
   H^{(1)}=\frac{m\dot{\bm{r}}^2}{2} = H_\perp^{(1)}
      + \frac{p_{z}^2}{2m},
\qquad
     H_\perp^{(1)} = \frac{\bm{\pi}_{\perp}^2}{2m},
\label{H1}
\eeq
where $m$ is the mass,  
\beq
  \bm\pi = m \dot{\bm{r}} = \bm{p} - \frac{Ze}{c}\,\bm{A}(\bm{r})
\label{r-dot}
\eeq
is the kinetic momentum,
$\bm{A}(\bm{r})$ is the vector potential of the field,
 and
$\bm{p}=-\ii\hbar\nabla$ is the canonical momentum operator 
conjugate to $\bm{r}$. In \req{H1} and hereafter,
``$\perp$'' denotes the ``transverse'' part,
related to the motion in the $xy$ plane, while $\bm{B}$ is along the
$z$-axis.

A classical particle moves along a
spiral around
the normal to the $xy$ plane at the \emph{guiding center}
$\bm{r}_{c}$. In quantum mechanics, $\bm{r}_{c}$ is an operator,
related to the \emph{pseudomomentum} operator $\bm{k}$:
\bea
&&
      \bm{k} = m\dot{\bm{r}} + \frac{Ze}{c}\,\bm{B}\times\bm{r}
   = \bm{p} - \frac{Ze}{c}\,\bm{A}(\bm{r}) +
 \frac{Ze}{c}\,\bm{B}\times\bm{r},
 \qquad
\label{k_i}
\\&&
      \bm{r}_{c}=\frac{c}{ZeB^2}\,\bm{k}\times\bm{B}.
\label{r_c}
\eea
The pseudomomentum is a constant of motion in a homogeneous magnetic
field (unlike the canonical momentum and the kinetic momentum, which
are not conserved).
Coordinate operators of $\bm{r}_{c}$ commute with
$H_\perp^{(1)}$, but do not commute with each other:
$[x_c,y_c]=-\ii\hbar c/(ZeB)$. 

Another constant of motion, which generalizes the parallel component
of the orbital momentum $\bm{l}$, is 
\beq
l_z =
\hat{\mathbf{z}}\cdot\left(\bm{r}\times\frac{\bm{k}+\bm{\pi}}{2}
\right)
   = \frac{c}{2ZeB}\,(k^2-\pi^2),
\label{lz}
\eeq
where $\hat{\mathbf{z}} = \bm{B}/B$ is the unit vector in the magnetic
field direction.

In general, $H_\perp^{(1)}$ should be supplemented by 
$(-\bm{B}\cdot\hat{\magmom})$, where
$
  \hat{\magmom}=g_\mathrm{mag}\,(e/2mc)\,\hat{\bm{S}}
$
is the intrinsic
magnetic moment of the particle,
$\hat{\bm{S}}$ is the spin operator, and
$g_\mathrm{mag}$ is the spin $g$-factor
($g_\mathrm{mag}=-2.0023$ and $5.5857$
for the electron and the proton, respectively).
We do not consider the spin term at the moment.

Quantum states with definite eigenvalue $\hbar k_z$ of the
longitudinal momentum $p_z$ in $H^{(1)}$ are described by the wave
functions 
\beq
\psi(\bm{r}) = C_\mathrm{norm} \erm{\ii k_z z}
\,\Phi^{(Z)}(\bm{r}_\perp),
\eeq
where $C_\mathrm{norm}$ is a normalization constant and
$\bm{r}_\perp=(x,y)=(r_\perp\cos\varphi,r_\perp\sin\varphi)$. The form
of the function $\Phi^{(Z)}(\bm{r}_\perp)$ depends
on a choice of the gauge for $\bm{A}(\bm{r})$. 
Let us consider the cylindrical gauge
\beq
\bm{A}=\frac12\bm{B}\times\bm{r}.
\label{Acyl}
\eeq
Then 
$\nabla\cdot\bm{A}=0$,  $A^2 = B^2 r_\perp^2$, and
$\bm{A}\cdot\bm{p}=\bm{p}\cdot\bm{A}=
\frac12(\bm{B}\times\bm{r})\cdot\bm{p}$. Therefore,
\bea
(\bm{p}\cdot\bm{A}+\bm{A}\cdot\bm{p})\,\psi
&=&
-\ii\hbar\left[
    \nabla\cdot(\bm{A}\psi)
       + \bm{A}\cdot\nabla\psi \right]
\nonumber\\ &=&
- \ii\hbar \left(
   \psi\,\nabla\cdot\bm{A}
       + 2\bm{A}\cdot\nabla\psi \right)
\nonumber\\ &=&
2\,\bm{A}\cdot\bm{p}\,  \psi,
\eea
so that
\bea
\pi^2 &=& \left(\bm{p} - \frac{Ze}{c}\bm{A} \right)^2
\nonumber\\ &=&
p^2 -
   \frac{Ze}{c} \big(\bm{p}\cdot\bm{A} + \bm{A}\cdot\bm{p} \big)
  + \left(\frac{Ze}{c}\right)^2A^2
\nonumber\\ &=&
 p^2 - \frac{Ze}{c}Bl_z + \left(\frac{Ze}{2c}\right)^2B^2r_\perp^2,
\eea
where $l_z$ is defined by \req{lz}.
In the cylindrical gauge $(\bm{k} + \bm{\pi})/2 = \bm{p}$, therefore
$l_z$ takes the same form as without the field.

For $H_\perp^{(1)}$ we have the following eigenvalue problem:
\beq
\left(
\frac{p_\perp^2}{2m} - \frac{\sigma\omc{}}{2}\,l_z
   + \frac{m\omc{}^2 r_\perp^2}{8} \right) \Phi = E_{\perp}\Phi,
\label{eigenproblem1}
\eeq
where $\omc{}=|Z|eB/mc$ is the cyclotron frequency
and $\sigma\equiv\mathrm{sign}\,Z$.
Since $l_z$ is an integral of motion, a solution to
\req{eigenproblem1} can be found separately for each its eigenvalue 
$\hbar\lambda$ with integer $\lambda$, such that it
can be written in the polar coordinate system $(r_\perp,\varphi)$,
in the $(x,y)$ plane as
\beq
\Phi^{(Z)}(r_\perp ,\varphi) =
 f(r_\perp )\,\frac{\erm{\ii\lambda\varphi}}{\sqrt{2\pi}}.
\label{psi_fact}
\eeq
Substitution of function (\ref{psi_fact}) into
\req{eigenproblem1} gives
\bea&&\hspace*{-1em}
-\frac{\hbar^2}{2m}\frac{1}{r_\perp }\frac{\dd}{\dd r_\perp }
   \left( r_\perp  \frac{\dd f}{\dd r_\perp } \right) + 
\left(\frac{\hbar^2\lambda^2}{2mr_\perp ^2}
 + \frac{m\omc{}^2 r_\perp ^2}{8} \right) f
\nonumber\\&&\qquad\qquad\qquad
 = \left(E_{\perp} + \sigma \frac{\hbar\omc{}}{2}\lambda\right) f.
\qquad
\label{radial1}
\eea
Let us introduce notations
\beq
\rho = \frac{r_\perp ^2}{2\amZ{Z}^2},
\quad
\varkappa = \frac{E_{\perp}}{\hbar\omc{}} +
 \sigma \frac{\lambda}{2} ,
\label{varkappa}
\eeq
where
\beq
\amZ{Z} = \sqrt{\frac{\hbar}{m\omc{}}} = \frac{\am}{\sqrt{|Z|}},
\eeq
and $\am=\sqrt{\hbar c/eB}$ is the \emph{magnetic length}.
Then \req{radial1} becomes
\beq
\frac{\dd^2 f}{\dd\rho^2} + \frac{1}{\rho}\frac{\dd f}{\dd\rho}
+\frac{\varkappa}{\rho}
 - \left( \frac{\lambda^2}{4\rho^2} + \frac14 \right) f =0.
\label{radial2}
\eeq
At $\rho\to0$, it turns into the Cauchy-Euler equation
\beq
\frac{\dd^2 f}{\dd\rho^2} + \frac{1}{\rho}\frac{\dd f}{\dd\rho}
 - \frac{\lambda^2}{4\rho^2} f =0,
\eeq
which has the explicit solution
 $f = A_1\rho^{\lambda/2} + A_2\rho^{-\lambda/2}$. 
Normalizability of $\psi(r_\perp,\varphi)$ requires that we select a
solution with non-negative power,
$
f \sim \rho^{|\lambda|/2} \sim r_\perp^{|\lambda|}
$ at $\rho\to0$.
Substitution of
$
f=\rho^{|\lambda|/2}\erm{-\rho/2} y(\rho) 
$
 in \req{radial2} gives
 the confluent hypergeometric equation
\beq
\rho\frac{\dd^2 y}{\dd\rho^2} + (|\lambda|+1-\rho)\frac{\dd y}{\dd\rho}
- \frac{|\lambda|+1}{2} + \varkappa = 0.
\eeq
Its solution is the Kummer function \cite{AS}, which
provides a finite $f$ at $\rho\to\infty$
only under the condition that
\beq
 \varkappa - \frac{|\lambda|+1}{2} = n_r,
\label{Kummercutoff}
\eeq
where $n_r$ is non-negative integer, called \emph{radial quantum
number}. 
Recalling \req{varkappa}, we
obtain
\beq
E_{\perp} = \hbar\omc{}\left( n_r + \frac12
+\frac{|\lambda|-\sigma \lambda}{2} \right)
\label{Eperp}
\eeq
and
\beq
  y(\rho) = C L_{n_r}^{|\lambda|}(\rho),
\eeq
where $L_{n_r}^{|\lambda|}(\rho)$ is a
generalized Laguerre polynomial \cite{AS}
and $C$ is a normalization constant.
Therefore, the normalized solutions to \req{eigenproblem1} can be
written as
\beq
   \Phi_{n,s}^{(Z)}(\bm{r}_\perp) = 
    \frac{\exp(\ii \lambda\varphi)}{\sqrt{2\pi}\,\amZ{Z}}\,
     I_{n+s,n}\left(\frac{r_\perp^2}{2\amZ{Z}^2}\right),
\label{Phi-Landau}
\eeq
and
\beq
   E_{\perp,n}^{(Z)} = \hbar\omc{}\,\left(n+\frac12\right),
\eeq
where
\beq
s = \sigma \lambda,
\qquad
n=n_r+\frac{|\lambda|-s}{2}
\label{nLandau}
\eeq
and $I_{n'n}(\rho)$ ($n'=n+s$)
are the normalized Laguerre functions \cite{SokTer},
which are proportional to
$
\erm{-\rho/2}
    \rho^{(n'-n)/2} L_n^{n'-n}(\rho)
$
and are normalized so that
\beq
\int_0^\infty I_{n,s}^2(\rho)\,\dd\rho = 1.
\eeq
Explicitly,
\begin{subequations}
\bea
   I_{n',n}(\rho) &=& \erm{-\rho/2} \rho^{(n'-n)/2}
       \sum_{k=0}^{n} 
\frac{(-1)^k \,\sqrt{{n!}{n'!}}\,\rho^k}{k!\,(n-k)!\,(n'-n+k)!} ,
\nonumber\\&&
\qquad\mbox{if $n'\geq n$}\,;
\\
I_{n',n}(\rho) &=& (-1)^{n'-n}I_{n,n'}(\rho),
\quad
\mbox{if $n' < n$}\,.
\label{I-}
\eea
\label{I-explicit}
\end{subequations}
The number
$n$ in \req{nLandau} 
enumerates the Landau energy levels (e.g., Ref.~\cite{LaLi-QM}). Its
definition implies that, for a given Landau level, the  quantum number
$s$ is bounded from below:
\beq
s= -n, -n+1, -n=2, \ldots.
\eeq

The functions $\Phi_{n,s}^{(-1)}(\bm{r}_\perp)$, which describe
electron motion perpendicular to magnetic field, are often
called \emph{Landau functions}.
They satisfy the condition of orthogonality
\beq
\int_{\mathbb{R}^2}\dd\bm{r}_\perp\,
\Phi_{n_1 s_1}^{(Z)*}(\bm{r}_\perp)\Phi_{n_2 s_2}^{(Z)}(\bm{r}_\perp) =
\delta_{n_1 n_2}\delta_{s_1 s_2}
\eeq
and completeness
\beq
\sum_{n,s}
\Phi_{ns}^{(Z)*}(\bm{r}_{\perp,1})\Phi_{ns}^{(Z)}(\bm{r}_{\perp,2}) =
\delta(\bm{r}_{\perp,1}-\bm{r}_{\perp,2}),
\eeq
therefore they form a complete basis in a Hilbert space.

By construction, $\Phi_{ns}^{(Z)}(\bm{r}_\perp)$ is an eigenfunction of
the orbital momentum projection operator $l_z$ with eigenvalue
$\lambda=\sigma s$ and an eigenfunction of the squared transverse kinetic momentum
operator $\pi_\perp^2=2mH_\perp^{(1)}$ with eigenvalue
$m\hbar\omc{}(2n+1)$. Therefore, according to \req{lz}, it is also an
eigenfunction of the squared transverse pseudomomentum
operator $k^2$ with eigenvalue $m\hbar\omc{}(2\tilde{n}+1)$, where
$\tilde{n}=n+\sigma\lambda=n+s$.
We can identify the eigenstates using
the pair of numbers $(n,\tilde{n})$ instead of $(n,s)$ and accordingly to
define the Landau functions with modified subscripts
(e.g., \cite{Bez95,BP17}), 
\beq
   \mathcal{F}_{n,\tilde{n}}^{(Z)}(\bm{r}_\perp) =
     \Phi_{n,\tilde{n}-n}^{(Z)}(\bm{r}_{\perp})
     = \frac{\erm{\ii \sigma (\tilde{n}-n)\varphi}}{\sqrt{2\pi}\,\amZ{Z}}\,
     I_{\tilde{n},n}\left(\frac{r_\perp^2}{2\amZ{Z}^2}\right)
\label{F-Landau}
\eeq
($\sigma\equiv\mathrm{sign}\,Z$, $ \amZ{Z}\equiv\am/\sqrt{|Z|}$).
 From \req{I-} we see that
\beq
   \mathcal{F}_{n,\tilde{n}}^{(Z)}(\bm{r}_\perp)
= (-1)^{\tilde{n}-n} \mathcal{F}_{\tilde{n},n}^{(-Z)}(\bm{r}_\perp)
= \mathcal{F}_{\tilde{n},n}^{(-Z)}(-\bm{r}_\perp).
\label{F-}
\eeq

Let us define cyclic components of any vector $\bm{a}$ as
\beq
   a_{\pm1} = \frac{a_x\pm \ii a_y}{\sqrt{2}},
\quad
   a_0 =a_z.
\label{cyclic}
\eeq
The transverse cyclic components of the kinetic momentum operator and
of the pseudomomentum operator
transform one Landau state $|n,\tilde{n}\rangle_\perp$, 
characterized by $\mathcal{F}_{n\tilde{n}}^{(Z)}(\bm{r}_\perp)$, 
into another Landau
state:
\bea
   \pi_\alpha
   |n,\tilde{n}\rangle_\perp
    &\!=&\!
     -\ii\sigma\alpha \frac{\hbar}{\am}
      \sqrt{n+1/2-\sigma\alpha/2}
      |n-\sigma\alpha,\tilde{n}\rangle,
\qquad
\label{pi_cyclic}
\\
   k_\alpha\,|n,\tilde{n}\rangle_\perp
    &\!=&
     \ii\sigma\alpha\frac{\hbar}{\am}\,
      \sqrt{\tilde{n}+1/2+\sigma\alpha/2}\,|n,\tilde{n}+\sigma\alpha\rangle,
\label{k_cyclic}
\eea
where $\alpha=\pm1$ and $\sigma=\mathrm{sign}\,Z$.

\subsection{Two charged particles in uniform magnetic field}
\label{sect:2part}

Motion of two particles with masses $m_-$ and $m_+$ and
charges $-e$ and $Ze$ ($e>0$, $Z>1$) in a homogeneous magnetic field
$\bm{B}=(0,0,B)$ is governed by the Hamiltonian
\beq
H_\mathrm{0} = \frac{\pi_-^2}{2m_-} + \frac{\pi_+^2}{2m_+} + V_\mathrm{C}
 = \frac{p_{-,z}^2}{2m_-} + \frac{p_{+,z}^2}{2m_+} +
H_\perp +V_\mathrm{C},
\label{H0}
\eeq
where
\beq
  H_\perp=\frac{\pi_{-,\perp}^2}{2m_-}+\frac{\pi_{+,\perp}^2}{2m_+}
\label{Hperp1}
\eeq
determines the motion of the two non-interacting particles
according to \req{H1},
\beq
\pi_-=\bm{p}_- + \frac{e}{2c}\bm{B}\times\bm{r}_-,
\quad
\pi_+ = \bm{p}_+ - \frac{Ze}{2c}\bm{B}\times\bm{r}_+
\label{pi+-}
\eeq
are the kinetic momenta,
and 
\beq
  V_\mathrm{C} = -\frac{Ze^2}{|\bm{r}_- - \bm{r}_+|}
\eeq
is the Coulomb potential.

Longitudinal motion of the system as a whole can be factorized out by
introducing the $z$-coordinate of center-of-mass 
$Z=(m_+ z_+ + m_- z_-)/M$, where $M=m_+ + m_-$, and the
relative coordinate $z=z_--z_+$. The change of variables
$(z_-,z_+)\to(z_\mathrm{c.m.},z)$ implies
\bea
\frac{\partial}{\partial z_\mp} &=&
 \frac{\partial z_\mathrm{c.m.}}{\partial z_\mp}
    \frac{\partial}{\partial z_\mathrm{c.m.}} +
 \frac{\partial z}{\partial z_\mp}
    \frac{\partial}{\partial z} =
  \frac{m_\mp}{M} \frac{\partial}{\partial z_\mathrm{c.m.}}
    \pm \frac{\partial}{\partial z};
\nonumber\\
\frac{\partial^2}{\partial^2 z_\mp} &=&
  \left( \frac{m_\mp}{M} \right)^2
    \frac{\partial^2}{\partial z_\mathrm{c.m.}^2}
  \pm 2 \frac{m_\mp}{M}\,
    \frac{\partial^2}{\partial z_\mathrm{c.m.}\partial z}
  + \frac{\partial^2}{\partial z^2}.
\nonumber
\eea
Substituting it into the original Hamiltonian (\ref{H0}), we see
that this change of coordinates concerns only the first and second
terms,
\bea
-\frac{\hbar^2}{2m_-}\,\frac{\partial^2}{\partial z_-^2} -
  \frac{\hbar^2}{2m_+}\,\frac{\partial^2}{\partial z_+^2} &=&
  -\frac{\hbar^2}{2M}\,\frac{\partial^2}{\partial z_\mathrm{c.m.}^2}
    - \frac{\hbar^2}{2\mred}\,\frac{\partial^2}{\partial z^2}
\nonumber\\
 &=&
  \frac{P_z^2}{2M} + \frac{p_z^2}{2\mred},
\eea
where $p_z = -\ii\hbar\partial/\partial z$ and $P_z =
-\ii\hbar\partial/\partial z_\mathrm{c.m.}$ are the relative and
total longitudinal momenta, respectively, and 
\beq
   \mred=m_+m_-/M
\label{mred}
\eeq
is the reduced mass. Therefore we can fix the eigenvalue of the total
longitudinal momentum $P_z$ and write the total wave function and energy
as
\bea
\Psi_{P_z}(\bm{r}_{+,\perp},\bm{r}_{-,\perp}, z, z_\mathrm{c.m.}) &=&
\erm{\ii P_z z_\mathrm{c.m.}/\hbar}\,\psi(\bm{r}_{+,\perp},\bm{r}_{-,\perp},z),
\nonumber
\\
E_\mathrm{tot} &=& E + \frac{\hbar^2 P_z^2}{2M}.
\eea
To find $\psi(\bm{r}_{+,\perp},\bm{r}_{-,\perp},z)$,
it is sufficient to consider the system that does not
move along $\bm{B}$. Thus we will assume $P_z=0$ hereafter.

The wave function $\psi$ and energy $E$ satisfies 
the Schr\"odinger equation with the Hamiltonian
\beq
H=\frac{p_z^2}{2\mred} + H_\perp + V_\mathrm{C}.
\label{H2}
\eeq
As follows from Sect.~\ref{sect:1part}, 
the Landau functions 
$
 \Phi_{n_-s_-}^{(-1)}(\bm{r}_{-,\perp})=
 (-1)^{s_-} \, \Phi_{n_-+s_-,-s_-}^{(1)}(\bm{r}_{-,\perp})
\label{Phi+-}
$
 and
$\Phi_{n_+s_+}^{(Z)}(\bm{r}_{+,\perp})$ are eigenfunctions of the
operators $\pi_{-,\perp}^2$ and $\pi_{+,\perp}^2$, so that
\bea
   \pi_{-,\perp}^2 \,\Phi_{n_-s_-}^{(-1)}(\bm{r}_{-,\perp})
   &=& \frac{\hbar^2}{\am^2}\,(2n_- + 1)\,
     \Phi_{n_-s_-}^{(-1)}(\bm{r}_{-,\perp}),
\qquad
\\
   \pi_{+,\perp}^2 \,\Phi_{n_+s_+}^{(Z)}(\bm{r}_{+,\perp})
   &=& \frac{\hbar^2}{\amZ{Z}^2}\,(2n_+ + 1)\,
     \Phi_{n_+s_+}^{(Z)}(\bm{r}_{+,\perp}),
\qquad
\eea
where $n_\pm\geq0$ and $s_\pm\geq-n_\pm$ are integer quantum numbers.
Therefore, the transverse energy of two
non-interacting particles, corresponding to the Hamiltonian
$H_\perp$, \req{Hperp1}, is 
\beq
E_{n_-,n_+}^\perp = \hbar\omc{+}\left(n_++\frac12\right) +
\hbar\omc{-}\left(n_- +\frac12\right),
\label{Eperp2}
\eeq
where 
\beq
    \omc{-} = \frac{eB}{m_-c},
\qquad
    \omc{+} = \frac{ZeB}{m_+c}.
\eeq
Since the eigenenergy 
(\ref{Eperp2}) is degenerate with respect to quantum numbers $s_+$ and
$s_-$, there is a manifold of representations of eigenfunctions, one
of the simplest being
\beq
\psi_{n_+,s_+,n_-,s_-}(\bm{r}_{+,\perp},\bm{r}_{-,\perp}) =
 \Phi_{n_+s_+}^{(Z)}(\bm{r}_{+,\perp})
 \Phi_{n_-s_-}^{(-1)}(\bm{r}_{-,\perp}).
\label{psi4}
\eeq
In addition, the wave function (\ref{psi4}) is an eigenfunction of
operators ${k}_{-,\perp}^2$ and ${k}_{+,\perp}^2$, where 
\beq
\bm{k}_{-} = \bm{p}_{-} - \frac{e}{2c}\,\bm{B}\times\bm{r}_-,
\quad
\bm{k}_{+} = \bm{p}_{+} + \frac{Ze}{2c}\,\bm{B}\times\bm{r}_+
\eeq
are pseudomomentum operators, and an eigenfunction of the operators of
the $z$-projections of the angular momenta
$l_{-,z}$ and $l_{+,z}$. Therefore, any set of commuting operators
$(\pi_{-,\perp}^2,\pi_{+,\perp}^2,k_{-}^2,k_{+}^2)$,
$(\pi_{-,\perp}^2,\pi_{+,\perp}^2,k_{-}^2,l_{+,z})$,
$(\pi_{-,\perp}^2,\pi_{+,\perp}^2,l_{-,z},k_{+}^2)$,
or
$(\pi_{-,\perp}^2,\pi_{+,\perp}^2,l_{-,z},l_{+,z})$ can be used for
determination of the quantum numbers $n_+,s_+,n_-,s_-$.

However, the basis (\ref{psi4}) is not optimal, because the above
sets of operators do not commute with our Hamiltonian (\ref{H2}),
which means that the four quantum numbers $n_+,s_+,n_-,s_-$ are not
``good'' in the presence of the Coulomb potential $V_\mathrm{C}$. 

On the other hand, there are operators, which commute with each
other and with the Hamiltonian $H$: the total pseudomomentum
\beq
\bm{k}_\mathrm{tot} = \bm{k}_{-}+\bm{k}_{+},
\label{pseudomom}
\eeq
which is conserved due to translational invariance of $H$ accompanied
by the translational gauge transformations of $\bm{A}$ \cite{LaLi-QM},
and the longitudinal component of the total angular momentum
$
L_z= l_{-,z}+l_{+,z}
$
(which is conserved because the potential $V_\mathrm{C}$ is rotationally
invariant).
For a charged system, the transverse Cartesian components of $\bm{k}_\mathrm{tot}$
($k_{\mathrm{tot},x}$ and $k_{\mathrm{tot},y}$) do not
commute with each other and cannot be fixed
simultaneously. Instead of selecting one of them, we
may consider the square of pseudomomentum $k_{\mathrm{tot}}^2$.
Thus we look for common eigenfunctions of commuting operators
$\pi_{-,\perp}^2$, $\pi_{+\perp}^2$, ${k}_\mathrm{tot,\perp}^2$, and $L_z$, which will serve as a
basis, corresponding to two integrals of motion of our system, because
two operators of this set commute with the Hamiltonian.

Unlike the system of particles without external fields, the
center-of-mass
coordinates of the system of charged particles in a  magnetic field
cannot be completely eliminated from the Hamiltonian. When an external
magnetic field is present, the collective behaviors of a neutral system
of charged particles, such as an atom, and of a charged system, such as
an atomic ion, are very different. In the former case, the collective
motion is free whereas in the latter case, a cyclotron motion arises.
This difference appears clearly in the detailed mathematical study of
Avron et al. \cite{AvronHS}. For a neutral system, one can perform so
called pseudoseparation of the collective motion, after which
the resulting Hamiltonian for the internal degrees of freedom depends on
the eigenvalues of the collective pseudomomentum. For a charged system,
the number of integrals of motion is less than the number of degrees of
freedom, therefore the collective and individual coordinates and momenta cannot be
separated. Thus the set of operators ($K^2$, $L_z$) is not exclusively
associated with the collective motion but involves both the collective and internal
degrees of freedom.  

One can, however, perform an \emph{approximate} separation in the form
$H=H_1+H_2+H_3$, where Hamiltonians $H_1$ and $H_2$ describe the motion
of quasiparticles corresponding to the collective and internal degrees
of freedom, respectively, and operator $H_3$ couples the internal and
collective phase coordinates. An example of such approximate separation
was presented by Schmelcher \& Cederbaum
\cite{SchmelcherCederbaum91,SchmelcherCederbaum00}, who applied a canonical
transformation of variables to the Hamiltonian written in terms of the
center-of-mass and relative coordinates of the particles.
Baye and Vincke \cite{BayeVincke90}
developed a general framework for approximate separations. 
They introduced a parametrized approximate collective integral of
motion, which
in the two-particle case has the form
$
\bm{C}(\alpha_1) = \bm{k}_{\mathrm{tot},\perp}
 - (Z-1)(e/c)\bm{B}\times
 (\alpha_0\,\bm{r}_{+,\perp} + \alpha_1\,\bm{r}_{-,\perp}),
$
where $\alpha_0=1-\alpha_1$ and 
$\alpha_1$ is a free parameter. 

In the following we will assume that $m_+\gg m_-$.  Baye and Vincke
\cite{BayeVincke90} have shown that in this case any choice of
$\alpha_1$ will provide the approximate separation as long as
$\alpha_1=O(m_-/m_+)$. The simplest choice $\alpha_1=0$ has
been introduced by Baye \cite{Baye82} and successfully used by Baye and
Vincke \cite{BayeVincke86} in the calculations of the collective motion corrections for
atomic ions.

However, since the separation of the collective motion is only
approximate, there is no particular advantage in selecting the center of
mass for describing the coordinate of the quasiparticle corresponding to
the collective motion. An equally reasonable choice can be just the
coordinate of the heavy particle (the nucleus). Using the latter choice,
Bezchastnov \cite{Bez95} derived a basis of eigenstates of the
transverse Hamiltonian $H_\perp$, whose elements are also eigenstates of
squared total pseudomomentum $K^2$, total $z$-projection of the angular
momentum $L_z$, and squared kinetic momenta of each particle,
$\pi_{+,\perp}^2$ and $\pi_{-,\perp}^2$. In Sect.\,\ref{sect:basis} we
present a more detailed and physically transparent derivation of the
same basis with some corrections.

\section{Transverse basis}
\label{sect:basis}

\subsection{Canonical transformations}
\label{sect:varichange}

Let us pass from variables $(\bm{r}_{+,\perp},\bm{r}_{-,\perp})$ to variables 
$(\bm{R}_{\perp},\bm{r}_\perp)$, where 
\beq
   \bm{R}_{\perp}=\bm{r}_{+,\perp},
\quad
\bm{r}_\perp = \bm{r}_{-,\perp} - \bm{r}_{+,\perp}.
\label{change_r}
\eeq
Hereafter we will also use the three-dimensional vectors $\bm{R}$ and
$\bm{r}$, assuming the $z$-components $R_z=0$ and $r_z = z$, so that
$\bm{r} = \bm{r}_- - \bm{r}_+$.
The canonical momenta transform as
\beq
\bm{p}_{+} = \bm{P} - \bm{p},
\quad
\bm{p}_{-} = \bm{p}.
\label{change_p}
\eeq
In the new variables, the transverse Hamiltonian (\ref{Hperp1})
becomes
\bea
H_\perp &=& \frac{1}{2m_+}\left( \bm{P}_{\perp} - \bm{p}_\perp -
\frac{Ze}{2c}\,\bm{B}\times\bm{R} \right)^2
\nonumber\\
&+&
\frac{1}{2m_-}\,\left( \bm{p}_\perp + \frac{e}{2c}\,\bm{B}\times\bm{r} +
\frac{e}{2c}\,\bm{B}\times\bm{R} \right)^2.
\eea
The total pseudomomentum
[\req{pseudomom}] becomes
\beq
\bm{k}_\mathrm{tot}= \bm{P} +
\frac{(Z-1)e}{2c}\,\bm{B}\times\bm{R} -
\frac{e}{2c}\,\bm{B}\times\bm{r}.
\label{P0new}
\eeq

We look for a unitary transformation, which will allow us
to separate motion of quasiparticles with the total
charge of the system, $(Z-1)e$, and with the electron charge, $-e$.
 For this aim, we should transform
the first bracket so as to add $(e/2c)\bm{B}\times\bm{R}$ in it, and the
second bracket so as to subtract $(e/2c)\bm{B}\times\bm{R}$.
The unitary
transformation operator can be written
in the form $U=\exp(-\ii\phi)$, where $\phi$
is a Hermitian operator to be determined. The wave function is
transformed as $\psi=U\psi'$, $\psi'=U^\dag\psi$, and the Hamiltonian
as $H=UH'U^\dag$, $H'=U^\dag HU$ (symbol $U^\dag$ denotes the Hermitian
adjoint to $U$). It is easy to check that for our purpose we can take
\bea
U &=& \exp\left(
-\frac{\ii e}{2\hbar c}\,(\bm{B}\times\bm{r})\cdot\bm{R}
 \right)
\nonumber\\
 &=&
 \exp\left(
-\frac{\ii e}{2\hbar
c}\,(\bm{B}\times\bm{r}_{-})\cdot\bm{r}_{+} 
 \right)
\nonumber\\
 &=& \exp\left(\frac{\ii e}{2\hbar
c}\,(\bm{B}\times\bm{r}_{+})\cdot\bm{r}_{-}
\right).
\label{Utransform}
\eea
This operator performs a gauge transformation and shifts
the momenta:
\beq
U^\dag\bm{P} U = \bm{P} + \frac{e}{2c}\,\bm{B}\times\bm{r},
\qquad
U^\dag\bm{p} U = \bm{p} - \frac{e}{2c}\,\bm{B}\times\bm{R}.
\label{Unishift}
\eeq
This shift cancels the
third term in \req{P0new}, and the total 
pseudomomentum becomes
\beq
\bm{K} = U^\dag\bm{k}_\mathrm{tot} U =
  \bm{P} + \frac{(Z-1)e}{2c}\,\bm{B}\times\bm{R} .
\label{P0'}
\eeq
The total angular momentum
\beq
\bm{L} = \bm{l}_{-} + \bm{l}_{+}
 = \bm{r}_{+}\times\bm{p}_{+} + \bm{r}_{-}\times\bm{p}_{-}
 = \bm{R}\times\bm{P} + \bm{r}\times\bm{p}
\label{L}
\eeq
retains its form after the transformation:
$
  U^\dag \bm{L} U = \bm{L}.
$

The operators of kinetic momenta of the nucleus and the electron become
respectively
\beq
   U^\dag \bm{\pi}_+ U = \bm{\Pi} - \bm{k},
\qquad
   U^\dag \bm{\pi}_- U = \bm{\pi}.
\label{pi+-'}
\eeq
where
\bea
   \bm{\Pi} = \bm{P} -
   \frac{(Z-1)e}{2c}\,\bm{B}\times\bm{R},
&&
   \bm{\pi} = \bm{p} + \frac{e}{2c}\,\bm{B}\times\bm{r},
\nonumber\\
   \bm{k} = \bm{p} - \frac{e}{2c}\,\bm{B}\times\bm{r}.
\eea
Therefore the transformed transverse Hamiltonian equals
\beq
H_\perp' = U^\dag H_\perp U = \frac{1}{2m_+}\,
\,(\bm{\Pi}_{\perp}
- 
\bm{k}_\perp )^2
+
\frac{\pi_\perp^2}{2m_-}\,.
\label{Hperp'}
\eeq
We will consider $\bm{\Pi}$ and $\bm{\pi}$
as kinetic momenta of
quasiparticles with charges $(Z-1)e$ and $(-e)$, and
$\bm{k}$ as a pseudomomentum of the latter  quasiparticle.
By analogy with \req{lz}, the $z$-projection of 
the total angular momentum [\req{L}] equals
\beq
L_z = \frac{c}{2(Z-1)\,eB}\,(K_\perp^2 - \Pi_\perp^2 )
     + \frac{c}{2eB}\,(\pi_\perp^2 - k_\perp^2 ).
\label{Lz}
\eeq
Expanding the brackets in \req{Hperp'} and substituting $k_\perp^2 =
\pi_\perp^2 - (2eB/c)\,l_z$, we obtain
\beq
H_\perp' = H_1' + H_2' + H_3',
\label{H123'}
\eeq
where
\bea&&
H_1' = \frac{\Pi_\perp^2}{2m_+},
\quad
H_2' = \frac{\pi_\perp^2}{2\mred},
\quad
\label{H1'H2'}
\\&&
H_3' = \! - \frac{1}{m_+}\bm{\Pi}_\perp
\cdot
\bm{k}_\perp
\! 
- \frac{eB}{m_+ c}\,l_z.
\label{H3'}
\eea
Here, we have introduced the $z$-projection $l_z$ of the angular
momentum of the second (negative) quasiparticle by analogy with
\req{lz}. The term $m_+^{-1}\bm{\Pi}_\perp \cdot \bm{k}_\perp$ in $H_3'$
couples together the motion of the two quasi-particles.

Equivalently, $\bm{k}$ and $\bm\pi$ might be considered as the kinetic
momentum and pseudomomentum of a quasiparticle with charge $+e$.
Accordingly, the transverse Hamiltonian can be rewritten as
\beq
H_\perp' = H_1'+H_2''+H_3'',
\label{H123''}
\eeq
 where $H_1'$ is the same as in \req{H1'H2'} and
\beq
H_2'' = \frac{k_\perp^2}{2\mred},
\quad
H_3'' = - \frac{1}{m_+}\bm{\Pi}_\perp
\cdot
\bm{k}_\perp
+ \frac{eB}{m_-c}\,l_z.
\eeq 
Comparing the last formula with \req{H3'}, we see that $H_3'' = H_3' +
({eB}/{\mred c})\,l_z$. The latter decomposition (\ref{H123''}) was used
in Ref.\,\cite{Bez95}. We prefer to use the former decomposition,
\req{H123'}, because $H_3'\to0$ at $m_+\to\infty$, ensuring asymptotic
decoupling for massive ions.

Operators $H_1'$ and $H_2'$ ($H_2''$) have the form of the Hamiltonian of free
quasiparticles with charges $(Z-1)e$ and $e$ $(-e)$ and masses $m_+$ and
$\mred$, respectively.  and $H_3'$ ($H_3''$) couples them together. According to
Sect.~\ref{sect:1part}, the eigenfunctions of $H_1'$ are
$\Phi_{N,S}^{(Z-1)}(\bm{R}_{\perp})$, $N\geq0$, $S \geq -N$, and the
eigenenergies are independent of $S$,
\beq
 E_{1,N} = \hbar\Omc\left( N+\frac12 \right),
\eeq
where
\beq
\Omc = \frac{(Z-1)eB}{m_+ c} = \frac{Z-1}{Z}\,\omc{+}.
\eeq
Analogously,  the eigenfunctions of $H_2'$ are
$\Phi_{n,s}^{(1)}(\bm{r}_\perp)$, $n\geq0$, $s \geq -n$, and the
eigenenergies are independent of $s$,
\beq
 E_{2,n} = \hbar\omcmu\left( n+\frac12 \right),
\eeq
where
\beq
\omcmu = \frac{eB}{\mred c} = \left( 1+\frac{m_-}{m_+} \right) \omc{-}.
\eeq

\subsection{Creation and annihilation operators}
\label{sect:operators}

Instead of $S$ and $s$, it is sometimes convenient to use quantum
numbers $\tilde{N}=N+S$ and $\tilde{n}=n+s$. As follows from
\req{Phi-Landau}, an interchange of $n$ with $\tilde{n}$ or $N$ with
$\tilde{N}$ does not affect the modulus of a single-particle
eigenfunction. We will use these quantum numbers to specify quantum
states $|N,\tilde{N}\rangle_1$ and $|n,\tilde{n}\rangle_2$ of the two
quasiparticles, described by Hamiltonians $H_1'$ and $H_2'$,
respectively, in the $(xy)$-plane.

Let us consider the cyclic components (\ref{cyclic}) of the operators
of the kinetic momenta and pseudomomenta of the quasiparticles.
According to Eqs.\,(\ref{pi_cyclic}) and (\ref{k_cyclic}),
the operators
\begin{subequations}
\bea
\hat{a} = \ii\frac{\am}{\hbar}\,\pi_{-1},
&&
\hat{\tilde{a}} =  -\ii\frac{\am}{\hbar} \,k_{+1},
\\
\hat{b} = \ii\frac{\amZ{Z-1}}{\hbar}\,\Pi_{+1},
&&
\hat{\tilde{b}} = -\ii\frac{\amZ{Z-1}}{\hbar}\,K_{-1}
\eea
\label{aa'bb'def}
\end{subequations}
 lower the quantum numbers
$n,\tilde{n},N,\tilde{N}$ by one, as follows:
\begin{subequations}
\bea
 \hat{a}\,|n,\tilde{n}\rangle_2 &=& \sqrt{n}\,|n-1,\tilde{n}\rangle_2,
\label{aa'1}
\\
  \hat{\tilde{a}}\,|n,\tilde{n}\rangle_2 &=& \sqrt{\tilde{n}}\,|n,\tilde{n}-1\rangle_2,
\label{aa'2}
\\
 \hat{b}\,|N,\tilde{N}\rangle_1 &=& \sqrt{N}\,|N-1,\tilde{N}\rangle_1,
\label{bb'1}
\\
 \hat{\tilde{b}}\,|N,\tilde{N}\rangle_1 &=& \sqrt{\tilde{N}}\,|N,\tilde{N}-1\rangle_1.
\label{bb'2}
\eea
\label{aa'bb'}
\end{subequations}
Their Hermitian adjoint operators
\begin{subequations}
\bea
\hat{a}^\dag = -\ii\frac{\am}{\hbar}\,\pi_{+1},
&&
\hat{\tilde{a}}^\dag =  \ii\frac{\am}{\hbar} \,k_{-1},
\\
\hat{b}^\dag = -\ii\frac{\amZ{Z-1}}{\hbar}\,\Pi_{-1},
&&
\hat{\tilde{b}}^\dag = \ii\frac{\amZ{Z-1}}{\hbar}\,K_{+1}
\eea
\label{a+a+'b+b+'}
\end{subequations}
raise the quantum numbers
$n,\tilde{n},N,\tilde{N}$ by one:
\begin{subequations}
\bea
 \hat{a}^\dag\,|n,\tilde{n}\rangle_2 &=& \sqrt{n+1}\,|n+1,\tilde{n}\rangle_2,
\label{aa'+1}
\\
  \hat{\tilde{a}}^\dag\,|n,\tilde{n}\rangle_2 &=& \sqrt{\tilde{n}+1}\,|n,\tilde{n}+1\rangle_2,
\label{aa'+2}
\\
 \hat{b}^\dag\,|N,\tilde{N}\rangle_1 &=& \sqrt{N+1}\,|N+1,\tilde{N}\rangle_1,
\\
\label{bb'+1}
 \hat{\tilde{b}}^\dag\,|N,\tilde{N}\rangle_1 &=& \sqrt{\tilde{N}+1}\,|N,\tilde{N}+1\rangle_1.
\label{b'dag}
\label{bb'+2}
\eea
\label{aa'+bb'+}
\end{subequations}
As far as  $\hat{a},\hat{\tilde{a}},\hat{b},\hat{\tilde{b}}$ can be
considered as annihilation operators for excitations in
$n,\tilde{n},N,\tilde{N}$, their Hermitian adjoint operators can be
considered as the creation operators.

It is noteworthy that
\bea
   \pi_{\pm1}&=&p_{\pm1} \pm \frac{\ii e B}{2c}\,r_{\pm1},
\\
   k_{\pm1}&=&p_{\pm1} \mp \frac{\ii e B}{2c}\,r_{\pm1},
\\
   \Pi_{\pm1}&=&P_{\pm1} \mp \frac{\ii (Z-1)e B}{2c}\,R_{\pm1},
\\
   K_{\pm1}&=&P_{\pm1} \pm \frac{\ii (Z-1)e B}{2c}\,R_{\pm1}.
\eea
Therefore,
\bea
   r_{\pm1} &=& \pm \frac{\ii c}{eB}\,(k_{\pm1}-\pi_{\pm1}),
\\
   R_{\pm1} &=& \mp \frac{\ii c}{(Z-1)eB}\,(K_{\pm1}-\Pi_{\pm1}),
\eea
so that 
\begin{subequations}
\bea
   r_{+1} = \am\,(\hat{a}^\dag-\hat{\tilde{a}}),
&&
   r_{-1} = \am\,(\hat{a}-\hat{\tilde{a}}^\dag),
\\
   R_{+1} = \amZ{Z-1}\,(\hat{b}-\hat{\tilde{b}}^\dag),
&&
   R_{-1} = \amZ{Z-1}\,(\hat{b}^\dag-\hat{\tilde{b}}).
\qquad
\eea
\label{rRpm}
\end{subequations}

It is also useful to consider the circular components of the kinetic
momentum of the nucleus $\bm{\pi}_+$. According to \req{pi_cyclic}, these
operators change the nucleus Landau number $n_+$ by $\pm1$:
\begin{subequations}
\bea
   \pi_{+,+1}\,|n_+, \tilde{n}_+ \rangle &=& 
     - \, \frac{\ii\hbar}{\amZ{Z}}\,\sqrt{n_+}\,
       |n_+ -1, \tilde{n}_+ \rangle_\perp,
\\
   \pi_{+,-1}\,|n_+, \tilde{n}_+ \rangle &=& 
     \frac{\ii\hbar}{\amZ{Z}}\,\sqrt{n_+ + 1}\,
       |n_+ +1, \tilde{n}_+ \rangle_\perp.
\qquad
\eea
\label{pi+cyclic}
\end{subequations}
The canonical transformation of these operators with account of
\req{pi+-'} and Eqs.~(\ref{aa'bb'def}), (\ref{a+a+'b+b+'}) gives
\begin{subequations}
\bea
   U^\dag \pi_{+,+1} U &\!=&\! \Pi_{+1} - k_{+1} \!=\!
       -\, \frac{\ii\hbar}{\am}\,
       \left(\sqrt{Z-1}\,\,\hat{b} + \hat{\tilde{a}} \right),
\\
   U^\dag \pi_{+,-1} U &\!=&\! \Pi_{-1} - k_{-1} \!=\!
       \frac{\ii\hbar}{\am}\,
       \left(\sqrt{Z-1}\,\,\hat{b}^\dag + \hat{\tilde{a}}^\dag \right).
\qquad
\eea
\label{pi+ab}
\end{subequations}

\subsection{Good quantum numbers}
\label{sect:qnum}

The effective quasiparticle Hamiltonians
can be written in terms of the creation and annihilation operators
(Sect.~\ref{sect:operators}) as
\beq
H_1' = \hbar\Omc\left( \hat{b}^\dag \hat{b} + \frac12 \right),
\quad
H_2' = \hbar\omcmu
\left(  \hat{a}^\dag \hat{a} + \frac12 \right).
\label{H12'}
\eeq
Using the expressions
$l_z = \hbar\,(\hat{\tilde{a}}^\dag \hat{\tilde{a}}
 - \hat{\tilde{a}}^\dag \hat{\tilde{a}})$
 and
\beq
 \bm{\Pi}_\perp\cdot\bm{k}_\perp  
  = - \frac{\hbar^2}{\am^2}\,\sqrt{Z-1}
  \,(\hat{\tilde{a}}^\dag \hat{b} + \hat{\tilde{a}} \hat{b}^\dag ),
\label{Pik}
\eeq
we can write the coupling operator $H_3'$ in \req{H3'} as
\beq
H_3' =
 \frac{\hbar\Omc}{\sqrt{Z-1}}\,(\hat{\tilde{a}}^\dag \hat{b}
  + \hat{b}^\dag \hat{\tilde{a}} )
+
\hbar\omc{+}\,
(\hat{\tilde{a}}^\dag \hat{\tilde{a}} - \hat{a}^\dag \hat{a}) .
\label{H_3'}
\eeq

Equations (\ref{H12'}) and (\ref{H_3'})
do not contain operators $b'$ and $b'^\dag$, which
determine the square of total transverse pseudomomentum
\beq
 K_\perp^2 =
\frac{2\hbar^2}{\am^2}\,(Z-1)\left( \hat{\tilde{b}}^\dag \hat{\tilde{b}}
+ \frac12 \right),
\eeq
which confirms that $K_\perp^2$ is an
integral of motion ($[H_\perp',K_\perp^2]=0$) and the related
quantum number $\tilde{N} = \langle \hat{\tilde{b}}^\dag \hat{\tilde{b}} \rangle$ is
conserved. In other words, $\tilde{N}$ is a good quantum number.

For the $z$-projection of the total angular momentum,
\req{Lz} gives
\beq
L_z = \hbar \,(\hat{\tilde{b}}^\dag \hat{\tilde{b}}
 - \hat{b}^\dag \hat{b}
  + \hat{a}^\dag \hat{a} - \hat{\tilde{a}}^\dag \hat{\tilde{a}}).
\label{L_z'}
\eeq
The eigenvalues of $L_z$ equal $\hbar L=\hbar(S-s)$. It is easy to
check that $[H_\perp',L_z']=0$. Therefore, $L$ is a good quantum
number. From \req{L_z'} we see
that
\beq
   L=\tilde{N}-N-\tilde{n}+n.
\label{LN'Nn'n}
\eeq

Finally, using Eqs.\,(\ref{H12'}), (\ref{H_3'}), and the expression
\beq
\pi_\perp^2 = 
\frac{\hbar^2}{\am^2}\,(2 \hat{a}^\dag \hat{a} + 1 ),
\label{p0'a}
\eeq
we can check that $[\pi_\perp^2,H_\perp']=0$.
It means that $n=\langle a^\dag a\rangle$
 is a good quantum number (as long as we disregard
$V_\mathrm{C}$).

\subsection{Transverse basis states}
\label{sect:states}

The results of Sect.~\ref{sect:qnum} allow us to consider
eigenstates of the transverse Hamiltonian $H_\perp'$ with fixed numbers
$\tilde{N},n,L$. On the other hand,  the quantum numbers $S$ and
$s=S-L$, or equivalently $N=\tilde{N}-S$ and $\tilde{n}=n+s$ are not
well defined,  because $l_z$ does not commute with $H_3'$. Since the
eigenfunctions 
$
   \mathcal{F}_{N,\tilde{N}}^{(Z-1)}(\bm{R}_{\perp})\, 
    \mathcal{F}_{n,\tilde{n}}^{(-1)}(\bm{r}_\perp)
$
of the states 
$|N,\tilde{N},n,\tilde{n}\rangle_\perp
=|N,\tilde{N}\rangle_1\otimes|n,\tilde{n}\rangle_2$ with fixed
$\tilde{N}$ and $n$
(recall that $|N,\tilde{N}\rangle_1$ and $|n,\tilde{n}\rangle_2$
are the eigenstates of $H_1'$ and $H_2'$, respectively)
form a complete basis in the Hilbert space of functions of
$(\bm{R}_{\perp},\bm{r}_\perp)$, the eigenstates of $H_\perp'$ can be
looked as superpositions of states
$|N,\tilde{N},n,\tilde{n}\rangle_\perp$ with different
$N$ and
$\tilde{n}$. Taking into account the constraint
$N+\tilde{n}=\mathcal{N}$, where
$\mathcal{N}\equiv
\tilde{N}-L+n$, we can write the eigenfunction $\Psi' = U\Psi$ of
$H_\perp'$ as
\beq
\Psi_{\tilde{N},n,L}'(\bm{R}_{\perp},\bm{r}_\perp) = 
\sum_{\tilde{n}=0}^\mathcal{N}
C_{\tilde{n}}\,\mathcal{F}_{\mathcal{N}-\tilde{n},\tilde{N}}^{(Z-1)}(\bm{R}_{\perp})
\mathcal{F}_{n,\tilde{n}}^{(-1)}(\bm{r}_\perp),
\label{Psi'}
\eeq
where $C_{\tilde{n}}$ are some constants, which may depend on
 $\tilde{N},n,L$. The
eigenfunctions of the initial Hamiltonian $H_\perp$ are
\begin{widetext}
\beq
\Psi_{\tilde{N},n,L}(\bm{r}_{+,\perp},\bm{r}_{-,\perp}) = 
U
\sum_{\tilde{n}=0}^\mathcal{N}
C_{\tilde{n}}\,\mathcal{F}_{\mathcal{N}-\tilde{n},\tilde{N}}^{(Z-1)}(\bm{r}_{+,\perp})\,
\mathcal{F}_{n,\tilde{n}}^{(-1)}(\bm{r}_{-,\perp} - \bm{r}_{+,\perp}),
\label{Psi}
\eeq
\end{widetext}
where $U$ is given by \req{Utransform}.

By construction, 
$\Psi_{\tilde{N},n,L}(\bm{r}_{+,\perp},\bm{r}_{-,\perp})$ is an
eigenfunction of $L_z$ and $K^2$ for any coefficients
$C_{\tilde{n}}$. Let us consider its transformation under the action
of operators $\pi_{\pm,\perp}^2$ [\req{pi+-}]. 
Equation~(\ref{p0'a}) gives
\beq
\pi_\perp^2\,|N,\tilde{N},n,\tilde{n}\rangle = 
\frac{\hbar^2}{\am^2} \left( 2n+ 1
\right) |N,\tilde{N},n,\tilde{n} \rangle.
\eeq
On the other hand, according to \req{pi+-'},
 $\pi_\perp^2=U^\dag\pi_-^2 U$. Therefore,
$\Psi_{\tilde{N},n,L}(\bm{r}_{+,\perp},\bm{r}_{-,\perp})$ 
is an eigenfunction of $\pi_{-,\perp}^2$ with the appropriate
eigenvalue $(\hbar/\am)^2 (2n+1)$ for any set of $C_{\tilde{n}}$, which
means that $n_- = n$. Therefore, we can write
\beq
   \mathcal{N} = \tilde{N}-L+n_-
\label{calN}
\eeq

The operator $\pi_{+,_\perp}^2$, being transformed according to \req{pi+-'}, can
be expressed using \req{Pik} as
\begin{widetext}
\beq
(\bm{\Pi}_\perp - \bm{k}_\perp)^2 = \frac{\hbar^2}{\am^2}\,\left[
2(Z-1)\left(\hat{b}^\dag \hat{b} + 1 \right)
 + 2\hat{\tilde{a}}^\dag \hat{\tilde{a}} + 1 +
2\sqrt{Z-1}\,(\hat{\tilde{a}}^\dag \hat{b}
 + \hat{b}^\dag \hat{\tilde{a}}) \right].
\label{Pimk2}
\eeq
From this equation, 
taking into account that $N=\mathcal{N}-\tilde{n}$, we obtain
\bea
(\bm{\Pi}_\perp - \bm{k}_\perp)^2 
\,|N,\tilde{N},n,\tilde{n}\rangle_\perp
&=&
\frac{2\hbar^2}{\am^2}\bigg[
\left( (Z-1)\mathcal{N} - (Z-2) \tilde{n} + \frac{Z}{2}\right)\,
|\mathcal{N}-\tilde{n},\tilde{N},n,\tilde{n}\rangle_\perp
\nonumber\\&&
+ \sqrt{Z-1} \sqrt{(\mathcal{N}-\tilde{n})(\tilde{n}+1)}
|\mathcal{N}-\tilde{n}-1,\tilde{N},n,\tilde{n}+1 \rangle_\perp
\nonumber\\&&
+ \sqrt{Z-1} \sqrt{(\mathcal{N}-\tilde{n}+1)\,\tilde{n}}
|\mathcal{N}-n+1,\tilde{N},n,\tilde{n}-1\rangle_\perp
\bigg].
\eea
Using this relation with \req{Psi'} and
changing the summation index $\tilde{n}$ so as to collect together the 
homogeneous terms with 
$\mathcal{F}_{\mathcal{N}-\tilde{n},\tilde{N}}^{(Z-1)}(\bm{R}_{\perp})
\mathcal{F}_{n,\tilde{n}}^{(-1)}(\bm{r}_\perp)$, 
we obtain
\bea
  (\bm{\Pi}_\perp - \bm{k}_\perp)^2
   \Psi_{\tilde{N},n,L}'(\bm{R}_{\perp},\bm{r}_\perp)
  & = & \frac{2\hbar^2}{\am^2}
   \sum_{\tilde{n}=0}^\mathcal{N}
\bigg[
\left( (Z-1)\mathcal{N} - (Z-2) \tilde{n} + \frac{Z}{2}\right)
\, C_{\tilde{n}}
\nonumber\\&&
+ \sqrt{Z-1} \sqrt{(\mathcal{N}-\tilde{n}+1)\,\tilde{n}}
\, C_{\tilde{n}-1}
\nonumber\\&&
+ \sqrt{Z-1} \sqrt{(\mathcal{N}-\tilde{n})(\tilde{n}+1)}
\, C_{\tilde{n}+1}
\bigg]\,
\mathcal{F}_{\mathcal{N}-\tilde{n},\tilde{N}}^{(Z-1)}(\bm{R}_{\perp})
\mathcal{F}_{n,\tilde{n}}^{(-1)}(\bm{r}_\perp).
\label{Psi''}
\eea
Comparing \req{Psi''} with \req{Psi'}, we see that
$\Psi_{\tilde{N},n,L}'(\bm{R}_{\perp},\bm{r}_\perp)$
will be an eigenfunction of $(\bm{\Pi}_\perp - \bm{k}_\perp)^2=
U^\dag\pi_{+,\perp}^2 U$
under the condition that the coefficients $C_{\tilde{n}}$ satisfy
the relation \footnote{Equation~(\ref{recurC})
is similar to Eq.~(32) in Ref.~\cite{Bez95}, but with the opposite sign
on the right-hand side.}
\beq
\sqrt{Z-1}\left[
\sqrt{(\mathcal{N}-\tilde{n}+1)\tilde{n}}\,C_{\tilde{n}-1}+
\sqrt{(\mathcal{N}-\tilde{n})(\tilde{n}+1)}\,C_{\tilde{n}+1}
\right]
=
 \big[(Z-2)\,\tilde{n}  - (Z-1)\mathcal{N}+ Z n_+ \big]\,C_{\tilde{n}},
\label{recurC}
\eeq
where $n_+ = 0,1,2,\ldots,\mathcal{N}$.
Imposing this relation, we see from \req{Psi''} that 
\beq
\pi_{+,\perp}^2
 \Psi_{\tilde{N},n,L}(\bm{r}_{+,\perp},\bm{r}_{-,\perp})
=
\frac{2\hbar^2}{\amZ{Z}^2}\left( n_+ + \frac12 \right)
\Psi_{\tilde{N},n,L}(\bm{r}_{+,\perp},\bm{r}_{-,\perp}),
\label{pi+2X}
\eeq
\end{widetext}
Each value of $n_+$ corresponds to an eigenvector
$\{C_{\tilde{n}}\}$ ($\tilde{n}=0,1,\ldots,\mathcal{N}$).
 We see that the numbers
$\tilde{N}$, $n=n_-$, and $L$ affect the
eigenvalue problem only in combination $\mathcal{N}=\tilde{N}+n_--L$.
 Therefore each
eigenvector may be marked by only two numbers $\mathcal{N}$ and $n_+$.
These eigenvectors are orthonormal,
\beq
\sum_{k=0}^\mathcal{N}
C_{k}^{(\mathcal{N},n_+')}C_{k}^{(\mathcal{N},n_+)}
 = \delta_{n_+',n_+},
\label{orthoC}
\eeq
and satisfy the completeness condition,
\beq
\sum_{n_+=0}^\mathcal{N} C_{k'}^{(\mathcal{N},n_+)}
C_{k}^{(\mathcal{N},n_+)} = \delta_{k'k},
\eeq
therefore they form an orthonormal basis in a $(\mathcal{N}+1)$-dimensional
vector space.

Thus we have built a basis of eigenstates of four operators
 $K_\perp^2$, $L_z$, $\pi_+^2$, and
$\pi_-^2$, such that
\beq
  |\tilde{N},L,n_-,n_+\rangle_{\perp,0} =
\sum_{{k}=0}^\mathcal{N} C_{{k}}^{(\mathcal{N},n_+)}
\, |\mathcal{N}-{k},\tilde{N},n_-,k \rangle_\perp ,
\label{bastat}
\eeq  
described by wave functions
\begin{widetext}
\beq
\Psi_{\tilde{N},L,n_-,n_+}(\bm{r}_{+,\perp},\bm{r}_{-,\perp}) = 
U^\dag
\sum_{{k}=0}^\mathcal{N} C_{{k}}^{(\mathcal{N},n_+)}
\,\mathcal{F}_{\mathcal{N}-{k},\tilde{N}}^{(Z-1)}(\bm{r}_{+,\perp})
\,\mathcal{F}_{n_-,{k}}^{(-1)}(\bm{r}_{-,\perp}-\bm{r}_{+,\perp}).
\label{Psi-sum}
\eeq
They are characterized by 4 discrete
quantum numbers, related to the four degrees of freedom
for motion of the 2 charged particles perpendicular to the magnetic
field:
\bea&&
   K_\perp^2\, |\tilde{N},L,n_-,n_+\rangle_{\perp,0}
     = \frac{\hbar^2}{\amZ{Z-1}^2}\,(2\tilde{N}+1)\,
       |\tilde{N},L,n_-,n_+\rangle_{\perp,0},
\\&&
   L_z\,|\tilde{N},L,n_-,n_+\rangle_{\perp,0} = 
     \hbar L\,|\tilde{N},L,n_-,n_+\rangle_{\perp,0},
\\&&
   \pi_-^2\,|\tilde{N},L,n_-,n_+\rangle_{\perp,0} =
      \frac{\hbar^2}{\am^2}\,(2n_- +1)
        \,|\tilde{N},L,n_-,n_+\rangle_{\perp,0},
\label{pi-2}
\\&&
      \pi_+^2\,|\tilde{N},L,n_-,n_+\rangle_{\perp,0} =
      \frac{\hbar^2}{\amZ{Z}^2}\,(2n_+ +1)
        \,|\tilde{N},L,n_-,n_+\rangle_{\perp,0},
\label{pi+2}
\\&&
   \tilde{N} \geq0,
\quad
   n_- \geq 0,
\quad
   L \leq \tilde{N}+n_-,
\quad
   0 \leq n_+ \leq \mathcal{N}=\tilde{N}+n_- - L.
\eea
According to Eqs.~(\ref{Hperp1}), (\ref{pi-2}), and (\ref{pi+2}),
\beq
   H_\perp\,|\tilde{N},L,n_-,n_+\rangle_{\perp,0} =
    E_{n_-,n_+}^\perp\,|\tilde{N},L,n_-,n_+\rangle_{\perp,0},
\label{HperpE}
\eeq
where $E_{n_-,n_+}^\perp$ is defined by \req{Eperp2}.
This basis is orthonormal,
\beq
\langle \tilde{N}',L',n_-',n_+'|\tilde{N},L,n_-,n_+\rangle_{\perp,0} = 
\delta_{\tilde{N}'\tilde{N}}\delta_{L'L}\delta_{n_-'n_-}\delta_{n_+'n_+}
\label{normPsi}
\eeq
and complete,
\beq
\sum_{\tilde{N},L,n_-,n_+}
\Psi_{\tilde{N},L,n_-,n_+}^*(\bm{r}_{+,\perp}',\bm{r}_{-,\perp}')
 \Psi_{\tilde{N},L,n_-,n_+}(\bm{r}_{+,\perp},\bm{r}_{-,\perp})
= \delta(\bm{r}_{+,\perp}' -
\bm{r}_{+,\perp})\,\delta(\bm{r}_{-,\perp}' - \bm{r}_{-,\perp}).
\label{completePsi}
\eeq

Using \req{F-}, one can rewrite \req{Psi-sum} in the form
\footnote{The factor $(-1)^{n_--\tilde{n}}$ 
in \req{Psi_alt} is
responsible for the opposite sign
in \req{recurC} compared to the analogous relation in
Ref.\,\cite{Bez95}.}
\beq
\Psi_{\tilde{N},L,n_-,n_+}(\bm{r}_{+,\perp},\bm{r}_{-,\perp}) = 
U^\dag
\sum_{\tilde{n}=0}^\mathcal{N} (-1)^{n_- -
\tilde{n}}\,C_{\tilde{n}}^{(\mathcal{N},n_+)}\,
 \mathcal{F}_{\mathcal{N}-\tilde{n},\tilde{N}}^{(Z-1)}(\bm{r}_{+,\perp})
  \mathcal{F}_{\tilde{n},n_-}^{(1)}(\bm{r}_\perp),
\label{Psi_alt}
\eeq
which is equivalent to the representation used in
Ref.\,\cite{Bez95}.

\subsection{Recurrence relations for the basis coefficients}
\label{sect:coefficients}

Equation (\ref{recurC}) can be rewritten in the form
\beq
   \sqrt{(\mathcal{N}-\tilde{n}+1)\,\tilde{n}}\,C_{\tilde{n}-1}^{(\mathcal{N},n_+)} +
   \sqrt{(\mathcal{N}-\tilde{n})(\tilde{n}+1)}\,C_{\tilde{n}+1}^{(\mathcal{N},n_+)} = 
   \frac{Z(\tilde{n} + n_+ -\mathcal{N})-2\tilde{n}+\mathcal{N}}{\sqrt{Z-1}}
   \,C_{\tilde{n}}^{(\mathcal{N},n_+)}
\label{Crecur}
\eeq
and used as a recurrence relation to calculate the coefficients
$C_{\tilde{n}}^{(\mathcal{N},n_+)}$ for given values of $\mathcal{N}$
and $n_+$.  One can start the recurrent procedure from an arbitrary
value of the initial coefficient to (for instance
$C_{0}^{(\mathcal{N},n_+)}=1$ for the upward recurrence of
$C_{\mathcal{N}}^{(\mathcal{N},n_+)}=1$ for the downward recurrence),
and then scale them by a single number factor to satisfy the
normalization relation (\ref{orthoC}),
\beq
\sum_{\tilde{n}=0}^\mathcal{N}
\left[C_{\tilde{n}}^{(\mathcal{N},n_+)}\right]^2 = 1,
\label{normC}
\eeq
The upward recurrence (that is, starting from $\tilde{n}=0$ to higher $\tilde{n}$
is stable under the condition that it is
performed  for $n_+\geq \mathcal{N}-Z$, whereas the downward recurrence
(starting from $\tilde{n}=\mathcal{N}$ to lower
$\tilde{n}$) is stable for $n_+ < \mathcal{N}-Z$ \footnote{The opposite 
statement was mistakenly done in Ref.~\cite{Bez95}.}.

Supplementary recurrence relations for coefficients
$C_{\tilde{n}}^{(\mathcal{N},n_+)}$ are derived in
Appendix~\ref{sect:suppl}.

\section{Solution of the Schr\"odinger equation}
\label{sect:solution}

\subsection{Expansion on the transverse basis}

We are looking for eigenfunctions of Hamiltonian $H$ in \req{H2}.
The potential $V_\mathrm{C}$ commutes with the transformed total
pseudomomentum $\bm{K}$ [\req{P0'}] and the total angular momentum
$\bm{L}$ [\req{L}]. Therefore, we may consider states with definite
$K^2$ and $L_z$, that is to fix $\tilde{N}$ and $L$. However, $V_\mathrm{C}$ does not
commute with squared kinetic momenta $\pi_\pm^2$. Let us expand
the states with fixed $\tilde{N}$ and $L$ over the complete basis of states
constructed in Sect.~\ref{sect:states}:
\beq
\psi_{\kappa}(\bm{r}_{+,\perp},\bm{r}_{-,\perp},z) =  \sum_{n_-,n_+} 
\Psi_{\tilde{N},L,n_-,n_+}(\bm{r}_{+,\perp},\bm{r}_{-,\perp})
\,g_{n_-n_+;\,\kappa}(z).
\label{psiPsi}
\eeq
Here $\kappa$ is a compound quantum number, which is assigned to the considered
quantum state and includes $\tilde{N}$ and $L$.
Let us substitute \req{psiPsi} into the Schr\"odinger equation
\beq
H\psi_{\kappa} = E_\kappa \psi_{\kappa},
\label{Schr}
\eeq
 multiply both sides by 
$\Psi_{\tilde{N},L,n_-,n_+}^*(\bm{r}_{+,\perp},\bm{r}_{-,\perp})$, and
integrate over $\bm{r}_{+,\perp},\bm{r}_{-,\perp}$. Using
Eqs.~(\ref{H0}), (\ref{HperpE}), and (\ref{normPsi}), we obtain
\beq
   \left( \frac{p_z^2}{2\mred} - E_\kappa^\| \right)
    g_{n_- n_+;\,\kappa}(z) = 
       - \sum_{n_-'=0}^\infty \sum_{n_+'=0}^{\mathcal{N}}
            V_{n_-',n_+';\,n_-,n_+}^{(\tilde{N}-L)}(z)\,\,
    g_{n_-' n_+';\,\kappa}(z),
\label{ODEsystem}
\eeq
where 
\beq
   E_\kappa^\| = E_\kappa - E_{n_-,n_+}^\perp
\label{Elong}
\eeq
 is the energy
corresponding to the relative motions along $z$ and
\bea
V_{n_-',n_+';\,n_-,n_+}^{(\tilde{N}-L)}(z) &=& \sum_{k'=0}^{\mathcal{N}'}\sum_{k=0}^\mathcal{N}
C_{k'}^{(\mathcal{N}',n_+')}C_{k}^{(\mathcal{N},n_+)}
\,\int_{\mathbb{R}^2}\mathcal{F}_{\mathcal{N}'-k',\tilde{N}}^{(Z-1)*}(\bm{R}_{\perp})
\,\mathcal{F}_{\mathcal{N}-k,\tilde{N}}^{(Z-1)}(\bm{R}_{\perp})\,\dd\bm{R}_{\perp}
\nonumber\\&&\times
\,\int_{\mathbb{R}^2}\mathcal{F}_{n_-',k'}^{(-1)*}(\bm{r}_\perp)\,
\frac{-Ze^2}{\sqrt{r^2 + z^2}}\,
\mathcal{F}_{n_-,k}^{(-1)}(\bm{r}_\perp)\,\dd\bm{r}_\perp
\label{V4init}
\eea
is an effective one-dimensional potential,
$\mathcal{N}' = \tilde{N}-L+n_-'$, $\mathcal{N}=\tilde{N}-L+n_-$.
Since the Coulomb potential  does not contain $\bm{R}_{\perp}$, the first
integral in \req{V4init} equals
$\delta_{\mathcal{N}'-k',\mathcal{N}-k} = \delta_{k',n_-'-n_- + k}$. 
Thus, using \req{F-Landau}.
we obtain
\beq
V_{n_-',n_+';\,n_-,n_+}^{(\tilde{N}-L)}(z) =
\sum_{k=k_\mathrm{min}}^{\mathcal{N}}
C_{n_-'-n_- +k}^{(\mathcal{N}',n_+')} C_{k}^{(\mathcal{N},n_+)}
\,
\int_{\mathbb{R}^2}\Phi_{n'_-,s}^{(-1)*}(\bm{r}_\perp)\,
 \frac{-Ze^2}{\sqrt{r_\perp^2 + z^2}}\,
 \Phi_{n_-,s}^{(-1)}(\bm{r}_\perp)\,\dd^2 r_\perp,
\label{Vnks}
\eeq
$k_\mathrm{min}=\mathrm{max}(0,n_--n_-')$.

Since the transverse basis is complete, the infinite system
(\ref{ODEsystem}) is equivalent to the Schr\"odinger equation
(\ref{Schr}). Truncating the sum
(\ref{psiPsi}), one obtains a finite system, which solves \req{Schr}
approximately. The same finite system of equations can be obtained
from the variational principle on the truncated basis.

\subsection{Calculation of effective potentials}
\label{sect:Veff}

It is convenient to define reduced (dimensionless) potentials through
the relations
\bea
V_{n_-',n_+';\,n_-,n_+}^{(\tilde{N}-L)}(z)
& = &
\frac{-Ze^2}{\am\sqrt2}\,v_{n_-',n_+';\,n_-,n_+}^{(\tilde{N}-L)}(\zeta),
\qquad
   \zeta \equiv \frac{z}{\am\sqrt{2}},
\label{zeta}
\\
  v_{n_-',n_+';\,n_-,n_+}^{(\tilde{N}-L)}(\zeta) &=&
 \sum_{s=s_\mathrm{min}}^{\tilde{N}-L}
C_{n_-'+s}^{(\mathcal{N}',n_+')} C_{n_-+s}^{(\mathcal{N},n_+)}
\,v_{n_-',n_-;\,s}(\zeta),
\hspace*{2em}
\label{V4viaV3}
\eea
where 
$s_\mathrm{min}=-\mathrm{min}(n_-,n_-')$, and
\beq
v_{n',n;\,s}(\zeta) =
\int_0^\infty \frac{I_{n+s,n}(\rho)\,I_{n'+s,n'}(\rho)}{
\sqrt{\rho+\zeta^2}}\,\dd\rho.
\label{vnks}
\eeq
 The functions $v_{n',n;\,s}(\zeta)$
belong to the class of effective potentials studied in
Ref.~\cite{P94}. 
Thanks to the relation $v_{n',n;\,s}(\zeta) =
v_{n'_r,n_r;\,|s|}(\zeta)$, where $n'_r=n+(s-|s|)/2$ and
$n_r=n+(s-|s|)/2$ are non-negative integers corresponding to radial
quantum numbers of the states $|n',s\rangle$ and $|n,s\rangle$, it is
sufficient to consider only non-negative subscripts of these functions.
Using \req{I-explicit}, we obtain
\beq
   v_{n,k;\,s}(\zeta) =
   \sqrt{n!(n+s)!k!(k+s)!}\,
\sum_{l=0}^n \frac{1}{l!(n-l)!(l+s)!}
 \,\sum_{m=0}^k \frac{(-1)^{m+l} (s+l+m)!}{m!(k+m)!(s+m)!}\,
   v_{s+l+m}(\zeta),
\eeq
and
\beq
   v_s(\zeta)
   = \frac{1}{s!}\,\int_0^\infty x^{s}
     \,\frac{\erm{-x}\,\dd x}{\sqrt{\zeta^2+x}}.
\eeq
A change of variable brings the latter integral to
\beq 
v_s(\zeta) = \frac{1}{s!}\,\int_{\zeta^2}^\infty (x-\zeta^2)^s
\,\erm{\zeta^2-x}\,\frac{\dd x}{\sqrt{x}} =
 \sum_{l=0}^s (-1)^l \frac{s!}{l!(s-l)!}\,\zeta^{2l}\, \erm{\zeta^2}
\Gamma(s-l+1/2,\zeta^2),
\eeq
\end{widetext}
where $\Gamma(a,x)$ is the incomplete gamma function. 
At small or moderate $\zeta$, $v_s(\zeta)$ can be calculated using 
the recurrence relation
\bea
   v_{s+1}(\zeta) &=&\!\! (2s+1)\,v_s(\zeta)
\!
 +
\!
 \frac{\zeta^2}{s+1}\,
     \big[v_{s-1}(\zeta) - v_s(\zeta) \big],
\qquad
\\
v_0(\zeta) &=& \sqrt\pi\,\erm{\zeta^2}\mathrm{erfc}(|\zeta|),
\\
v_1(\zeta) &=& \frac{1 - 2\zeta^2}{2}\,v_0(\zeta) + |\zeta|.
\label{recur}
\eea
Here, $\mathrm{erfc}(\zeta)$ is the complementary error function, which
can be calculated using, e.g., an expansion in power series at small
$|\zeta|$ and continued fractions at $|\zeta|\gtrsim1$ (\cite{AS}, 7.1.5
and 7.1.14). At
large $|\zeta|$ or $s$, however, the
recurrence relation (\ref{recur}) fails because of round-off errors in
positive and negative terms, which nearly annihilate. In
this case, one can use the asymptotic formula
\beq
v_s(\zeta) \sim \frac{1}{\sqrt{\zeta^2+1+s}}\left(
    1+\frac38\,\frac{1+s}{(\zeta^2+1+s)^2} \right).
\eeq

\subsection{Adiabatic approximation}
\label{sect:adiabatic}

In strong magnetic fields, such that $\hbar\omc{-}\gg Z$ Ha (where
Ha$=m_- e^4/\hbar^2$ is the atomic unit of energy), the system of
equations (\ref{ODEsystem}) approximately splits into separate subsystems
with fixed $n_-$. Indeed, since the magnetic length $\am$ is small at
large $B$, the denominator in \req{Vnks} remains nearly constant over
the range where the Landau functions $\Phi$ in the integral
are not small. Since the Landau functions are orthogonal,
the integral is small for $n'_-\neq n_-$. Therefore the potentials
$V_{n_-',n_+';\,n_-,n_+}^{(\tilde{N}-L)}(z)$ with
$n_-'\neq n_-$ only weakly couple the subsystems with different
fixed $n_-$ numbers. 

\begin{figure}
\centering
\includegraphics[width=\columnwidth]{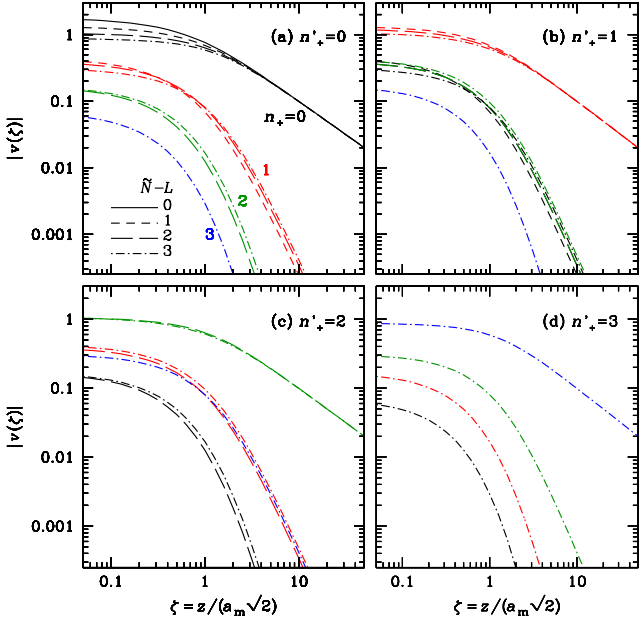}
\caption{Absolute values of the reduced effective potentials 
$v_{n_-',n_+';\,n_-,n_+}^{(\tilde{N}-L)}(\zeta)$  for He$^+$ ($Z=2$),
$n_-'=n_-=0$ and four lowest values of $\mathcal{N}=\tilde{N}-L$ (different line
styles, explained by the legend in panel (a)), $n_+$ (different colors,
as marked in panel (a)), and $n'_+=0,1,2,3$ (panels (a)--(d),
respectively).
}
\label{fig:vef}
\end{figure}

A solution with fixed $n_-$ may be called adiabatic approximation with
respect to the electron motion, or ``e-adiabatic approximation'' for
short. By analogy with the well known adiabatic approximation for
electron motion in a strong magnetic field with a stationary Coulomb
potential, it assumes that the Coulomb potential affects only the motion
along the magnetic field, whereas the relatively fast motion in the
transverse plane is governed by the magnetic field alone. In each
subsystem, the effective potential is given by \req{V4viaV3} with
$n_-'=n_-$.
Examples of these effective potentials for $n_-=0$ are shown in
Fig.~\ref{fig:vef}. We see that the effective potentials
$v_{0,n_+';\,0,n_+}^{(\mathcal{N})}(\zeta)$ with $n_+'\neq n_+$ are
relatively small. Neglecting  these small potentials results in the full
adiabatic approximation, where the system of equations (\ref{ODEsystem})
is split into separate equations with fixed $n_-$ and $n_+$:
\beq
   \left( -\frac{\hbar^2}{2\mred}\frac{\dd^2}{\dd z^2}
    - E_\kappa^\| \right)
    g_{\kappa}(z) = 
\frac{Ze^2}{\am\sqrt2}
\,v_{n_-,n_+}^{(\tilde{N}-L)}(\zeta)\,
    g_{\kappa}(z),
\label{Schradiab}
\eeq
where $\zeta\equiv z/\am\sqrt2$,
\beq
  v_{n_-,n_+}^{(\tilde{N}-L)}(\zeta) \equiv
   v_{n_-,n_-;\,n_+,n_+}^{(\tilde{N}-L)}(\zeta)
\eeq
and $E_\kappa^\|$ is given by by \req{Elong}.
Due to the symmetry of the effective potential, $v(\zeta)=v(-\zeta)$,
it is
sufficient to solve \req{Schradiab} for $z>0$ for the even and odd wave
functions separately.

The finite solutions, with $E_\kappa^\| <0$, form
the discrete spectrum. We number these solutions
by $\nu=0,1,2,\ldots$, with even
and odd $\nu$ corresponding to the symmetric and antisymmetric
longitudinal wave functions, respectively: $g_{\kappa}(z) =
(-1)^\nu g_{\kappa}(z)$. In the continuum, with
$E_\kappa^\|>0$, we have two linearly independent solutions
$g_{\kappa}(z)$ at any energy. For example, we can
use the real solutions, whose behaviors at
$|z|\to\infty$ is analogous to the usual Coulomb functions (e.g.,
\cite{Seaton83})
 \beq
   g_{\kappa}^\mathrm{real}(z) \sim
   \sin[\phi_{n_-,n_+,E}(z)]
   + \mathcal{R}_{E,\pm}
   \cos[\phi_{n_-,n_+,E}(z)],
\label{asymp_ad}
\eeq
where
\beq
   \phi_{n_-,n_+,E}(z)=k_\kappa z +
       \frac{\mred e^2}{\hbar^2 k_\kappa}
            \,\ln(k_\kappa z)
\label{phase}
\eeq
is the $z$-dependent part of the phase of the wave function
at $z\to+\infty$,
\beq
   k_\kappa=k_{n_-,n_+,E}=(2\mred E_\kappa^\|)^{1/2}/\hbar
\label{k_kappa}
\eeq
is the wavenumber, and $E_\kappa^\|=E-E_{n_-,n_+}^\perp$
[\req{Elong}].

\subsection{Coupled channel formalism}

In a strong magnetic field, the adiabatic approximation is a convenient
starting point for solving the full system (\ref{ODEsystem}) by
iterations, so that the leading longitudinal wave function in
\req{ODEsystem} remains the one corresponding to the starting adiabatic
solution (cf.{} Ref.~\cite{P94}). Then the numbering of the quantum
states can be the same as in the adiabatic approximation,
$|\kappa\rangle = |\tilde{N},L,n_{0-},n_{0+},\nu\rangle$ for the
discrete spectrum and $|\kappa\rangle =
|\tilde{N},L,n_{0-},n_{0+},E,\pm\rangle$  for the continuum. Here,
$n_{0-}$ and $n_{0+}$ are the values of $n_-$ and $n_+$ for
the leading term in expansion (\ref{psiPsi}).

\subsubsection{Bound-state wave functions}

For the
bound states, the energies $E_\kappa$ are determined as the
eigenenergies of the system of equations (\ref{ODEsystem}), and
$\nu=0,1,2,\ldots$ corresponds to the longitudinal degree of freedom and
controls the parity of the wave function.
For the continuum, the energy $E$ and the parity ($\pm$) are fixed
arbitrarily. Since
$\tilde{N}$ and $L$ enter \req{ODEsystem} only in combination
$(\tilde{N}-L)$, the
``longitudinal wave functions'' $g_{n_- n_+;\,\kappa}(z)$ can be numbered
as $g_{n_- n_+;\,\mathcal{N},\nu}(z)$ for the discrete spectrum and 
$g_{n_- n_+;\,\mathcal{N},E,\pm}(z)$ for the continuum. The degeneracy
in $\tilde{N}$, at a fixed $\mathcal{N}$, reflects the translational
invariance. Indeed, different $\tilde{N}$ correspond to different mean values of
the squared total pseudomomentum projection on the $(xy)$-plane,
$\langle k_{\mathrm{tot},\perp}^2 \rangle$, which according to
Eqs.\,(\ref{pseudomom}) and (\ref{r_c}) is proportional to
the squared sum of the guiding centers,
 $\langle (r_{\mathrm{c},-}+r_{\mathrm{c},+})^2 \rangle$, measured from
the chosen gauge axis, which can be freely changed by the
gauge transformation $\bm{A}(\bm{r}) \to (1/2)\,
\bm{B}\times(\bm{r}-\bm{r}_A)$ with arbitrary $\bm{r}_A$
(cf.{} Ref.~\cite{P94}).

Accordingly, for the physical problems that do not involve explicit
positions of the guiding centers in space, including the present case of
a single ion in a uniform field, one can identify the ion states by
four quantum numbers instead of five:
$|\kappa\rangle =|\mathcal{N},n_{0-},n_{0+},\nu\rangle$.

\subsubsection{Continuum wave functions}
\label{sect:wfc}

For the continuum, we construct the basis by analogy with the $R$-matrix
formalism \cite{Seaton83}.
Let
$I_\mathrm{o}$ be the total
number of \emph{open channels} at given $E$, i.e., number of such pairs
$n_-,n_+$ that $E_{n_-,n_+}^\perp < E$.
In this case, numbers $n_{0\pm}$ mark a selected
open channel, defined for 
$E > E_{n_{0-},n_{0+}}^\perp$
by asymptotic conditions
at $z\to+\infty$
\begin{widetext}
\beq
   g_{n_-^\mathrm{o},n_+^\mathrm{o};\,n_{0-},n_{0+},\mathcal{N},E,\pm}^\mathrm{real}(z) 
   \sim C_\mathrm{norm}\,\Big\{
    \delta_{n_-^\mathrm{o}n_{0-}}\delta_{n_+^\mathrm{o}n_{0+}}
   \sin[\phi_{n_-^\mathrm{o},n_+^\mathrm{o};\,\kappa}(z)]
   + \mathcal{R}_{n_-^\mathrm{o},n_+^\mathrm{o};\,n_{0-},n_{0+};\,\mathcal{N},E,\pm}
   \cos[\phi_{n_-^\mathrm{o},n_+^\mathrm{o};\,\kappa}(z)]\,\Big\},
\label{real_asymp}
\eeq
\end{widetext}
where the pairs $(n_-^\mathrm{o},n_+^\mathrm{o})$
relate to different open channels 
($E>E^\perp_{n_-^\mathrm{o},n_+^\mathrm{o}}$),
$\phi_{n_-^\mathrm{o},n_+^\mathrm{o};\,E}(z)$
is the $z$-dependent part of the phase of the wave function
at $z\to+\infty$, defined in \req{phase}, 
and $C_\mathrm{norm}$ is a normalization constant. 
The quantities
 $\mathcal{R}_{n_-^\mathrm{o},n_+^\mathrm{o};\,n_-,n_+;\,\mathcal{N},E,\pm}$
with different pairs of $(n_-^\mathrm{o},n_+^\mathrm{o})$ and
 $(n_-,n_+)$, corresponding to the open channels, constitute the 
reactance matrix $\mathcal{R}_{\mathcal{N},E,\pm}$, which
has dimensions
$I_\mathrm{o}\times I_\mathrm{o}$.
For the \emph{closed channels}, defined by the
inequality $E<E^\perp_{n_-^\mathrm{c},n_+^\mathrm{c}}$, one should
impose the boundary conditions
$g_{n_-^\mathrm{c},n_+^\mathrm{c};\,n_{0-},n_{0+},\mathcal{N},E,\pm}(z) \to0$
 at $z\to\infty$. 
 
The set of solutions,
defined by Eqs.~(\ref{ODEsystem}) and (\ref{real_asymp}),
constitute a complete set of $I_o$ independent
real basis functions. 
 If the wave functions are normalized by the condition
\beq
   \int_{\mathbb{R}^2}\dd\bm{r}_{+,\perp}
   \int_{\mathbb{R}^2}\dd\bm{r}_{-,\perp}
   \int_{-\zmax}^{\zmax}\dd z
    \, |\psi_{\kappa}(\bm{r}_{+,\perp},\bm{r}_{-,\perp},z)|^2 = 1,
\label{z_norm}
\eeq
where $\zmax$ is half the normalization length,
then the orthonormality of the transverse basis
[\req{normPsi}] leads to the condition
\beq
   \sum_{n_-,n_+} \int_{-\zmax}^{\zmax}
      |g_{n_-n_+;\,\kappa}(z)|^2 \dd z =1.
\label{gz_norm}
\eeq
In the adiabatic approximation, a single term of the sum
is retained in \req{gz_norm} for each open channel.

One can compose the basis of outgoing waves, appropriate to
photoionization, by analogy with the
case of the H atom in Refs.\,\cite{PPV97,PP97}. 
The basis of outgoing waves with definite $z$-parity
is defined by the asymptotic
conditions at $z\to\infty$
\begin{widetext}
\beq
   g_{n_-^\mathrm{o},n_+^\mathrm{o};\,n_{0-},n_{0+},\mathcal{N},E,\pm}^\mathrm{out}(z)
    \sim C_\mathrm{norm}\,\Big\{
   \delta_{n_-^\mathrm{o}n_{0-}}\delta_{n_+^\mathrm{o}n_{0+}}\,
   \erm{\ii\phi_{n_{0-},n_{0+},E}(z)}
   - \mathcal{S}_{n_-^\mathrm{o}n_+^\mathrm{o};\,n_{0-},n_{0+};\,\mathcal{N},E,\pm}^*\,
   \erm{-\ii\phi_{n_-^\mathrm{o}n_+^\mathrm{o}}(z)}
   \,\Big\},
\label{asout}
\eeq
where $\mathcal{S}_{n_-^\mathrm{o}n_+^\mathrm{o};n_- n_+;\,\mathcal{N},E,\pm}$
 are the elements of the unitary scattering matrix
$\mathcal{S}_{\mathcal{N},E,\pm}=
(1+\mathrm{i}\mathcal{R}_{\mathcal{N},E,\pm})
(1-\mathrm{i}\mathcal{R}_{\mathcal{N},E,\pm})^{-1}$.
The 
basis of outgoing waves is obtained from the
real basis by transformation
\beq
   g_{n_-'' n_+'';\,n_- n_+;\,\mathcal{N},E,\pm}^\mathrm{out}(z)
    = 2\mathrm{i}
   \sum_{n_-' n_+'}\big[
        (1+\mathrm{i}\mathcal{R}_{\mathcal{N},E,\pm})^{-1}\big]_{n_- n_+;\,n_-' n_+'}
   \,g_{n_-'' n_+'';\,n_-' n_+';\,\mathcal{N},E,\pm}^\mathrm{real}(z).
\label{g_as_out}
\eeq
Here, pairs $(n_-, n_+)$ and $(n_-', n_+')$ run over
open channels, but $(n_-'',n_+'')$ run over all 
(open and closed) channels. As follows
from the orthonormality of the transverse basis, the normalization
integral on the left-hand side of \req{z_norm} equals
\beq
   2\zmax\,|C_\mathrm{norm}|^2 \left[ 1+ 
\mathcal{S}_{\mathcal{N},E,\pm}\mathcal{S}_{\mathcal{N},E,\pm}^\dag
\right] = 
   4\zmax\,|C_\mathrm{norm}|^2,
\eeq
where we have used the unitarity of the S-matrix.
Then according to \req{z_norm}
\beq
|C_\mathrm{norm}|=(4\zmax)^{-1/2}.
\label{Cnorm}
\eeq

After the orthonormal basis of outgoing waves have been obtained at
given $\mathcal{N}$ and $E$ for each $z$-parity, with  $g_{n_-'
n_+';\,n_- n_+;\,\mathcal{N},E,\pm}^\mathrm{out}(z) = \pm  g_{n_-'
n_+';\,n_- n_+;\,\mathcal{N},E,\pm}^\mathrm{out}(-z)$, we can easily
construct solutions for electron waves propagating at $z\to\pm\infty$ in
a definite open channel $(n_{0-},n_{0+})$ for an arbitrary
$L\leq\mathcal{N}-n_-$. These solutions are given by \req{psiPsi} with
coefficients 
\beq
   g_{n_-'n_+';\,\kappa}(z) = \frac{1}{\sqrt{2}}
   \left(
   g_{n_-' n_+';\,n_{0-},n_{0+};\,\mathcal{N},E,+}^\mathrm{out}(z)
   \pm
   g_{n_-' n_+';\,n_{0-},n_{0+};\,\mathcal{N},E,-}^\mathrm{out}(z)
   \right),
\label{outgoing}
\eeq
where the sign $+$ or $-$ represents electron escape in the positive or
negative $z$ direction, respectively. Waves incoming from
$z\to\pm\infty$ are given by the complex conjugate of \req{outgoing}.
 
\subsection{Geometric sizes of the ions}

In order to evaluate populations of different bound levels and the
collision frequency $\nu_\mathrm{coll}$ in the plasma environment, it is
useful to calculate the geometric sizes of the ions. For this aim we use
the root-mean-square longitudinal and transverse sizes,
$|\langle\kappa|z^2|\kappa\rangle|^{1/2}$  and
$|\langle\kappa|r_\perp^2|\kappa\rangle|^{1/2}$, by analogy with
Ref.\,\cite{PCS99}. 
Let us use the basis expansion (\ref{psiPsi}) 
for calculation of these matrix elements.
For the longitudinal mean squared size, using the
orthonormality condition (\ref{normPsi}), we obtain
\beq
   \langle\kappa|z^2|\kappa\rangle = 
   \sum_{n_-,n_+} \int_{-\zmax}^{\zmax}
   g_{n_-n_+;\,\kappa}^*(z)\,g_{n_-n_+;\,\kappa}(z)
   \,z^2\,\dd z.
\label{longsize}
\eeq
For the mean squared transverse size,
using \req{rRpm}, we obtain
\beq
   \langle\kappa|r_\perp^2|\kappa\rangle =
   2\langle\kappa|r_{+1}r_{-1}|\kappa\rangle = 
   2\am^2\, \langle\kappa|(\hat{a}^\dag - \hat{\tilde{a}})\,
    (\hat{a} - \hat{\tilde{a}}^\dag)|\kappa\rangle.
\eeq
Then, using Eqs.\,(\ref{aa'1}), (\ref{aa'2}), (\ref{aa'+1}), and
(\ref{aa'+2}), we obtain
\bea
   \langle\kappa|r_\perp^2|\kappa\rangle &= &
2\am^2
   \sum_{n_-=0}^\infty
   \sum_{n_+'=0}^{\mathcal{N}} \sum_{n_+=0}^{\mathcal{N}}
  \Bigg[
   \sum_{k=0}^{\mathcal{N}}
 (n_-+k+1)\, C_{k}^{(\mathcal{N},n_+)}C_{k}^{(\mathcal{N},n_+')}
  \int_{-\zmax}^{\zmax}
   g_{n_-n_+';\,\kappa}^*(z)\,g_{n_-n_+;\,\kappa}(z)
   \,\dd z
\nonumber\\&&
    - \sum_{k=0}^{\mathcal{N}} \sqrt{(n_-+1)(k+1)}\,
    C_{k}^{(\mathcal{N},n_+)}
C_{k}^{(\mathcal{N}+1,n_+')}
  \int_{-\zmax}^{\zmax}
   g_{n_-+1,n_+';\,\kappa}^*(z)\,g_{n_-n_+;\,\kappa}(z)
   \,\dd z
\nonumber\\&&
    - \sum_{k=0}^{\mathcal{N}} \sqrt{n_-k}\,
    C_{k}^{(\mathcal{N},n_+)}
C_{k-1}^{(\mathcal{N}-1,n_+')}
  \int_{-\zmax}^{\zmax}
   g_{n_--1,n_+';\,\kappa}^*(z)\,g_{n_-n_+;\,\kappa}(z)
   \,\dd z
  \Bigg].
\label{perpsize}
\eea
\end{widetext}

\section{Interaction with radiation}
\label{sect:rad}

\subsection{Radiative transitions}

In presence of an electromagnetic wave, its vector potential
$\bm{A}_\mathrm{rad}$ should be added to the vector potential
of the constant magnetic field $\bm{A}$ in Eqs.\,(\ref{r-dot}) and
(\ref{k_i}).
Then the total Hamiltonian of the system of an electron, a nucleus, a
constant magnetic field, and radiation can be written as
\bea
   H_\mathrm{tot} &=& \frac{1}{2m_+}\left(\bm{\pi}_+ -
\frac{Ze}{c}\,\bm{A}_\mathrm{rad,\perp}(\bm{r}_+)
\right)^2
\nonumber\\&&
+
\frac{1}{2m_-}\left(\bm{\pi}_- +
\frac{e}{c}\,\bm{A}_\mathrm{rad,\perp}(\bm{r}_-)
\right)^2
\nonumber\\&&
-\frac{Ze^2}{\sqrt{r^2+z^2}} + \sum_{\bm{q}\gamma}
\hbar\omega_q\,\hat{c}_{\bm{q}\gamma}^\dag \hat{c}_{\bm{q}\gamma},
\label{Htot}
\eea
where $\bm{q}$,  $\alpha$, and $\omega_q$ are the photon wavevector,
polarization index, and frequency, respectively,
$c^\dag_{\bm{q}\gamma}$ and $\hat{c}_{\bm{q}\gamma}$ are the photon creation
and annihilation operators,
\beq
   \bm{A}_\mathrm{rad}(\bm{r})
   =
   \sum_{\bm{q}\gamma} \left(\frac{2\pi\hbar
   c^2}{\omega_q \mathcal{V}}\right)^{1/2}
   \left( \bm{e}_{\bm{q}\gamma} \hat{c}_{\bm{q}\gamma}
   \erm{\ii\bm{q}\cdot\bm{r}}
+
   \bm{e}_{\bm{q}\gamma}^* \hat{c}_{\bm{q}\gamma}^\dag
   \erm{-\ii\bm{q}\cdot\bm{r}}
   \right)
\label{Arad}
\eeq
is the electromagnetic field operator (in the Schr\"odinger
representation), the subscripts $\perp$ and $z$ mark the transverse and
longitudinal vector components with respect to the magnetic field
direction, and $\mathcal{V}$ is the normalization volume (see, e.g.,
Refs.~\cite{SokTer,LaLi-QED}).

The operator of
interaction with radiation is obtained by expanding the brackets in \req{Htot}. 
We will use the Coulomb gauge,
$\nabla\cdot\bm{A}_{\mathrm{rad}} = 0$, and the transverse approximation
assuming $\bm{q}\cdot\bm{e}_{\bm{q}\gamma}=0$. Then operator
$\bm{A}_\mathrm{rad}$ commutes with $\bm{\pi}_\pm$, so that
$\bm{A}_\mathrm{rad}\cdot\bm{\pi}_\pm+\bm{\pi}_\pm\cdot\bm{A}_\mathrm{rad} =
2\bm{A}_\mathrm{rad}\cdot\bm{\pi}_\pm$. 

Neglecting nonlinear effects (i.e., terms proportional to
$A_\mathrm{rad}^2$), we obtain the following operator that couples
internal degrees of freedom to radiation:
\bea
   V_\mathrm{int} &=&
\frac{e}{m_-c}\bm{A}_\mathrm{rad,\perp}(\bm{r}_-)\cdot\bm{\pi}_-
 -
\frac{Ze}{m_+c}\bm{A}_\mathrm{rad,\perp}(\bm{r}_+)\cdot\bm{\pi}_+
\nonumber\\
&+&\!\!\!
\frac{e}{m_-c}{A}_\mathrm{rad,z}(\bm{r}_-)\,p_{-,z}
-
\frac{Ze}{m_+c}{A}_\mathrm{rad,z}(\bm{r}_+)\,p_{+,z}.
\qquad
\label{Vint}
\eea

Using \req{Arad}, we can rewrite \req{Vint} as 
\beq
 V_\mathrm{int} =
\sqrt{\frac{2\pi\hbar}{\omega_q\mathcal{V}}}
\sum_\mathrm{\bm{q}\gamma}
\left( 
\hat{c}_\mathrm{\bm{q}\gamma}
\,\bm{e}_\mathrm{\bm{q}\gamma}\cdot\bm{j}_\mathrm{eff}  +
\bm{e}_\mathrm{\bm{q}\gamma}^*\cdot\bm{j}_\mathrm{eff}^\dag\,\hat{c}_\mathrm{\bm{q}\gamma}^\dag 
\right),
\eeq
where
\beq
 \bm{j}_\mathrm{eff} = 
\erm{\ii\bm{q}\cdot\bm{r}}
\left(
 e\,\dot{\bm{r}}_-  - Ze\,\dot{\bm{r}}_+
  \right)
\label{jeff}
\eeq
is the effective current operator ($\dot{\bm{r}}_\pm = \bm{\pi}_\pm/m_\pm$).

Let us consider an absorption of a photon with a given wavevector
$\bm{q}$ and polarization $\alpha$. The initial state is
$|i\rangle=f_{\bm{q}\gamma}(1)|\psi_i\rangle$ and the final state is
$|f\rangle=f_{\bm{q}\gamma}(0)|\psi_f\rangle$, where $|\psi_{i,f}\rangle$ denotes the
state of the system of charged particles and $f_{\bm{q}\gamma}(N)$ is the function of
photon number \cite{SokTer}.
According to the Fermi's golden rule, probability of transition, per
unit time, from initial quantum state $f_{\bm{q}\gamma}(1)|i\rangle$ to final state
$f_{\bm{q}\gamma}(0)|f\rangle$ with absorption of one photon is given by
\bea
  \dd w_{i\to f} &=& \frac{2\pi}{\hbar}\,\big|\langle \psi_f
   |f_{\bm{q}\gamma}^\dag(0) V_\mathrm{int}
  f_{\bm{q}\gamma}(1) | \psi_i \rangle \big|^2\,
\nonumber\\&&\times
  \delta(E_f-E_i-\hbar\omega_q)\,\dd\nu_f,
\label{w_if}
\eea
where
$E_i$, $E_f$, and $\hbar\omega_q$ are the energies of the initial state,
final state, and absorbed photon, respectively, and $\dd\nu_f$ is the
density of final states. Taking into account the properties of the
photon creation and annihilation operators \cite{SokTer},
$\hat{c}_\mathrm{\bm{q}\gamma} f_{\bm{q}\gamma}(N) =
\sqrt{N}f_{\bm{q}\gamma}(N-1)$, $\hat{c}_\mathrm{\bm{q}\gamma}^\dag
f_{\bm{q}\gamma}(N) = \sqrt{N+1}f_{\bm{q}\gamma}(N+1)$, where
$f_{\bm{q}\gamma}(N)$ is the function of photon number, and performing
normalization of transition rate (\ref{w_if}) by the photon flux
$c/\mathcal{V}$, we arrive at the differential cross section
\beq
   \dd\sigma_{i\to f,\bm{q}\gamma} = \frac{4\pi^2}{\omega_q
   c}\,\big| \langle\psi_f| 
   \bm{e}_{\bm{q}\gamma}\cdot\bm{j}_\mathrm{eff} | \psi_i\rangle
   \big|^2
   \,\delta(E_f-E_i-\hbar\omega_q)\,\dd\nu_f.
\label{dsigma}
\eeq

If the final state belongs to the discrete spectrum, the final energies
$E_f$ are distributed within a narrow band around the value
$
   E_{f0} = E_i + \hbar\omega{fi},
$
where $\omega{fi}$ is a central value of the transition frequency.
Then integration of \req{dsigma} over the final states gives
\beq
    \sigma_{i\to f,\bm{q}\gamma} = \frac{4\pi^2}{\hbar\omega_q
   c}\,\big| \langle\psi_f| 
   \bm{e}_{\bm{q}\gamma}\cdot\bm{j}_\mathrm{eff} | \psi_i\rangle
   \big|^2
   \,\Delta_{fi}(\omega_q - \omega_{fi}),
\label{sigma_bb}
\eeq
where $\Delta_{fi}(\omega_q - \omega_{fi})$ describes the profile of the
spectral line, which is normalized so that
$
   \int \Delta_{fi}(\omega)\,\dd\omega = 1.
$

If the final state belongs to the continuum and 
wave functions are normalized according to \req{z_norm}, then
\bea&&
   \dd\nu_f = \frac{\zmax}{\pi}\,\frac{\mred}{
   \hbar^2 k_f}\,\dd E_f,
\\&&
   k_f = \hbar^{-1}\sqrt{2\mred E_f^\|},
\quad
   E_f^\| \equiv E_f - E_{n_{0-,f}n_{0+,f}}^\perp,
\qquad
\label{k_f}
\eea
with $E_f - E_{n_{0-,f}n_{0+,f}}^\perp > 0$.
In this case the cross section of photoabsorption takes the form
\beq
   \sigma_{i\to f,\bm{q}\gamma} = \frac{4\pi \zmax\mred}{
   \hbar^2 k_f \omega_q c}\,\big| \langle\psi_f| 
   \bm{e}_{\bm{q}\gamma}\cdot\bm{j}_\mathrm{eff} | \psi_i\rangle
   \big|^2.
\label{sigma_bf}
\eeq

The above equations do not
include the photon interaction with
magnetic moments $\hat{\magmom}_\pm$ of the particles. 
For transitions without spin-flip, the latter interaction
can be taken into account by supplementing
the operator $\bm{e}_{\bm{q}\gamma}\cdot\bm{j}_\mathrm{eff}$ by the term
$
     -\ii
   (\bm{q}\times\bm{e})\cdot(\hat{\magmom}_-
   + \hat{\magmom}_+)
$
(cf.~\cite{KVH}), whereas operators
$(\bm{q}\times\bm{e})\times\hat{\magmom}_\pm$
are responsible for spin-flip transitions (cf.~\cite{Wunner_ea83}).
The corresponding contributions to the transition matrix elements are
proportional to $q$ and prove
to be of the same order of magnitude as the first-order corrections
($\propto q$) to the dipole approximation that we use below. 
Therefore, these terms will be neglected in the dipole approximation.
\\

\subsection{Dipole approximation}
\label{sect:dipole_appr}

In the dipole approximation for the
matrix elements of radiative transitions, 
$\erm{\ii\bm{q}\cdot\bm{r}}$ is replaced by 1. For bound-bound and bound-free
transitions this is justified, provided that the mean bound-state size
$l$ is much smaller than $q^{-1}=\omega_q/c$. Since the binding energy
$E_\mathrm{b}$ is by the order of magnitude $\sim Ze^2/l$, this requirement
translates into $\hbar\omega_q\ll\alphaf^{-1}E_\mathrm{b}/Z$, where
$\alphaf=e^2/\hbar c$ is the fine structure constant.
In this approximation, the effective current (\ref{jeff})
takes the form
\beq
   \bm{j}_\mathrm{eff} = \frac{e}{m_-}\,\bm{\pi}_- -
   \frac{Ze}{m_+}\,\bm{\pi}_+.
\eeq
Using the commutation relation
\beq
   \frac{\bm{\pi}_\pm}{m_\pm} = \frac{\ii}{\hbar}\,[H_0,\bm{r}_\pm],
\eeq
where $H_0$ is the field-free Hamiltonian given by \req{H0}, we can
transform  ``velocity form'' of the matrix element in \req{dsigma} to
the ``length form'' (cf., e.g., Ref.\,\cite{BetheSalpeter})
\beq
   \langle\psi_f| \bm{j}_\mathrm{eff} | \psi_i\rangle
   = -\frac{\ii}{\hbar} (E_f-E_i)\,\bm{D}_{fi} 
   = \ii\,\omega_q\bm{D}_{fi} .
\label{j_dip}
\eeq
Here, $\bm{D}_{fi} = \langle\psi_f | \bm{D} | \psi_i\rangle$
is the matrix element of the electric dipole moment,
\beq 
\bm{D} = Ze\,\bm{r}_+-e\bm{r}_- 
   = (Z-1)e\,\bm{R} - e\,\bm{r}.
\label{D}
\eeq

For the bound-bound transitions,
substitution of \req{j_dip} into \req{sigma_bb} gives
\beq
    \sigma_{i\to f,\bm{q}\gamma} = \frac{4\pi^2\omega_q}{\hbar
   c}\,| \bm{e}_{\bm{q}\gamma}\cdot\bm{D}_{fi} |^2
   \,\Delta_{fi}(\omega_q - \omega_{fi}).
\label{sigma_bb_dip}
\eeq
For the bound-free transitions, substitution  of \req{j_dip}
 into \req{sigma_bf} gives
\beq
   \sigma_{i\to f,\bm{q}\gamma} = 4\zmax\,
    \frac{\pi\mred\omega_q}{
   \hbar^2 k_f c}\,| \bm{e}_{\bm{q}\gamma}\cdot\bm{D}_{fi} |^2.
\label{sigma_bf_dip}
\eeq
In the cyclic coordinates
(\ref{cyclic}), we have
\beq
   \bm{e}_{\bm{q}\gamma}\cdot\bm{D}_{fi} = \sum_{\alpha=-1}^1
         e_{\bm{q}\gamma,-\alpha} D_{fi,\alpha}.
\eeq
Using \req{rRpm}, we can write the transverse (right and left)
cyclic components of the
dipole operator $\bm{D}$ as
\begin{widetext}
\bea
   D_{+1} &=& (Z-1)e \,\amZ{Z-1} (\hat{b}-\hat{\tilde{b}}^\dag)
    - e \am (\hat{a}^\dag - \hat{\tilde{a}})
    =
    e \,\am \,\big[ \sqrt{Z-1}\,(\hat{b}-\hat{\tilde{b}}^\dag)
    - \hat{a}^\dag + \hat{\tilde{a}} \big],
\label{Dp1}
\\
   D_{-1} &=& (Z-1)e \amZ{Z-1} (\hat{b}^\dag-\hat{\tilde{b}})
    - e \am (\hat{a}-\hat{\tilde{a}}^\dag)
    =
    e \,\am \,\big[ \sqrt{Z-1}\,(\hat{b}^\dag-\hat{\tilde{b}})
    - \hat{a}+\hat{\tilde{a}}^\dag \big],
\label{Dm1}
\eea
whereas the longitudinal component $D_0=-ez$ does not affect the
transverse states of motion. 
From these equations and Eqs.~(\ref{aa'bb'}), (\ref{aa'+bb'+}),
and (\ref{pi+ab}) we see that $D_\alpha$ transforms each
pure transverse state $|N,\tilde{N},n,\tilde{n}\rangle_\perp$ into the
superposition of such states,
\bea
   D_{+1}\,|N,\tilde{N},n,\tilde{n}\rangle_\perp &=& e\am\,
   \sqrt{Z-1}\,\big[\sqrt{N}\,|N-1,\tilde{N}\rangle_1 -
   \sqrt{\tilde{N}+1}\,|N,\tilde{N}+1\rangle_1
   \big]\otimes|n,\tilde{n}\rangle_2
\nonumber\\&&
   -  e\am\, |N,\tilde{N}\rangle_1\otimes
   \big[\sqrt{n+1}\,|n+1,\tilde{n}\rangle_2
    - \sqrt{\tilde{n}}\,|n,\tilde{n}-1\rangle_2
   \big],
\label{D+1}
\\
   D_{-1}\,|N,\tilde{N},n,\tilde{n}\rangle_\perp &=& e\am\,
   \sqrt{Z-1}\,\big[\sqrt{N+1}\,|N+1,\tilde{N}\rangle_1 -
   \sqrt{\tilde{N}}\,|N,\tilde{N}-1\rangle_1
   \big]\otimes|n,\tilde{n}\rangle_2
\nonumber\\&&
   -  e\am\, |N,\tilde{N}\rangle_1\otimes
   \big[\sqrt{n}\,|n-1,\tilde{n}\rangle_2
    - \sqrt{\tilde{n}+1}\,|n,\tilde{n}+1\rangle_2
   \big].
\label{D-1}
\eea
\end{widetext}
It follows that $D_\alpha$ transforms a state with
a definite $L=\tilde{N}-N+n-\tilde{n}$ into a state with $L'=L+\alpha$.
This entails the selection rule
\beq
   L_f = L_i + \alpha.
\label{Lselect}
\eeq 
It reflects conservation of the total angular momentum of the
entire system, comprising an electron, a nucleus, and a photon.
As a consequence, in the right-hand side of \req{sigma_bf_dip} we can
use the expansion
\beq
   | \bm{e}_{\bm{q}\gamma}\cdot\bm{D}_{fi} |^2 =
     \sum_{\alpha=-1}^1
         |e_{\bm{q}\gamma,-\alpha}|^2 |D_{fi,\alpha}|^2.
\eeq
Terms $e_{-\alpha}e_{-\alpha'}^*D_{fi,\alpha}D_{fi,\alpha'}^*$
with $\alpha\neq\alpha'$
are absent, because $D_\alpha$ and $D_{\alpha'}$ transform a pure
quantum state $|\psi_i\rangle$
into states
$|\psi_f\rangle$ and
$|\psi_{f'}\rangle$ with different $z$-projections of the angular
momentum, so that $D_{fi,\alpha}$ and $D_{fi,\alpha'}$ cannot
be non-zero simultaneously.
Thus
Eqs.\,(\ref{sigma_bb_dip})
and (\ref{sigma_bf_dip}) can be written as
\beq
   \sigma_{i\to f,\bm{q}\gamma} = 
     \sum_{\alpha=-1}^1
         |e_{\bm{q}\gamma,\alpha}|^2\,\sigma_{i\to
                f,\alpha}^\mathrm{(bb,bf)}(\omega_q)
\eeq
where
\bea
   \sigma_{i\to f,\alpha}^\mathrm{bb}(\omega) &=& \frac{4\pi^2\omega}{\hbar c}\,
     |D_{fi,-\alpha}|^2
   \,\Delta_{fi}(\omega - \omega_{fi}),
\label{sigma_bb_dip1}
\\
   \sigma_{i\to f,\alpha}^\mathrm{bf}(\omega) &= &
    4\zmax\,
    \frac{\pi \mred\omega}{
   \hbar^2 c \,k_f}\,|D_{fi,-\alpha}|^2
\label{sigma_bf_dip1}
\eea
for the bound-bound and bound-free transitions, respectively. 

Neglecting the Doppler broadening, we can model $\Delta_{fi}(\omega)$ in
\req{sigma_bb_dip1} by the Lorentz-Cauchy profile,
\beq
   \Delta_{fi}(\omega) = \frac{1}{\pi}\,
   \frac{\nu_\mathrm{eff}}{(\omega-\omega_{fi})^2+\nu_\mathrm{eff}^2},
\eeq
where $\nu_\mathrm{eff}$ is an effective damping frequency. In the simplest
approximation,
$
 \nu_\mathrm{eff} = \nu_\mathrm{coll} + \nu_\mathrm{rad}$,
 where $\nu_\mathrm{coll}$ is an effective frequency of collisions of a
 given ion with plasma particles, and
\beq
    \nu_\mathrm{rad} = \frac{4\omega_{fi}^3}{3\hbar c^3}
    \sum_{\alpha=-1}^1 |D_{fi,\alpha}|^2
\eeq
is the natural radiative width (cf., e.g., \cite{LaLi-QED}).

For practical computations, it is convenient to consider the continuum
wave functions normalized  so that the amplitude of the outgoing wave in
a selected open channel equals 1 at infinity. Then the factor
$4\zmax$ in \req{sigma_bf_dip1} drops out from the numerical
code, being canceled by the squared normalization constant
(\ref{Cnorm}).

Kopidakis et al. \cite{KVH} defined
a dimensionless interaction operator, which can be written as
\beq
   \hat{\bm{M}} = \frac{2\hbar}{e^3}\,
   \bm{e}_{\bm{q}\gamma}\cdot\bm{j}_\mathrm{eff} .
\eeq
The correspondence between the ``velocity form'' and ``length form'' of
this operator has been discussed in Ref.~\cite{PPV97} regarding the
problem of a hydrogen atom in a strong magnetic field. It was also
employed in
Ref.~\cite{PP97} for treatment of the bound-free transitions of a
hydrogen atom moving in the magnetic field. Using the
dipole approximation, we can rewrite \req{sigma_bf_dip1} in the same
form as Eq.\,(7) of \cite{PP97}:
\bea
   \sigma_{i\to f,\alpha}^\mathrm{bf}(\omega) &=&
    \frac{\pi\alphaf \zmax }{k_f}\,\frac{\mbox{Ha}}{\hbar\omega}\,
      \frac{\mred}{\mel}\,
      |M_{fi,-\alpha}|^2
\nonumber\\
      &=& 2\pi\alphaf\,\frac{\mbox{Ry}}{\hbar\omega}\,
        \sqrt{\frac{\mbox{Ry}}{E_f^\|}\frac{\mred}{\mel}}\,
         \zmax\aB\,
      |M_{fi,-\alpha}|^2,
\qquad
\label{sigma_bf_M}
\eea
where $M_{fi,\alpha} = \ii (\hbar\omega/\mbox{Ry})\,
 D_{fi,\alpha}/e\aB$ is the respective cyclic
component of the dimensionless matrix element
$\langle\psi_f| 
   \bm{e}_{\bm{q}\gamma}\cdot\hat{\bm{M}} | \psi_i\rangle$,
$\aB=\hbar^2/\mel e^2$ is the Bohr radius,
$\mbox{Ha}=2\mbox{Ry}=e^2/\aB$ is the Hartree energy unit,
Ry being the Rydberg energy
and $\mel$ the electron mass.

For the bound-bound transitions, it is customary to define dimensionless
oscillator strengths (e.g., \cite{HasegawaHoward61})
\beq
   f_{fi,\alpha} = \frac{\hbar\omega}{\mbox{Ry}}\,
       \left| \frac{D_{fi,-\alpha}}{e\aB} \right|^2.
\label{osc}
\eeq
In these notations, \req{sigma_bb_dip1} can be written as
\beq
   \sigma_{i\to f,\alpha}^\mathrm{bb}(\omega) = 2\pi^2
     \frac{e^2}{\mel c}\,f_{fi,\alpha}\,\Delta_{fi}(\omega).
\label{sigma_bb_f}
\eeq

\subsection{Expansion on the transverse basis}
\label{sect:dipole_expan}

Let us use the basis expansion (\ref{psiPsi}) 
for calculation of the matrix elements
$D_{fi,\alpha} = \langle\psi_f | D_\alpha | \psi_f \rangle$.
For the longitudinal polarization ($\alpha=0$), using the
orthonormalization condition (\ref{normPsi}), we obtain
\beq
   D_{fi,0} \!=\! -e\delta_{\tilde{N}_f \tilde{N}_i}\delta_{L_f L_i}
   \!\!\!\!
   \sum_{n_-,n_+}
   \!\!\!
    \int_{-\zmax}^{\zmax}
   \hspace*{-2em}
   g_{n_-n_+;\,\kappa_f}^*(z)\,g_{n_-n_+;\,\kappa_i}(z)
   \,z\,\dd z.
\label{D_0}
\eeq
In the adiabatic approximation, we are left with the only term with
$n_-=n_{-,i}=n_{-,f}$ and $n_+=n_{+,i}=n_{+,f}$, while transitions
with $n_{-,i}\neq n_{-,f}$ or $n_{+,i}\neq n_{+,f}$ are forbidden. Beyond
the adiabatic approximation, the latter transitions are allowed, but
$|D_{fi,0}|$ is small compared to the case without changing $n_-$ and
$n_+$. For the initial and final states with definite $z$-symmetry, the
corresponding selection rule follows: $D_{fi,0}$ is non-zero only for
transitions between the state of opposite $z$-symmetry. In particular,
bound-bound transitions between states with $\nu_i$ and $\nu_f$ of
the same parity (both even or both odd) are dipole forbidden.

For the circular polarizations
$\alpha=\pm1$, using
Eq.~(\ref{pi+ab}), we can transform Eqs.\,(\ref{Dp1}) and (\ref{Dm1})
 to
\bea
   \frac{D_{+1}}{e\am} &=& \frac{\ii\am}{\hbar} U^\dag\pi_{+,+1}U
                  - \hat{a}^\dag - \sqrt{Z-1}\,\hat{\tilde{b}}^\dag ,
\\
  \frac{D_{-1}}{e\am} &=& - \frac{\ii\am}{\hbar} U^\dag\pi_{+,-1}U
                  - \hat{a}  - \sqrt{Z-1}\,\hat{\tilde{b}}  .
\qquad
\eea
The first term in each of these equations shifts the quantum number
$n_+$ by $\pm1$ according to \req{pi+cyclic}, the second term shifts the
number $n_-$ according to Eqs.\,(\ref{aa'1}), (\ref{aa'2}), (\ref{aa'+1})
and (\ref{aa'+2}), while the
last term shifts the number $\tilde{N}$ according to Eqs.\,(\ref{bb'1}),
(\ref{bb'2}),  (\ref{bb'+1}),
and (\ref{bb'+2}). 
In all the cases the selection rule (\ref{Lselect}) holds.
Thus the transverse basis states (\ref{bastat}) are
transformed as
\begin{widetext}
\bea\hspace*{-2em}
   D_{+1} | \tilde{N},L,n_-,n_+ \rangle_{\perp,0}/e\,\am &=& 
    \sqrt{Z}\,\sqrt{n_+}\,| \tilde{N},L+1,n_-,n_+ -1 \rangle_{\perp,0}
    - \sqrt{n_- +1} \,| \tilde{N},L+1,n_- +1 ,n_+ \rangle_{\perp,0}
\nonumber\\&&
    - \sqrt{Z-1}\,\sqrt{\tilde{N}+1}\,
     | \tilde{N}+1,L+1,n_-,n_+ \rangle_{\perp,0}  ,
\label{Dp1tr}
\\
   D_{-1} | \tilde{N},L,n_-,n_+ \rangle_{\perp,0}/e\,\am &=& 
    \sqrt{Z}\,\sqrt{n_+ + 1}\,| \tilde{N},L-1,n_-,n_+ +1
     \rangle_{\perp,0}
    - \sqrt{n_-} \,| \tilde{N},L-1,n_- -1 ,n_+ \rangle_{\perp,0}
\nonumber\\&&
    - \sqrt{Z-1}\,\sqrt{\tilde{N}}\,
     | \tilde{N}-1,L-1,n_-,n_+ \rangle_{\perp,0} .
\label{Dm1tr}
\eea
Let us substitute the basis expansion (\ref{Psi-sum}) into
$D_{fi,\alpha}=\langle\psi_f|D_\alpha|\psi_i\rangle$ with $\alpha=\pm1$
 and use relations
(\ref{Dp1tr}) and (\ref{Dm1tr}). Then for transitions with
$L_f=L_i\pm1$ and $\tilde{N}_f = \tilde{N}_i$,
taking into account the orthogonality relation (\ref{normPsi}), 
we obtain
\bea
  \frac{D_{fi,+1}}{e\am} &=& \sqrt{Z}\,
   \sum_{n_+=0}^{\mathcal{N}} \sqrt{n_+}\,\sum_{n_-=0}^\infty
    \mathcal{L}_{n_-,n_+ -1;\,n_-,n_+}(\kappa_f|\kappa_i)
    - \sum_{n_-=0}^\infty \sqrt{n_- +1} \,\sum_{n_+=0}^{\mathcal{N}}
    \mathcal{L}_{n_- +1,n_+;\,n_-,n_+}(\kappa_f|\kappa_i),
\label{D+}
\\
  \frac{D_{fi,-1}}{e\am} &=& \sqrt{Z}\,
   \sum_{n_+=0}^{\mathcal{N}} \sqrt{n_+ +1}\,\sum_{n_-=0}^\infty
    \mathcal{L}_{n_-,n_+ +1;\,n_-,n_+}(\kappa_f|\kappa_i)
    - \sum_{n_-=0}^\infty \sqrt{n_-} \,\sum_{n_+=0}^{\mathcal{N}}
    \mathcal{L}_{n_- -1,n_+;\,n_-,n_+}(\kappa_f|\kappa_i),
\label{D-}
\eea
\end{widetext}
where $\mathcal{N}\equiv\tilde{N}_i-L_i+n_-$
and $\mathcal{L}$ denotes the longitudinal overlap integral,
\beq
   \mathcal{L}_{n_-',n_+';\,n_-,n_+}(\kappa'|\kappa) =\!\!
     \int_{-\zmax}^{\zmax}\!\!\!
       g_{n_-',n_+';\,\kappa'}^*g_{n_-,n_+;\,\kappa}\,\dd{}z.
\label{LongInt}
\eeq

The last term in each of equations (\ref{Dp1tr}) and (\ref{Dm1tr})
corresponds to transitions with
$L_f=L_i+\alpha$ and $\tilde{N}_f = \tilde{N}_i+\alpha$, which leave
$(\tilde{N}-L)$ unchanged. The corresponding dipole
matrix elements equal
$e\am\sqrt{(Z-1)\mathrm{max}(N_i,N_f)}\,\langle\psi_f|\psi_i\rangle =0$,
because $\psi_i$ and $\psi_f$ are orthogonal.
The absence of such transitions agrees
with the degeneracy of the problem in $\tilde{N}$,
discussed above. Thus it is sufficient to study only the transitions
between states with $\tilde{N}=0$; the results for non-zero $\tilde{N}$
are then obtained by adding $\tilde{N}$ to both $L_i$ and $L_f$.

\section{Approximate solutions for ultra-strong fields}
\label{sect:approx}

In this section we consider an approximate treatment of the hydrogenlike
ion in the full adiabatic approximation (Sect.~\ref{sect:adiabatic}),
using the method previously developed by \citet{HasegawaHoward61} for a
strongly magnetized H atom.

\subsection{Wave functions and eigenenergies}
\label{sect:approxWF}

Let us find an approximate solution to the Schr\"odinger equation in the
adiabatic approximation, \req{Schradiab}, for the bound states.
Following \citet{HasegawaHoward61}, we find asymptotic solutions
at small and large $z$
and match them at an intermediate point 
\beq
  z_0 = C\am^\lambda, \quad 0 < \lambda < 2/3,
  \label{z_0}
\eeq
where $C$ and $\lambda$ are constants, independent of $z$ and $\am$. 
The matching is provided by
equating the logarithmic derivatives 
\beq
 \eta(z) =
\frac{g'_\kappa(z)}{g_\kappa(z)} 
\label{logderiv}
\eeq 
of the interior and exterior
solutions at $z=z_0$.  The bounds on $\lambda$ in \req{z_0} ensure that
$z/\am\to\infty$ and $z^2\,|\ln\am|\to0$ with $\am\to0$ for all $z>z_0$,
which is needed for validation of the approximate exterior solution
(Sect.\,\ref{sect:exterior}), while $z^3/\am^2\to0$  for all $z<z_0$, as
required for the validity of the approximate exterior solution
(Sect.\,\ref{sect:even}).

\subsubsection{Exterior solution for bound states}
\label{sect:exterior}

At $\zeta\to\infty$ all effective potentials
$V_{0,n_+}^{(\mathcal{N})}(z)$ converge to the 1D Coulomb potential,
so that \req{Schradiab} is replaced by
\beq
   -\,\frac{\hbar^2}{2\mred}\frac{\dd^2 g_\kappa}{\dd z^2}
     \,-\,\frac{Ze^2}{|z|}g_\kappa=E_\kappa^\|g_\kappa.
\label{1DSchr}
\eeq
The well known solution to this equation for a bound state (i.e., for
$E_\kappa^\|<0$, so that $\lim_{z\to\infty} g_\kappa(z) =0$) is
\beq
   g_\mathrm{ext}(z) = C_W W_{\efqm,\frac12}(2z/\efqm\aBred),
\label{Whit}
\eeq
where $W_{\efqm,\frac12}(x)$ is the Whittaker function \cite{AS},
 $C_W$ is a normalization constant,
\beq
   \aBred = \frac{\hbar^2}{\mred Z e^2},
\label{aBred}
\eeq
is the ``effective Bohr radius'', $\efqm$ is the ``effective principal
quantum number'' defined through the relation
\beq
    \epsilon_\kappa \equiv \frac{|E_\kappa^\||}{\mathrm{Ry_*}}
     \equiv \frac{1}{\efqm^2},
\label{efqunum}
\eeq
where
\beq
   \mathrm{Ry_*} \equiv \frac{Z^2 e^4\mred}{2\hbar^2}
=\frac{Z^2\mred}{\mel}\,\mbox{Ry}
\label{effRy}
\eeq
is the ``effective Rydberg energy''.

At small (but non-zero) argument $x=2z/\efqm\aBred$ and non-integer
$\efqm$, the Whittaker function can be expanded as
\cite{HasegawaHoward61}
\bea&&\hspace*{-1em}
   W_{\efqm,\frac12}(x) = - \Gamma(\efqm) 
\bigg\{ \left[\efqm x- (\efqm x)^2\right] \cos\efqm\pi 
\nonumber\\&&\hspace*{-1em}
  + \left[ -1+ \efqm x\left(\ln\frac{x}{\efqm} + \frac{1}{2\efqm} +
\psi(\efqm) - \psi(1) - \psi(2) \right)\right]
\frac{\sin\efqm\pi}{\pi}
\nonumber\\&&
 + O(\efqm\, x\ln x) \bigg\} ,
\label{Whit1}
\eea
where 
\beq
   \psi(x)\equiv \frac{\dd\ln\Gamma(x)}{\dd x}
\label{digamma}
\eeq
is the digamma
function \cite{AS}; $\psi(1)=\psi(2)-1=-\gamma_\mathrm{E}$, where
$\gamma_\mathrm{E}=0.5772\ldots$ is the Euler-Mascheroni constant.

If $z_0$ were zero, $\efqm$ would be integer. At non-zero $z_0$, with
$\am\to0$, $\efqm$ tends to integer values in such a way that
$\sin\efqm\pi$ tends to zero logarithmically (as
will be seen from the solutions below), that is slower than $z_0\to0$.
Therefore, $z_0\cot\efqm\pi\to0$, and from \req{Whit1} we obtain
\beq
   W_{\efqm,\frac12}\left(\frac{2z_0}{\efqm\aBred}\right) =
     \Gamma(\efqm)\,\frac{\sin\efqm\pi}{\pi}
     \left[1+O\left(\am^{1-\lambda}\cot\efqm\pi\right)\right].
\label{Whit0}
\eeq
Taking the derivative in \req{Whit1} and dividing by \req{Whit0}, we
obtain the logarithmic derivative in the form
\bea
   \eta_\mathrm{ext} &=& \frac{\dd}{\dd z}
    \ln W\left(\frac{2z}{\efqm\aBred}\right)
\nonumber\\ &\approx&
   -\frac{2}{\aBred}\left(\ln\frac{2z}{\aBred}
    + \pi\,\cot\efqm\pi +2\gamma_\mathrm{E}
  -\Theta(\efqm) \right),
\qquad
\label{eta_ext}
\eea
where $z \sim z_0 \ll \aBred$ and
\beq
   \Theta(\efqm) = \ln\efqm-\frac{1}{2\efqm}-\psi(\efqm)
   = \ln\efqm+\frac{1}{2\efqm}-\psi(1+\efqm).
\label{Theta}
\eeq

\begin{figure*}
\centering
\includegraphics[width=.52\textwidth]{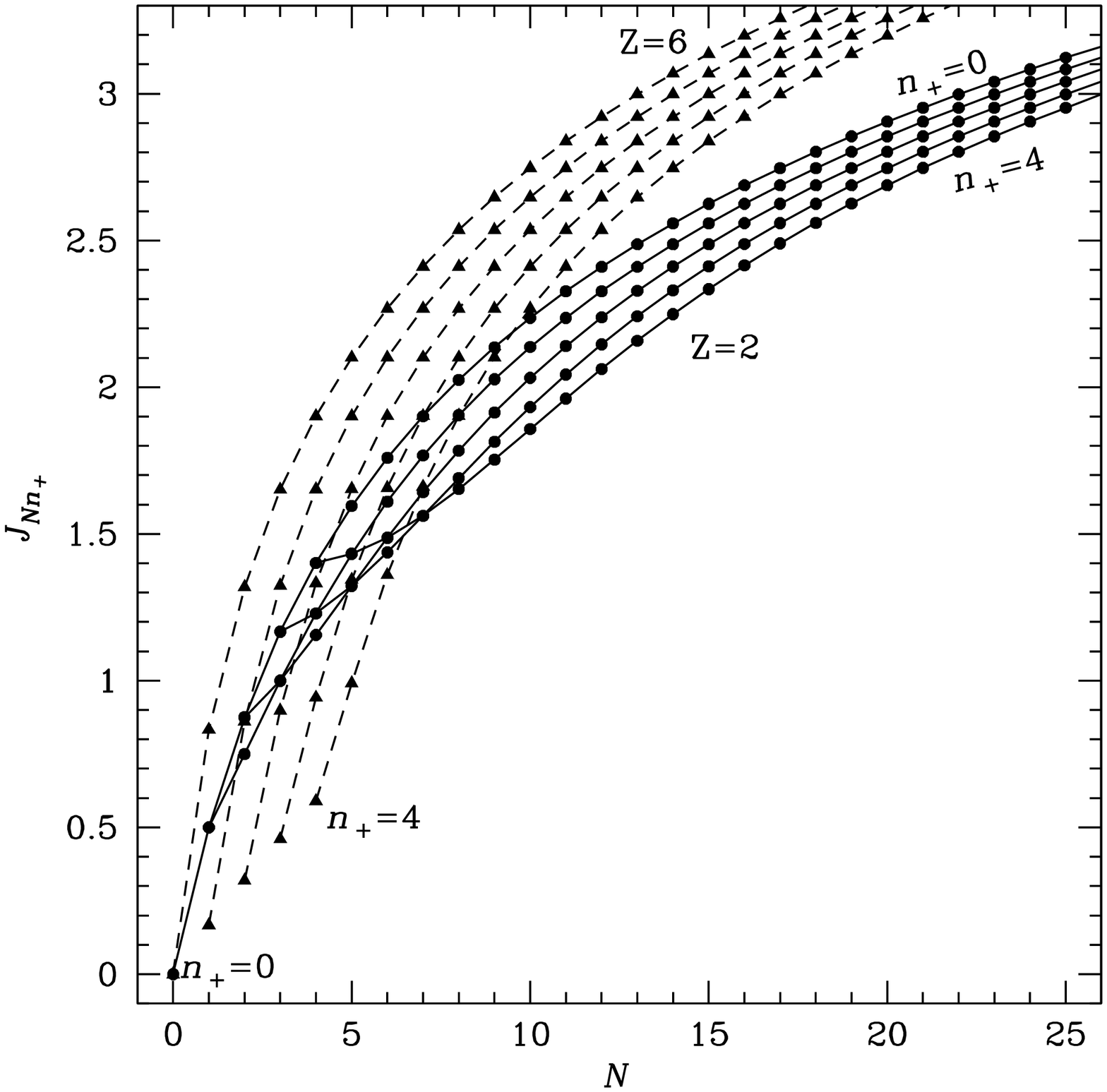}
\includegraphics[width=.46\textwidth]{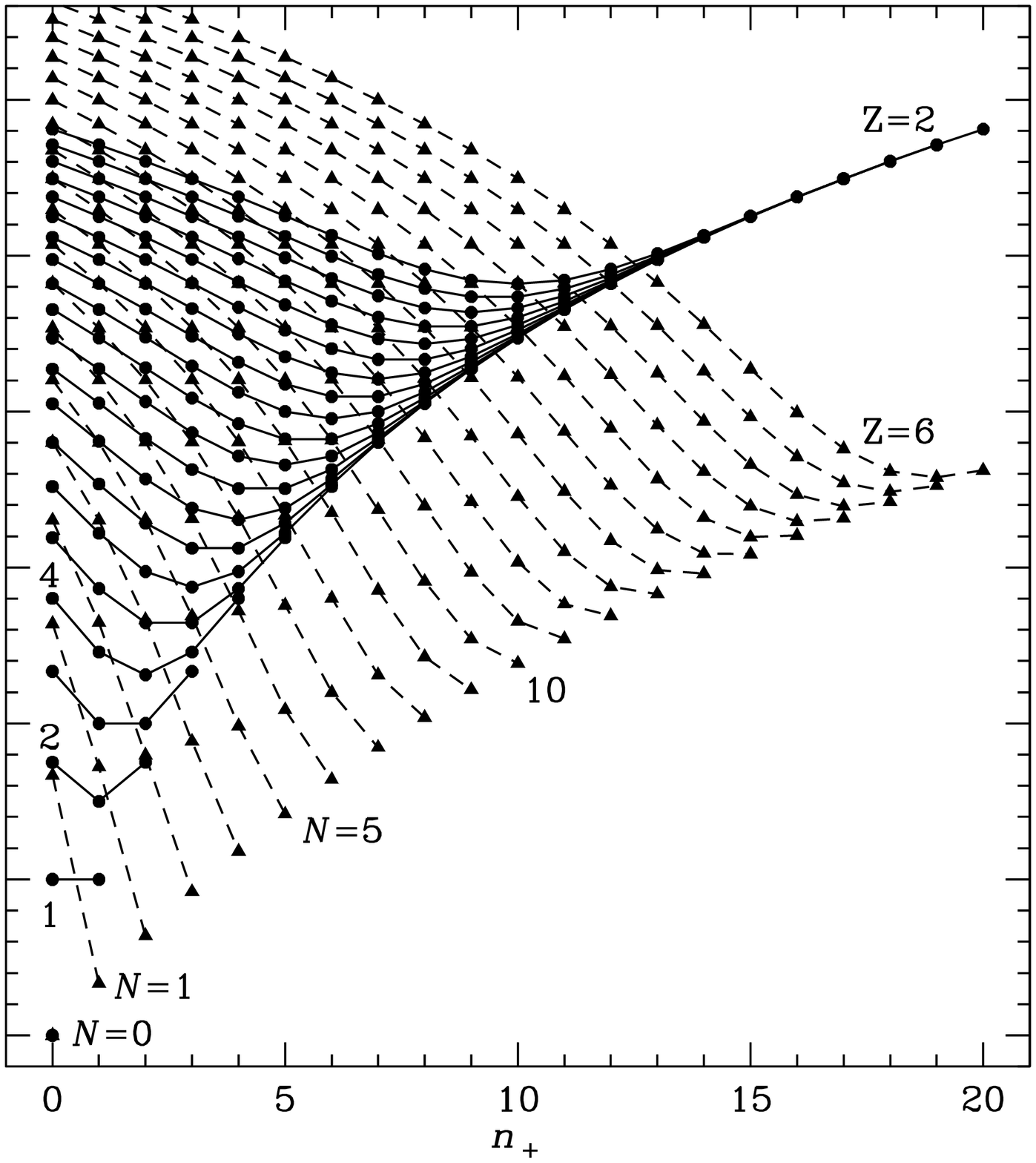}
\caption{Dependence of the quantities $J_{\mathcal{N}n_+}$ [\req{JNn}]
on quantum number $\mathcal{N}$ (left panel, for $n_+=0$, 1, 2, 3, and
4) and on quantum number $n_+$ (right panel, for $\mathcal{N}$ from 0 to
20) for $Z=2$ (dots connected by solid lines) and $Z=6$ (triangles
connected by dashed lines).
}
\label{fig:J}
\end{figure*}

\subsubsection{Interior solution for even states}
\label{sect:even}

Now let us consider $|z|\ll z_0$. In terms of the dimensionless
coordinate $\zeta$
(\ref{zeta}), the Schr\"odinger equation in the adiabatic approximation
[\req{Schradiab}] becomes
\beq
   \frac{\dd^2 g_\kappa}{\dd\zeta^2} +2^{3/2}\frac{\am}{\aBred}\,
   v_{0,n_+}^{(\mathcal{N})}(\zeta)\,g_\kappa -
   2\left(\frac{\am}{\aBred}\right)^2\epsilon_\kappa\,g_\kappa =0,
\label{Schred}
\eeq
where $\epsilon_\kappa$ is the
dimensionless eigenvalue defined in \req{efqunum}.
We solve this equation approximately, using
 the perturbation theory with respect to the small parameter
$\am/\aBred$ separately for the even and odd states. 
Since in this section we are interested in the bound states,
$E_\kappa^\|<0$, we can safely set $n_-=0$. Then
$\mathcal{N}=\tilde{N}-L$ and
\beq
   v_{0,n_+}^{(\mathcal{N})} = \sum_{k=0}^{\mathcal{N}}
   \left(C_k^{(\mathcal{N},n_+)}\right)^2
   \int_0^\infty \frac{I_{k,0}^2(\rho)}{\sqrt{\rho+\zeta^2}} \,\dd\rho,
\label{v0}
\eeq
where, according to \req{I-explicit}, $I_{k,0}(\rho) =
\rho^{k/2}\erm{-\rho/2}/k!$, and coefficients $C_k^{(\mathcal{N},n_+)}$
are given by the relation (\ref{Crecur}) and normalization
(\ref{normC}).

Let us consider even states.
In the first approximation (linear in $\am/\aBred$),
the last term in \req{Schred} drops out, and the solution is
proportional to
\[
   1-\frac{2^{3/2}\am}{\aBred}\int_0^\zeta\dd\zeta'
   \int_0^{\zeta'}\!\!\! \dd\zeta'' \,v_{0,n_+}^{(\mathcal{N})}(\zeta'')
\]
The logarithmic derivative for the interior solution becomes
\beq
   \eta_\mathrm{int}(z) \approx -\frac{2}{\aBred}\int_0^{z/\am\sqrt2} 
    v_{0,n_+}^{(\mathcal{N})}(\zeta)\,\dd\zeta.
\label{eta_int1}
\eeq
As shown in Appendix~\ref{sect:eta_int},
this expression leads to
\beq
   \eta_\mathrm{int} = -\frac{1}{\aBred} \left[ 2\ln\frac{z_0}{\am}
         + \ln2 + \gamma_\mathrm{E}
    - J_{\mathcal{N}n_+} + O\left(\frac{\am^2}{z_0^2}\right)
     \right],
\label{eta_int}
\eeq
where
$
  H_k=\sum_{n=1}^k n^{-1}
$
is the $k$th harmonic number, $J_{00}=0$, and for $\mathcal{N}\geq1$
\beq
   J_{\mathcal{N}n_+} = \sum_{k=1}^{\mathcal{N}} H_k
   \left(C_k^{(\mathcal{N},n_+)}\right)^2.
\label{JNn}
\eeq
In the particular case $\mathcal{N}=0$, which corresponds to the
non-moving ion, \req{eta_int} reduces to 
the result of Hasegawa \& Howard \cite{HasegawaHoward61}.
For $\mathcal{N}=1$, we obtain
\beq
   J_{10} = \frac{Z-1}{Z},
\quad
  J_{11} = \frac{1}{Z}.
\eeq
The dependence of the quantities $J_{\mathcal{N}n_+}$ 
on the quantum numbers $\mathcal{N}$ and $n_+$ is shown in
Fig.~\ref{fig:J}.

\subsubsection{Binding energies of even states}
\label{even}

By equating $\eta_\mathrm{int}$ (\ref{eta_int}) to $\eta_\mathrm{ext}$
(\ref{eta_ext}) at $z=z_0$ we obtain the following equation for the
effective quantum number $\efqm$:
\beq
   \pi\,\cot\efqm\pi + \Theta(\efqm) =
    \ln\frac{\aBred}{2\am} + \frac{\ln2-3\gamma_\mathrm{E}
    - J_{\mathcal{N}n_+}}{2}
\label{even_spectrum}
\eeq
    $(|\sin\efqm\pi|\ll1)$,
where $\Theta(\efqm)$ is defined by \req{Theta}.
Solution of \req{even_spectrum} gives the longitudinal
energies of the even states through \req{efqunum}.

\begin{figure*}
\centering
\includegraphics[width=.567\textwidth]{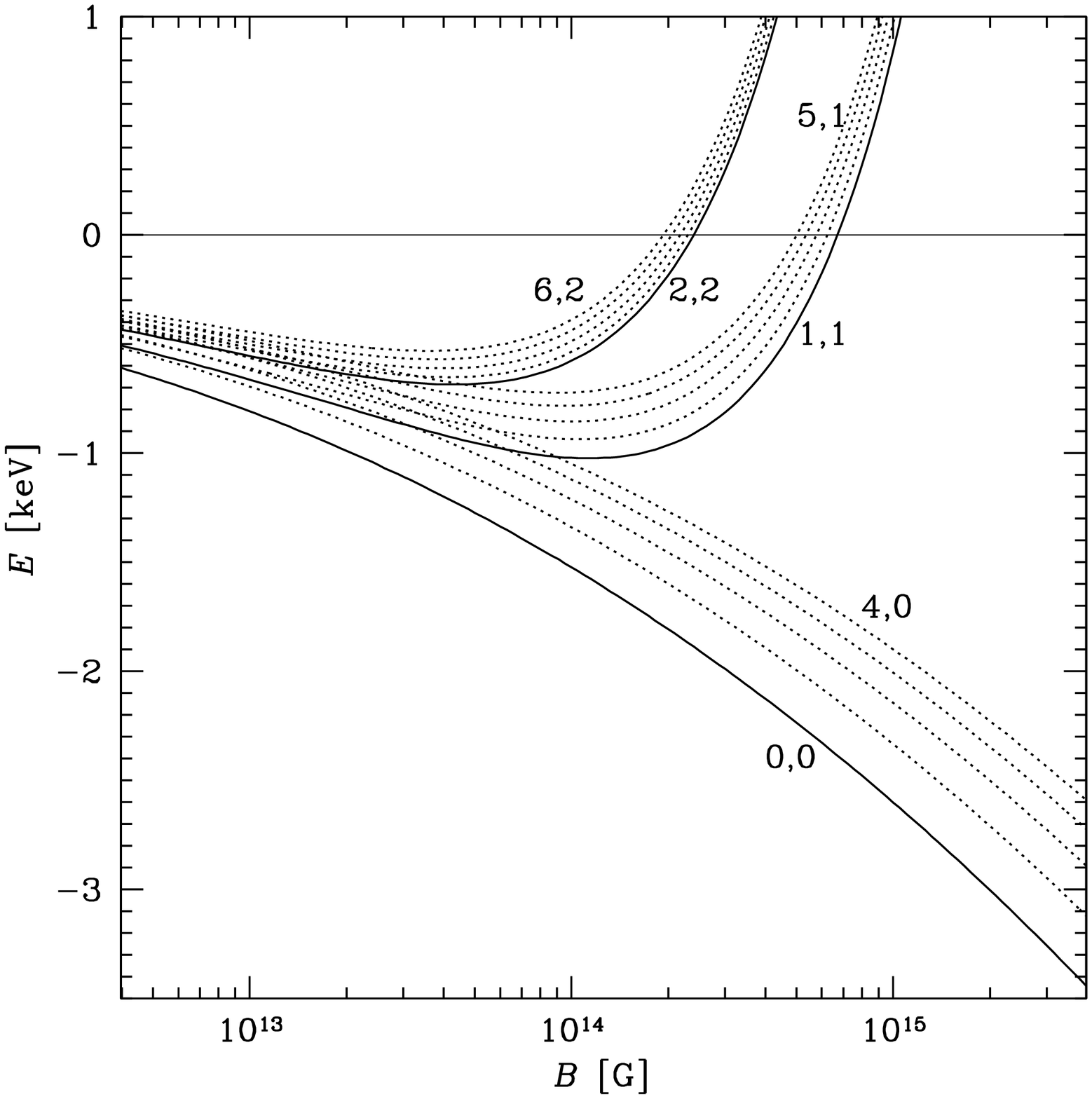}
\includegraphics[width=.32\textwidth]{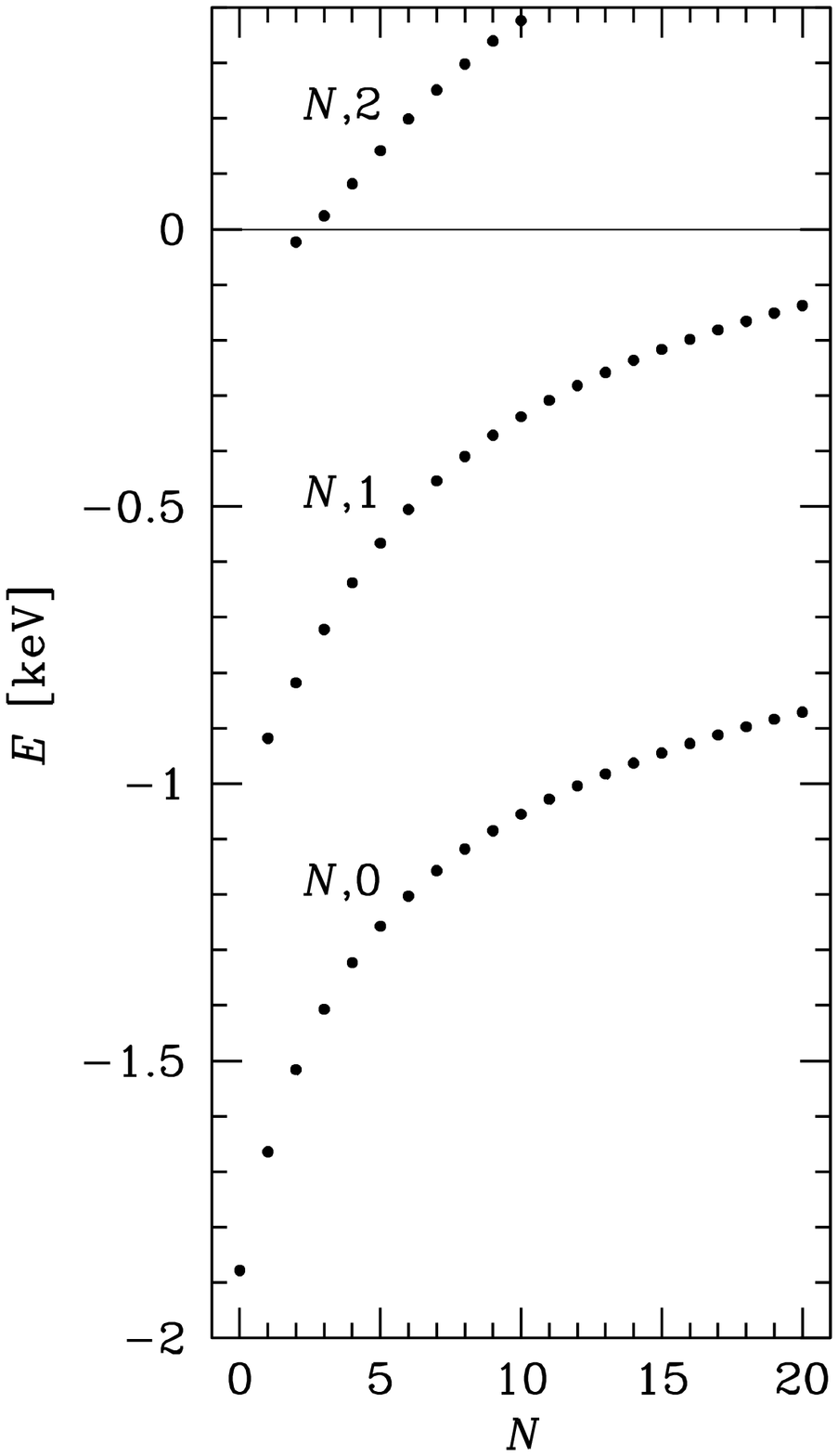}
\caption{Energy levels of the helium ion in strong magnetic fields.
\textit{Left panel}: Dependence of the energies on magnetic field
strength. The numbers at the curves are $\mathcal{N}$ and $n_+$, which
mark the discrete states $|\kappa\rangle =
|\mathcal{N},n_-,n_+,\nu\rangle$ with $n_-=0$ and
$\nu=0$ (i.e., only 
tightly-bound states are considered).
\textit{Right panel}: Dependence of the energies on $\mathcal{N}$
for the three smallest values of $n_+$ at $B=2.35\times10^{14}$~G.
}
\label{fig:energies}
\end{figure*}

The lowest-energy state for each $\mathcal{N}$
and $n_+$ (so called \emph{tightly bound state})
corresponds to $\efqm\ll1$. In this case, the
series expansion (\cite{AS}, 6.3.14, 23.2),
$
   \psi(1+\efqm) = -\gamma_\mathrm{E} + (\pi^2/6)\,\efqm -
    (\pi^4/90)\,\efqm^2+\ldots
$
leads to an approximate relation
$
   \Theta(\efqm)  \approx \ln\efqm+1/(2\efqm) + \gamma_\mathrm{E}
     - (\pi^2/6)\,\efqm.
$
To the same accuracy up to $O(\efqm^2)$, 
$\pi\cot\pi\efqm\approx \efqm^{-1} - \pi^2\efqm/3$.
Thus from \req{even_spectrum} we get approximately
\beq
   \frac{1}{\efqm} =
    \ln\frac{\gamma_*}{2} + 2\ln\efqm
      - \gamma_\mathrm{E} - J_{\mathcal{N}n_+} 
       + \frac{\pi^2}{3}\efqm,
\label{tightly}
\eeq
where we have introduced the magnetic-field parameter
\beq
   \gamma_* = \frac{\aBred^2}{\am^2}
    = \frac{\hbar^3 B}{Z^2\mred^2e^3c}
    =  \frac{B}{B_*},
\label{gamma*}
\eeq
which should be large in the considered approximation. Here,
$
  B_* \equiv
   Z^2\mred^2e^3c/\hbar^3
  \approx 2.35\times10^9\,(\mred/\mel)^2Z^2
$~G.
Equation~(\ref{tightly}) can be easily solved numerically.
For example, as long as 
$\ln(\gamma_*/2)>\gamma_\mathrm{E} + J_{\mathcal{N}n_+}$,
one can use iterations of the form
\bea
  \efqm_{i+1} &=& w\, \left(
    \ln\frac{\gamma_*}{2} + 2\ln\efqm_i
      - \gamma_\mathrm{E} - J_{\mathcal{N}n_+} 
       + \frac{\pi^2}{3}\efqm_i
       \right)^{-1}
\nonumber\\&&
       + (1-w)\,\efqm_i
\eea
with an appropriate weight $w$, starting from
\beq
   \frac{1}{\efqm_0} = \ln\frac{\gamma_*}{2} 
      - \gamma_\mathrm{E} - J_{\mathcal{N}n_+} .
\label{lowest_order}
\eeq
Having tried different values of $w$ at different $\gamma_*$
we found that a good choice of $w$, which provides a quick
convergence of the iterations, can be roughly approximated as
$w=(1+4\efqm_0^{3/2})/(1+15\efqm_0^{3/2})$.

In the ground state, $\mathcal{N}=n_+=0$, 
and \req{tightly} without the last (linear in $\efqm$) term 
reproduces the result of Hasegawa and Howard \cite{HasegawaHoward61}.
However, by means of a comparison with the numerical results from
Ref.~\cite{PB05} we found that even at the superstrong field strengths
$B\sim10^{15}$~G, encountered in magnetars, a solution of such a truncated equation
produces an unacceptable error  $\sim200$~eV in the ground-state energy.
The linear term in \req{tightly} considerably improves
the accuracy.

In Fig.~\ref{fig:energies} we show the dependence of energy
$E_\kappa=E_\kappa^\| + \hbar\omc{+} n_+$ on the magnetic field strength
$B$ and on the quantum number $\mathcal{N}$ (the left and right panels,
respectively). The ``electron cyclotron'' quantum number $n_-$ must
equal zero in the displayed energy range. The zero-point energy
$E_{0,0}^\perp$ is dismissed, because it does not affect the binding,
being the same for the bound and unbound states. It is of interest to
compare the $B$-dependences of our analytic estimates of energy shown in
the left panel  with Fig.~2 in Ref.~\cite{PB05}, which shows the
analogous dependences computed numerically. We can notice a qualitative
difference at $B < 10^{14}$~G, where different energy branches overlap
in our figure. This difference is caused by the fact that our analytic
estimates have a good accuracy only at superstrong fields. At such high
fields, however, the branches corresponding to non-zero $n_+$ merge into
continuum, because  $E_\kappa^\|$ increases slower (logarithmically)
than the  energy of transverse excitations $n_+\hbar\omc{+}$.
Ultimately, at $B\gtrsim7\times10^{14}$~G only the branch with $n_+=0$
survives, while the states with $n_+ > 0$ become metastable (i.e., turn
into continuum resonances).

The case where $\efqm$ is not small corresponds to the
``loosely-bound'' or ``hydrogenlike'' states. 
As long as the logarithm on the right-hand side of \req{even_spectrum}
is large, this equation can be satisfied only if $\cot\efqm\pi$ is
also large (tends to infinity at $\am\to0$).
Then we can write
\beq
   \efqm = \frac{\nupar}{2}+\delta_{\nupar},
\quad
   \nupar=2,4,6,\ldots,
\eeq
where the even numbers $\nupar$ enumerate the even states
and the quantum defect $\delta_{\nupar}$ is given by
\beq
   \frac{1}{\delta_{\nupar}} \approx
   \pi\,\cot\pi\delta_{\nupar}
\approx 
    \frac12\left(\ln\frac{\gamma_*}{2}  - 3\gamma_\mathrm{E}
    - J_{\mathcal{N}n_+}  \right)
   + \Theta\left(\frac{\nupar}{2}\right),
\label{loosely}
\eeq
where function $\Theta$ is defined by \req{Theta}.

The ``tightly-bound'' solution described above [\req{tightly}]
corresponds formally to $\nupar=0$. One should note that the
approximations (\ref{tightly}) and (\ref{loosely}) are based on the
condition that the right-hand side in \req{even_spectrum} is large,
implying that  $J_{\mathcal{N}n_+} \ll\ln\gamma_*$. One can show (see
Appendix~\ref{sect:Jlarge}) that $J_{\mathcal{N}0}\sim \ln\mathcal{N}$
at $\mathcal{N}\gg1$. Therefore, the necessary condition of the
applicability of the approximations (\ref{tightly}) and (\ref{loosely})
for the states with $n_+=0$ is $\ln(\gamma_*/\mathcal{N})\gg1$. If this
is not the case, one has to use \req{even_spectrum} instead of the
approximations (\ref{tightly}) or (\ref{loosely}). 

\subsubsection{Odd bound states}
\label{sect:odd}

For the odd-parity states, the longitudinal  wave function
$g_\kappa(z)$ tends to zero at $z\to0$. 
Therefore it is small in the region
$|z|\lesssim\am$ where the effective potentials 
$V_{0,n_+}^{(\mathcal{N})}(z)$ substantially differ from the 1D
Coulomb potential. In the first approximation with respect to the
small parameter $\am/\aBred=\gamma_*^{-1/2}$, 
one can use the 1D Coulomb potential instead of the
true effective potential. Thus the problem is reduced to finding
odd-parity solutions for the 1D H atom.
Formally it
corresponds to setting $z_0=0$ and $\efqm$ to integers in
Sect.\,\ref{sect:exterior}.
 The review of the theory of the
1D H atom is given in Ref.\,\cite{Loudon16}. 
The odd-parity solutions are well-behaved, because the singularity of
the potential term in the Schr\"odinger equation (\ref{1DSchr})
is finite for such wave functions. The Whittaker functions (\ref{Whit}) 
with integer $\efqm$ are expressed through the generalized 
Laguerre polynomials,
$W_{\efqm,\frac12}(x) = (-1)^\efqm
x\,\erm{-x/2}\,L_\efqm^1(x)/\efqm$, and the normalized
wave functions become
\beq
   g_{\nupar}(z) \approx g_{\nupar}^{(0)}(z) =  
   \sqrt{\frac{2}{\aBred\efqm^5\efqm!}}\,z\exp\left(
   -\frac{|z|}{\efqm\aBred} \right)
   L_\efqm^1\left(\frac{2|z|}{\efqm\aBred} \right),
\eeq
where $\efqm = (\nupar+1)/2 = 1,2,3,\ldots$, and
$\nupar$ is an odd longitudinal quantum number.
This solution does not depend on the state of motion of the ion. 
Such dependence can be revealed by the perturbation theory.
To the first order,
\beq
   E_\kappa^\| = - \frac{\hbar^2}{2\mred\aBred^2\,\efqm^2} +
   \int_{-\infty}^\infty \delta V_{n_+}^{(\mathcal{N})}(z) \left[
   g_{\nupar}^{(0)}(z) \right]^2 \dd z,
\eeq
where
\beq
   \delta V_{n_+}^{(\mathcal{N})}(z) 
      = V_{0,n_+}^{(\mathcal{N})}(z) + \frac{Ze^2}{|z|}.
\eeq
is the perturbation potential.
Since the integrand is non-negative, the energy levels are shifted
upwards.

\subsubsection{Continuum states}
\label{sect:cont}

For positive longitudinal energies, the exterior solution is obtained
by analogy with Sect.\,\ref{sect:exterior} with a replacement
$\efqm\to\ii\efqm$, where $\efqm$ is defined by \req{efqunum}
with positive
$E_\kappa^\|$.
According to this definition,
\beq
   \efqm=(\aBred k_\kappa)^{-1},
\label{efqunum'}
\eeq
where
$k_\kappa$
is the absolute value of the outgoing wavenumber
 given by \req{k_kappa}.
Instead of \req{Whit} we have two linearly independent solutions
with asymptotes
\beq
   W_{\pm\ii{\efqm}}\left(\mp\frac{2\ii z}{{\efqm}\aBred}\right)
    \sim \,\erm{{\efqm}\pi/2} \exp\left[\pm\ii \left(
   k_\kappa z +\frac{\ln|2k_\kappa z|}{k\aBred} \right) \right],
\eeq
at $z\to\infty$. Then the longitudinal part
$g_\kappa(z)$ of a real wave function of definite $z$-parity,
$\psi_\kappa(\bm{r}_{+,\perp},\bm{r}_{-,\perp},z)=\pm
\psi_\kappa(\bm{r}_{+,\perp},\bm{r}_{-,\perp},-z)$,  normalized
according to \req{gz_norm}, is approximated at $z>z_0$ by
\bea
  g_{\kappa}^\mathrm{real}(z) &=&
  \frac{\erm{-{\efqm}\pi/2}}{2\sqrt{\zmax}}
   \left[ W_{\ii{\efqm}}\left(-\frac{2\ii z}{{\efqm}\aBred}
   \right)\,\erm{\ii\theta}
\right.\nonumber\\&&\left. + 
       W_{-\ii{\efqm}}\left(\frac{2\ii z}{{\efqm}\aBred}
   \right)\,\erm{-\ii\theta} \right],
\label{gW}
\eea
where the phase factor $\erm{\pm\ii\theta}$ has to be determined from
the matching conditions at $z=z_0$. In \req{gW},
the prefactor is obtained assuming
that the normalization integral over
the interval $[-z_0,z_0]$ is negligibly
small compared with the total normalization integral over
$[-\zmax,\zmax]$.

\subsection{Overlap integrals between even states}
\label{sect:overlap}

In the exterior region $z>z_0$,
assuming that at least one of the longitudinal wave functions
$g_\kappa(z)$ or $g_{\kappa'}(z)$ belongs to discrete spectrum (tends to
zero at $z\to\infty$), the use of the Schr\"odinger equation and
integration by parts leads to the identity
\beq
   \int_{z_0}^\infty g_\kappa(z) g_{\kappa'}(z)\,\dd z = 
     \frac{\hbar}{2\mred}\, g_\kappa(z) g_{\kappa'}(z)
      \, \frac{\eta-\eta'}{E_\kappa^\|-E_{\kappa'}^\|}\,\Bigg|_{z=z_0},
\label{overlap}
\eeq
where $\eta$ and $\eta'$ are the corresponding logarithmic derivatives
(\ref{logderiv}).
In the limit $E_\kappa^\| \to E_{\kappa'}^\|$ we have
\beq
   \int_{z_0}^\infty g_\kappa^2(z) = 
     \frac{\hbar}{2\mred}\, g_\kappa^2(z)\,\dd z
       \,\frac{\partial\eta}{\partial E_\kappa^\|}\,\Bigg|_{z=z_0}.
\label{Int_g2}
\eeq
The latter equation is used for normalization of the overlap
integral (\ref{overlap}). 
At $\am\to0$, we have $z_0\to0$, hence
the integral over the interior region $|z|<z_0$
can be neglected. Then we have the normalized overlap
integral
\beq
   \mathcal{L}(\kappa'|\kappa)\approx
   \frac{\int_{z_0}^\infty g_\kappa(z) g_{\kappa'}(z)\,\dd z}{
   \left[\int_{z_0}^\infty g_\kappa^2(z)\,\dd z\right]^{1/2}
   \left[ \int_{z_0}^\infty g_{\kappa'}^2(z)\,\dd z 
     \right]^{1/2}
   }\,.
\eeq
Using \req{overlap} for the numerator and \req{Int_g2}
for the denominator on the right-hand side, we obtain 
\beq
    |\mathcal{L}(\kappa'|\kappa)| =
    \left|\frac{\eta-\eta'}{E_\kappa^\|-E_{\kappa'}^\|}\right|
    \,\left|\frac{\partial\eta}{\partial E_\kappa^\|}\,
      \frac{\partial\eta'}{\partial E_{\kappa'}^\|}\right|^{-1/2}\,
        \Bigg|_{z=z_0}.
\label{LongInt2}
\eeq
Here,  $\mathcal{L}$ is defined by \req{LongInt}, where the subscripts
are suppressed because only the terms with $n_\pm=n_{0\pm}$ and
$n_\pm'=n_{0\pm}'$ survive in the adiabatic approximation.

Equations (\ref{efqunum}) and (\ref{eta_ext}) give
\beq
   \frac{\partial\eta}{\partial E_{\kappa}} =
\frac{2\mred\aBred^2\efqm^3}{\hbar^2}\,\frac{\partial\eta}{\partial\efqm} =
    \frac{2\mred\aBred\efqm^3}{\hbar^2}
     \left(\frac{\pi^2}{\sin^2\efqm\pi}
  + \frac{\dd\Theta(\efqm)}{\dd\efqm} \right),
\label{deta1}
\eeq
where function $\Theta$ is defined by \req{Theta}.
From \req{deta1},
using Eqs.\,(\ref{eta_int}), (\ref{even_spectrum}) and the identity
$1/\sin^2x=1+\cot^2x$, we obtain
\beq
   \frac{\partial\eta}{\partial E_{\kappa}} =
   \frac{2\mred\aBred\efqm^3}{\hbar^2}
   \,\mathcal{G}_{\mathcal{N},n_+,\efqm},
\label{deta3}
\eeq
where
\bea
\mathcal{G}_{\mathcal{N},n_+,\efqm} &\equiv&
 \left[
     \left( \frac12\ln\frac{\gamma_*}{2} - \frac32 \gamma_\mathrm{E}
          - \frac12 J_{\mathcal{N}n_+} + \Theta(\efqm) \right)^2
\right.\nonumber\\&&\left.
   + \pi^2 + \frac{\dd\Theta(\efqm)}{\dd\efqm} \right].
\label{deta3a}
\eea
Although $\eta$ and $\eta'$ depend on $z_0$, their difference does not.
From \req{eta_int} we obtain the approximation
\beq
    \eta-\eta' = \frac{J_{\mathcal{N}'n_+'} -
             J_{\mathcal{N}n_+}}{\aBred}.
\label{deta2}
\eeq
In summary, $|\mathcal{L}(\kappa'|\kappa)|$ is given by
Eqs.~(\ref{LongInt2}), (\ref{deta2}), (\ref{deta3}),
and (\ref{efqunum}) as function of $\efqm$. In its turn, $\efqm$ is
determined by \req{even_spectrum}.

\subsubsection{The case of tightly bound states}

For the tightly bound states ($\nupar=\nupar'=0$), we have $0<\efqm\ll1$ and
$0<\efqm'\ll1$. In this particular case Eqs.~(\ref{eta_ext})
and (\ref{Theta}) yield
\beq
   \frac{\partial\eta}{\partial E_\kappa^\|}  =
      \frac{\mred\aBred}{\hbar^2}\left[\efqm + 2\efqm^2 + O(\efqm^3)
      \right].
\label{deta_tb}
\eeq
Substitution of Eqs.~(\ref{deta2}) and
(\ref{deta_tb}) into \req{LongInt2} with the use of
\req{efqunum} gives
\beq
   \mathcal{L}^2(\kappa'|\kappa) \approx
        \frac{4\,( J_{\mathcal{N}'n_+'} - J_{\mathcal{N}n_+} )^2 }{
\displaystyle
    \left( \frac{1}{(\efqm')^2}-\frac{1}{\efqm^2} \right)^{\!\!\!2}
           \left(\efqm+2\efqm^2\right) \left[(\efqm')+2(\efqm')^2\right]
            }.
\eeq
Using \req{lowest_order} for $\efqm$ and $\efqm'$, 
we substitute
\bea&&\hspace*{-1em}
  \frac{1}{\efqm^2} - \frac{1}{(\efqm')^2} \approx
      2  \left(
      J_{\mathcal{N}'n_+'} - J_{\mathcal{N}n_+} \right)
       \ln\frac{\gamma_*}{2},
\\&&\hspace*{-1em}
    \efqm+2\efqm^2 \approx
    \left( \ln\frac{\gamma_*}{2} \right)^{\!\!-1}
      + 2 \left( \ln\frac{\gamma_*}{2}
       \right)^{\!\!-2}\!\!\!
       \approx (\efqm')+2(\efqm')^2
\qquad
\eea
and thus obtain
\beq
 \mathcal{L}^2(\kappa'|\kappa) \approx
 \left( 1+\frac{2}{\ln(\gamma_*/2)} \right)^{\!\!-2} \sim
   1 - \frac{4}{\ln\gamma_*}.
\label{L2tb}
\eeq
We see that the overlap integral does not depend on quantum numbers in
this lowest-order approximation, valid at $\gamma_*\to\infty$. The
asymptotic fractional accuracy of this approximation can be estimated
from comparison of \req{tightly} with \req{lowest_order} as
$\sim[\ln(\ln\gamma_*)]/\ln\gamma_*$.

\subsubsection{Overlaps of tightly bound states with other even states}

For the overlap integral between a tightly bound state ($\nupar=0$,
$0<\efqm\ll1$) and an even loosely-bound state ($\nupar'=2$, 4, \ldots, 
$\efqm'\gtrsim1$), we keep using \req{deta_tb} for the first state
$|\kappa\rangle = |\mathcal{N},0,n_-,n_+,0\rangle$ and substitute
\req{deta3} for the second state  $|\kappa'\rangle =
|n_+',\mathcal{N}',\nu_\|'\rangle$.  Recalling \req{efqunum}, we finally
obtain
\beq
   \mathcal{L}^2(\kappa'|\kappa) =
      \frac{2}{(\efqm')^3}\,\frac{
       ( J_{\mathcal{N}'n_+'} - J_{\mathcal{N}n_+}
       )^2\,\sqrt{\epsilon_\kappa}}{
       \mathcal{G}_{\mathcal{N}',n_+',\efqm'}
         \sqrt{\epsilon_{\kappa}-\epsilon_{\kappa'}}},
\eeq
where $\mathcal{G}_{\mathcal{N}',n_+',\efqm'}$ and
$J_{\mathcal{N}n_+}$ are defined in
Eqs.~(\ref{JNn}) and (\ref{deta3a}), respectively.

\subsubsection{The bound-free case}

Equations (\ref{overlap}) and (\ref{Int_g2}) give the relation
\beq
\frac{\left(
   \int_{z_0}^\infty g_\kappa(z) g_{\kappa'}(z)\,\dd z
   \right)^2 
   }{
     \int_{z_0}^\infty g_\kappa^2(z)\,\dd z
   }
  \!  = \!
   \frac{\hbar^2}{2\mred} 
         \frac{
   g_{\kappa'}^2(z)}{
   \partial\eta/\partial E_\kappa}
   \!
    {\left( \frac{\eta'-\eta}{E_{\kappa'}^\| - E_\kappa^\|}
\right)\!\!}^2\Bigg|_{z=z_0}\!\!.
\label{L2cont}
\eeq
In the case where the initial state $\kappa$ belongs to the discrete
spectrum but the final state $\kappa'$ belongs to the continuum,
taking into account that
 $z_0\to0$ at $\gamma_*\to \infty$, we can use the leading term of
\req{gW} at small $z$. It is given by \req{Whit0} with $\efqm$
replaced by $\ii\efqm'$. Using the expression for the
gamma function of imaginary argument (cf.~\cite{AS}, 6.1.29)
\beq
   \Gamma(\ii\efqm') = \sqrt{\frac{\pi}{\efqm'\sinh\efqm'\pi}}\,
   \erm{-\ii\phi_{\efqm'}},
\eeq
where
\beq
   \phi_{\efqm'} = \frac{\pi}{2}\,\mathrm{sign}\,\efqm' + \frac{1}{2\ii}
\left[
    \ln\Gamma(1-\ii\efqm') - \ln\Gamma(1+\ii\efqm') \right],
\eeq
we obtain
\beq
   g_{\kappa'}^2(z_0) \approx g_{\kappa'}^2(0) =
    \frac{\erm{-\efqm'\pi}}{\zmax}\,
    \frac{\sinh\efqm'\pi}{\efqm'\pi}\,
    S_{\kappa'}
    =
     \frac{1-\erm{-2\pi\efqm'}}{2\pi\efqm'\zmax}\,
      ,
\label{g2cont}
\eeq
where
\beq
   S_{\kappa'} \equiv \sin^2(\phi_{\efqm'}-\theta)
\label{Sk}
\eeq
and $\theta$ is the undetermined phase factor in \req{gW}.

To find the factor
$S_{\kappa'}$, we use the
matching condition $\eta'_\mathrm{ext} = \eta'_\mathrm{int}$. 
The derivative of the wave function (\ref{gW} can be written as
\beq
   \frac{\dd g_{\kappa'}}{\dd z} =
    \frac{ \erm{-\pi\efqm'/2}}{ \sqrt{\zmax}}\,
      \mbox{Re}\left[\erm{\ii\theta}\, \frac{\dd}{\dd z}
        W_{\ii \efqm'}\left(-\frac{2\ii z}{\efqm'\aBred}\right)
        \right].
\label{dg'dz}
\eeq
At $z\to0$, 
\bea&&\hspace*{-1em}
   \frac{\dd}{\dd z}
        W_{\ii \efqm'}\left(-\frac{2\ii z}{\efqm'\aBred}\right) =
        \erm{-\ii\phi_{\efqm'}}
        \,\sqrt{\frac{\pi}{\efqm'\sinh\pi\efqm'}}
        \bigg[ \cosh\efqm'\pi
\nonumber\\&&
   +\frac{\ii}{\pi}\sinh\efqm'\pi
    \left( \ln\frac{2z}{\aBred} - \mbox{Re}\,\Theta(\ii\efqm') +
    2\gamma_\mathrm{E} \right)
    \bigg],
\eea
where function $\Theta$ is given by \req{Theta}.
Substitution of the last expression into \req{dg'dz} gives
\bea&&\hspace*{-1em}
\frac{\dd g_{\kappa'}}{\dd z} =
-\frac{4}{\aBred}\,\frac{\erm{-\efqm'\pi/2}}{\sqrt{\zmax}}
\bigg[
   \bigg( \cosh\efqm'\pi
      \mbox{Im}\,\Theta(\ii\efqm') \bigg) \cos(\pi_{\efqm'}-\theta)
\nonumber\\&&\hspace*{-1em}
 + \frac{\sinh\efqm'\pi}{\pi}
    \bigg( \!\! \ln\frac{2z_0}{\aBred}
 - \mbox{Re}\,\Theta(\ii\efqm') +
    2\gamma_\mathrm{E} \!\bigg) \sin(\pi_{\efqm'}-\theta)
    \bigg].
\qquad~
\label{dg'dz2}
\eea
Here,
\bea
   \mbox{Im}\,\Theta(\ii\efqm') & = & \mbox{Im}\left(\ln\ii\efqm' -
   \frac{1}{2\ii\efqm'} - \psi(\ii\efqm')\right)
\nonumber\\&=&
   \frac{\pi}{2}\,(1-\coth\pi\efqm'),
\label{ImTheta}
\eea
where in the last equality we have used the relation
$\psi(\ii\efqm') = 1/(2\efqm') + (\pi/2)\,\coth\pi\efqm')$
and have chosen the value of $\mbox{Im}\,\ln\ii\efqm')=\pi/2$.
Dividing \req{dg'dz2} by
\req{g2cont} and using \req{ImTheta},
we thus obtain the logarithmic derivative
$\eta_\mathrm{ext}'$ at small $z=z_0$:
\bea
   \eta_\mathrm{ext}' &=& - \frac{2}{\aBred} \left[
     \ln\frac{2z_0}{\aBred} +\frac{\pi\,\mathrm{sign}\,\efqm'}{
1-\erm{-2\pi\efqm'} } \, \cot(\phi_{\efqm'}-\theta)
\right.\nonumber\\&&\left. + 2\gamma_\mathrm{E} -
\mbox{Re}\,\Theta(\ii\efqm') \right],
\label{eta_ext1}
\eea
From \req{Theta} we obtain
\beq
\mbox{Re}\,\Theta(\ii\efqm') = \ln\efqm' - \mbox{Re}\,
\psi(1+\ii\efqm').
\label{ReTheta1}
\eeq
An efficient way of calculating $\mbox{Re}\,\Theta(\ii\efqm')$ is
presented in Appendix~\ref{sect:ReTheta}.

By matching of $\eta_\mathrm{ext}'$, which is given by \req{eta_ext1},
to 
$\eta_\mathrm{int}'$, which is given by \req{eta_int}, we obtain
\bea
  \mathrm{sign}\,\efqm'\,\cot(\phi_{\efqm'}-\theta)
      &=&  \frac{ 1 - \erm{-2\pi\efqm'} }{2\pi}
          \left[ \ln\frac{\gamma_*}{2} - 3\, \gamma_\mathrm{E}
\right.\nonumber\\&&\left.
          - J_{\mathcal{N}'n_+'} + 2\,\mbox{Re}\,\Theta(\ii\efqm')
            \right].
\label{cot}
\eea
Now the factor $S_{\kappa'}$ (\ref{Sk})
is provided in the explicit form by the identity
$1/\sin^2(\phi_{\efqm'}-\theta) = 1+\cot^2(\phi_{\efqm'}-\theta)$:
\bea
   S_{\kappa'} &=&
     \left\{ 1 +
       \left( \frac{ 1 - \erm{-2\pi\efqm'} }{2\pi} \right)^2
       \left[  \ln\frac{\gamma_*}{2}
\right.\right.\nonumber\\&&\left.\left.
 - 3\, \gamma_\mathrm{E}
          + J_{\mathcal{N}'n_+'} + 2\,\mbox{Re}\,\Theta(\ii\efqm')
            \right]^2
            \right\}^{-1}.
\label{sin2phi}
\eea

If the initial state is tightly bound ($\nupar=0$), then we can use
\req{deta_tb} for the factor $(\partial\eta/\partial E_\kappa)$ in
\req{L2cont}. In the lowest approximation with respect to
$\efqm\ll1$, we retain only the first (linear in $\efqm$) term in
this equation: 
$
{\partial\eta}/{\partial E_\kappa^\|} \approx
      {\mred\aBred\efqm}/{\hbar^2}.
$
For the
functions $g_\kappa$ and $g_{\kappa}'$ of equal parity, 
the denominator on the left-hand side of \req{L2cont} is approximately
$
\int_{z_0}^{\zmax} g_\kappa^2(z)\,\dd z \approx 
\frac12 \int_{-\infty}^\infty g_\kappa^2(z)\,\dd z =\frac12
$
because of the normalization condition (\ref{gz_norm}),
while the numerator is
\beq
\left(
   \int_{z_0}^\infty g_\kappa(z) g_{\kappa'}(z)\,\dd z
   \right)^2
 \approx 
  \left[ \frac12 \, \mathcal{L}(\kappa'|\kappa) \right]^2
  =\frac14\,\mathcal{L}^2(\kappa'|\kappa),
\eeq
for the same normalization.
Thus \req{L2cont} can be rewritten as
\beq
   \mathcal{L}^2(\kappa'|\kappa) = \frac{\hbar^4\,g_{\kappa'}^2(z_0)}{
      \mred^2\aBred\efqm}
        \left( \frac{(\eta'-\eta)|_{z=z_0}}{E_{\kappa'}^\| - E_\kappa^\|}
          \right)^2
\eeq
Using also \req{g2cont} for $g_\kappa^2(z_0)$ and \req{deta2}
for $(\eta'-\eta)$ in \req{L2cont}, 
we obtain 
\beq
   \mathcal{L}^2(\kappa'|\kappa) = 
     \frac{\hbar^4}{\mred^2\aBred^3}\,
  \frac{1-\erm{-2\pi\efqm'}}{2\pi \zmax\efqm'\efqm}
  \left( \frac{J_{\mathcal{N}'n_+'} - J_{\mathcal{N}n_+}}{
         E_{\kappa'}^\| - E_\kappa^\| } \right)^2
           S_{\kappa'}.
\label{L2bf}
\eeq
We recall that $\efqm$ and $\efqm'$ are defined
by \req{efqunum} and \req{efqunum'} respectively, the numbers
$J_{\mathcal{N}n_+}$ and $J_{\mathcal{N}'n_+'}$ are
given by \req{JNn}, $\aBred$ is defined by \req{aBred},
and $S_{\kappa'}$ is provided by Eqs.~(\ref{sin2phi}) and
(\ref{ReTheta1}).

\subsection{Transverse geometric size}
\label{sect:adiabgeom}

In the adiabatic approximation, \req{perpsize} reduces to
\beq
   \langle\kappa|r_\perp^2|\kappa\rangle =
   2\am^2 \sum_{k=0}^{\mathcal{N}} (k+1 + n_-)
     \left(C_k^{(\mathcal{N},n_+)}\right)^2.
\label{adtrans}
\eeq
This equation shows that the transverse size of the ion increases with
increasing $\mathcal{N}$. In classical physics, this increase 
corresponds to the action of the electric field, induced in the
reference frame comoving with the ion. The forces on the nucleus and the
electron, caused by this field, have opposite directions and therefore
tend to stretch the ion along the radius. Since on the average
$\mathcal{N}$ is proportional to the square of transverse momentum of
the transverse motion of the ion as a whole, this stretching
tends to enhance with an increase of $\mathcal{N}$.

One can show (see
Appendix~\ref{sect:d}) that
\beq
   \sum_{k=0}^{\mathcal{N}} k
     \left(C_k^{(\mathcal{N},n_+)}\right)^2
     = \frac{Z-1}{Z}\,(\mathcal{N} - n_+)
       + \frac{n_+}{Z}.
\label{kCk}
\eeq
In the most important case of the helium ion, the right-hand side of
\req{kCk} reduces to $\mathcal{N}/2$.

Using \req{kCk} and the normalization condition (\ref{normC}),
we can simplify \req{adtrans}.
Recalling that the true bound states in the ultrastrong fields exist
only for $n_-=0$, for these states we obtain
\beq
   \langle\kappa|r_\perp^2|\kappa\rangle = 
      2\am^2 \left(
         1 + \frac{Z-1}{Z}\,(\mathcal{N} - n_+)
       + \frac{n_+}{Z}
       \right)
\eeq
In particular, for the helium ion ($Z=2$) 
$ \langle\kappa|r_\perp^2|\kappa\rangle = ( 2 + \mathcal{N}) \am^2$
is independent of $n_+$.

\subsection{Radiative transitions for circular polarization}
\label{sect:adiabrad}

In the adiabatic approximation, expressions (\ref{D+}) and (\ref{D-})
for the circular components of the dipole matrix element for the
radiative transitions from state $|i\rangle=|\kappa\rangle$ to state
$|f\rangle=|\kappa'\rangle$ with $n_-=n_-'=0$
reduce to
\beq
     \frac{D_{fi,-\alpha}}{e\am} =
     \sqrt{Z\,n_{+}^{\mathrm{max}}}\,\mathcal{L}(\kappa'|\kappa)\,
     \delta_{n_+',n_+ +\alpha} \delta_{\mathcal{N}',\mathcal{N}+\alpha},
\label{D_ad}
\eeq
where $\alpha=\pm1$, $n_{+}^{\mathrm{max}}=\max(n_+,n_+')$, and
the overlap integral $\mathcal{L}(\kappa'|\kappa)$ is given by
\req{LongInt2}.

\begin{figure}
\centering
\includegraphics[width=\columnwidth]{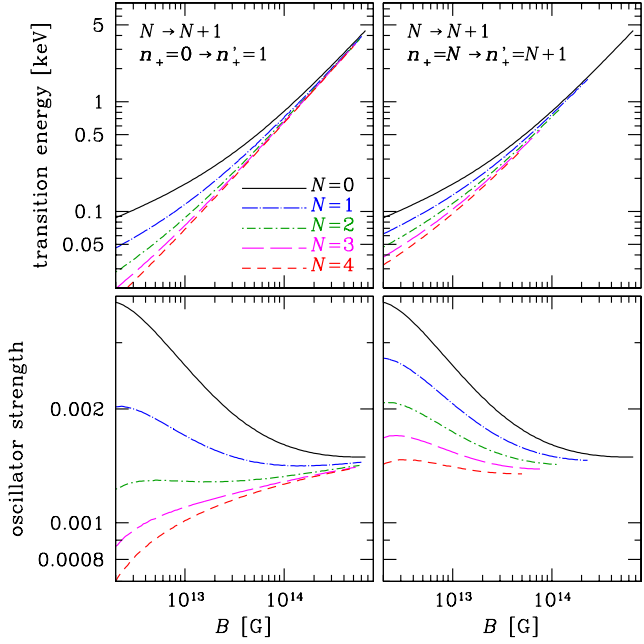}
\caption{Resonance transition energies $E_f-E_i$ (upper panels) and
oscillator strengths $f_{fi,\alpha}$ (lower panels) for transitions
between tightly-bound states $|i\rangle =
|\mathcal{N},n_-,n_+,\nu\rangle$  ($n_{-}=0$, $\nu=0$) and $|f\rangle =
|\mathcal{N}',n_-',n_+',\nu'\rangle$ ($n_{-}'=0$, $\nu'=0$) with
$\mathcal{N}'=\mathcal{N}+1$ and
$n_+'=n_++1$ in the adiabatic approximation according to \req{oscstr},
as functions of magnetic field strength $B$, for  $\mathcal{N}=0$ (solid
lines), 1 (long-dash-dot lines). 2 (short-dash-dot lines), 3 (long
dashes), and 4 (short dashes).  Left panels: $n_+=0$; right panels:
$n_+=\mathcal{N}$. The lines are terminated at the points where the
final state crosses the continuum and becomes autoionizing ($E_f=0$).
}
\label{fig:oscstrap}
\end{figure}

For the transitions between tightly-bound states ($\nupar'=\nupar=0$),
Eqs.\,(\ref{osc}), (\ref{D_ad}), and
(\ref{L2tb}) give the following approximate expression for the 
oscillator strength  to the leading order in $1/\ln\gamma_*$:
\beq
   f_{fi,\alpha} = \frac{\hbar\omega}{\mbox{Ry}}
     \frac{\am^2}{\aB^2} 
      Z\,n_{+}^{\mathrm{max}}\!
   \left(1-\frac{4}{\ln\gamma_*}\right)
   \delta_{n_+',n_+ + \alpha}
    \delta_{\mathcal{N}',\mathcal{N}+\alpha},
\label{oscstr}
\eeq
and \req{sigma_bb_f} yields
\begin{widetext}
\beq
   \sigma_{i\to f,\alpha}^\mathrm{bb}(\omega) =
   4\pi^2\alphaf\,\am^2\,Zn_{+}^{\mathrm{max}}
   \left(1-\frac{4}{\ln\gamma_*}\right)
   \omega\,\Delta_{fi}(\omega - \omega_{fi})
   \,\delta_{n_+',n_+ + \alpha}
   \,\delta_{\mathcal{N}',\mathcal{N}+\alpha},
\eeq
where $\alphaf$ is the fine structure constant.
Figure~\ref{fig:oscstrap} presents some examples of the transition
energies and oscillator strengths for radiative transitions between
different tightly-bound states in the dipole adiabatic approximation,
according to \req{oscstr}.
 
For the bound-free transitions, substitution of \req{D_ad}
 into Eq.\,(\ref{sigma_bf_dip1}) gives
 \beq
      \sigma_{i\to f,\alpha}^\mathrm{bf}(\omega) =
      \frac{4\pi \zmax\mred\omega Ze^2\am^2 n_+^\mathrm{max}}{
        \hbar^2 c k_f}\,
         | \mathcal{L}(\kappa'|\kappa) |^2
         \,\delta_{n_+',n_+ +\alpha}
         \,\delta_{\mathcal{N}',\mathcal{N}+\alpha},
\eeq
where $k_f\equiv k_{\kappa'}$ is the longitudinal wavenumber defined by
\req{k_f}. Using \req{L2bf}, we obtain
\beq
      \sigma_{i\to f,\alpha}^\mathrm{bf}(\omega) =
      2Zn_{+}^{\mathrm{max}}\,
    \frac{\hbar^2\omega\,\am^2 e^2}{
   \mred c \,k_f \aBred^3\efqm}\,
   \frac{1-\erm{-2\pi\efqm'}}{\efqm'}
  \,S_{\kappa'}
  \left( \frac{J_{\mathcal{N}'n_+'} - J_{\mathcal{N}n_+}}{
         E_f^\| - E_i^\| } \right)^2
   \,\delta_{n_+',n_+ + \alpha}
         \,\delta_{\mathcal{N}',\mathcal{N}+\alpha}.
\label{sigma_bf1}
\eeq
Taking into account definitions of $\gamma_*$ (\ref{gamma*}), 
$\efqm'$ (\ref{efqunum'}), and $\alphaf=e^2/\hbar c$,
we can write \req{sigma_bf1} in the form
\beq
      \sigma_{i\to f,\alpha}^\mathrm{bf}(\omega) =
      2\alphaf Zn_{+}^{\mathrm{max}}\,
    \frac{\hbar^3\omega}{\mred\gamma_*\efqm}\,
   \left[1-\exp\left(-\frac{2\pi}{\aBred k_f}\right)\right]
  \,S_{\kappa'}
  \left( \frac{J_{\mathcal{N}'n_+'} - J_{\mathcal{N}n_+}}{
         E_f^\| - E_i^\| } \right)^2
   \,\delta_{n_+',n_+ + \alpha}
         \,\delta_{\mathcal{N}',\mathcal{N}+\alpha}.
\label{sigma_bf2}
\eeq
Using notations (\ref{efqunum}) and (\ref{effRy}), 
we can also rewrite it as
\beq
    \sigma_{i\to f,\alpha}^\mathrm{bf}(\omega) =
      4\alphaf n_{+}^{\mathrm{max}}\am^2
      \frac{\hbar\omega}{\mathrm{Ry_*}}\,
   \left[1-\exp\left(-\frac{2\pi}{\sqrt{\epsilon_f}}\right)\right]
  \,S_f\,\sqrt{\epsilon_i }
  \left( \frac{J_{\mathcal{N}'n_+'} - J_{\mathcal{N}n_+}}{
         \epsilon_f + \epsilon_i } \right)^2
   \,\delta_{n_+',n_+ + \alpha}
         \,\delta_{\mathcal{N}',\mathcal{N}+\alpha}.
\label{sigma_bf_ad}
\eeq
\end{widetext}
Here we have used the relation $k_f\aBred=\sqrt{\epsilon_f}$, which
follows from Eqs.~(\ref{k_f}), (\ref{aBred}), (\ref{efqunum}), and
(\ref{effRy}). Note that both $\epsilon_i$ and $\epsilon_f$ are positive
by definition. The energy conservation law requires that
$E_{\kappa'}=E_\kappa+\hbar\omega$. Therefore, according to 
Eqs.~(\ref{Elong}) and (\ref{Eperp2}),
for the allowed dipole transitions ($n_+'=n_++\alpha$) we have
$\hbar\omega/\mathrm{Ry_*} = \epsilon_f + \epsilon_i +
2\alpha\gamma_*$. 

\begin{figure}
\centering
\includegraphics[width=\columnwidth]{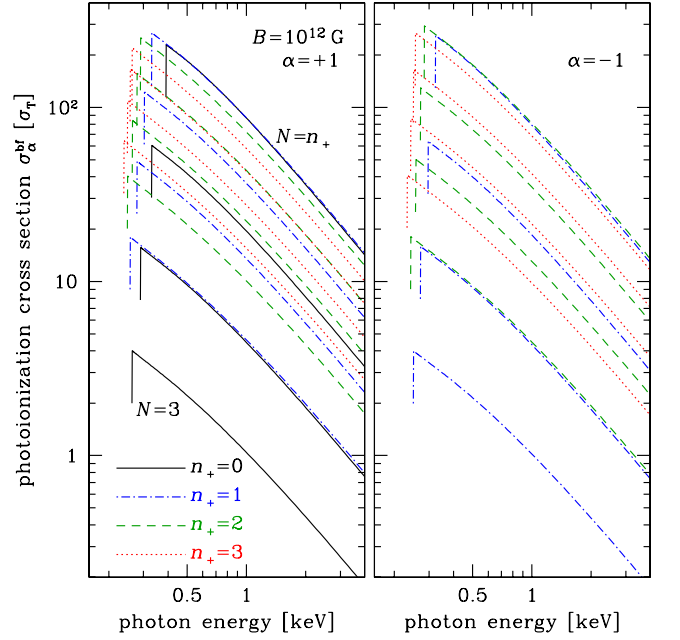}
\caption{Photoionization cross sections $\sigma_{i\to
f,\alpha}^\mathrm{bf}(\omega)$ for different initial tightly-bound
states $|i\rangle = |\mathcal{N},n_{-},n_{+},\nu\rangle$
($n_{-}=0$, $\nu=0$) at
$B=10^{12}$~G in the adiabatic approximation according to
\req{sigma_bf_ad} for the right ($\alpha=+1$, the left panel) and left
($\alpha=-1$, right panel) circular polarizations as functions of the
photon energy $\hbar\omega$ in units of Thomson cross section 
$\sigma_\mathrm{T}=(8\pi/3)(e^2/\mel c^2)^2$. The results are displayed
for initial states with quantum numbers $n_+=0$ (solid lines), 1
(dot-dashed lines), 2 (dashed lines), and 3 (dotted lines) and 
$\mathcal{N}=n_+,n_++1,n_++2,n_++3$ (lines of the same
type from top to bottom for each $n_+$).
}
\label{fig:bfap12}
\end{figure}

\begin{figure}
\centering
\includegraphics[width=\columnwidth]{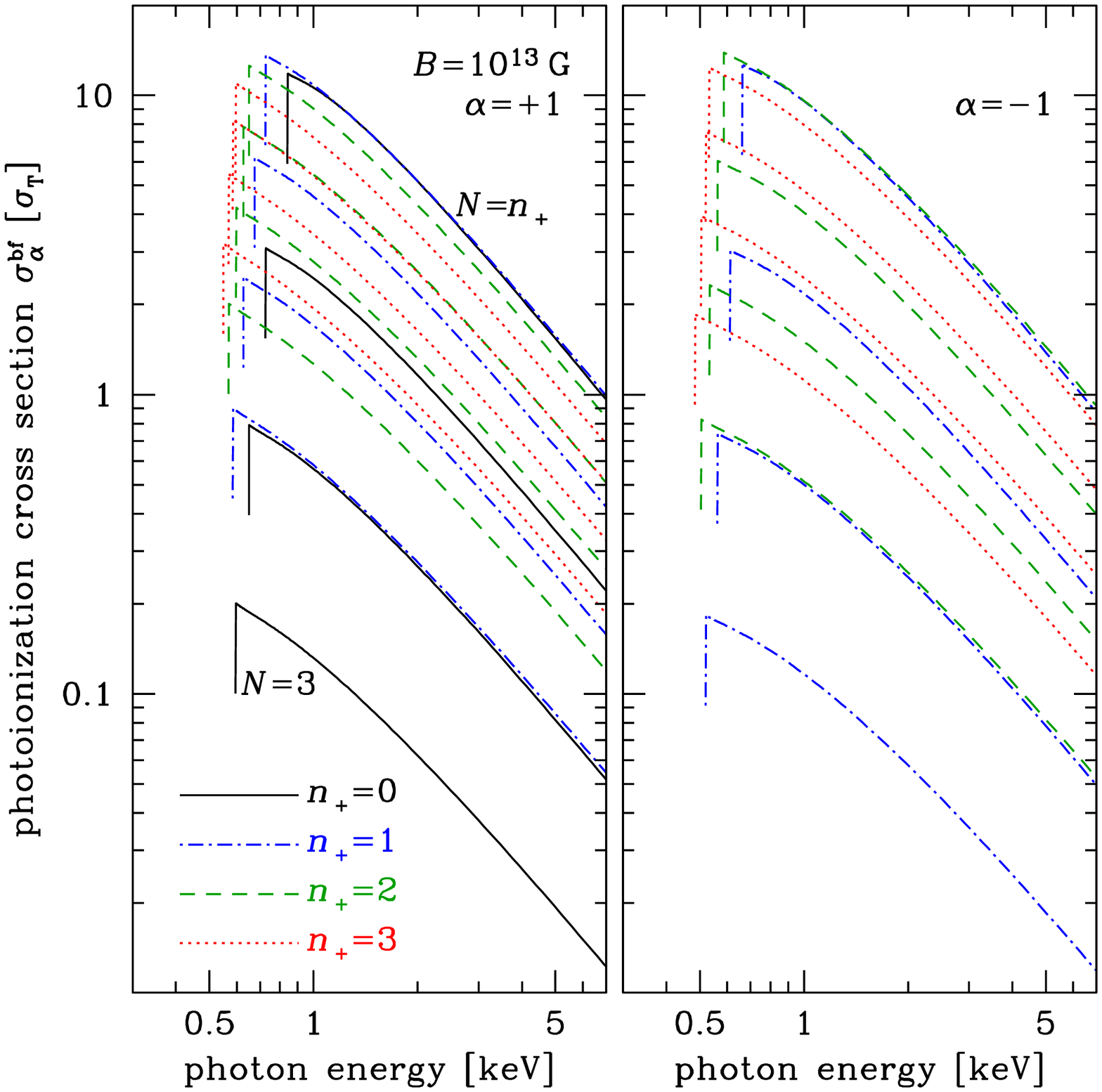}
\caption{The same as in Fig.~\ref{fig:bfap12}, but for $B=10^{13}$~G.
}
\label{fig:bfap13}
\end{figure}
\begin{figure}
\centering
\includegraphics[width=\columnwidth]{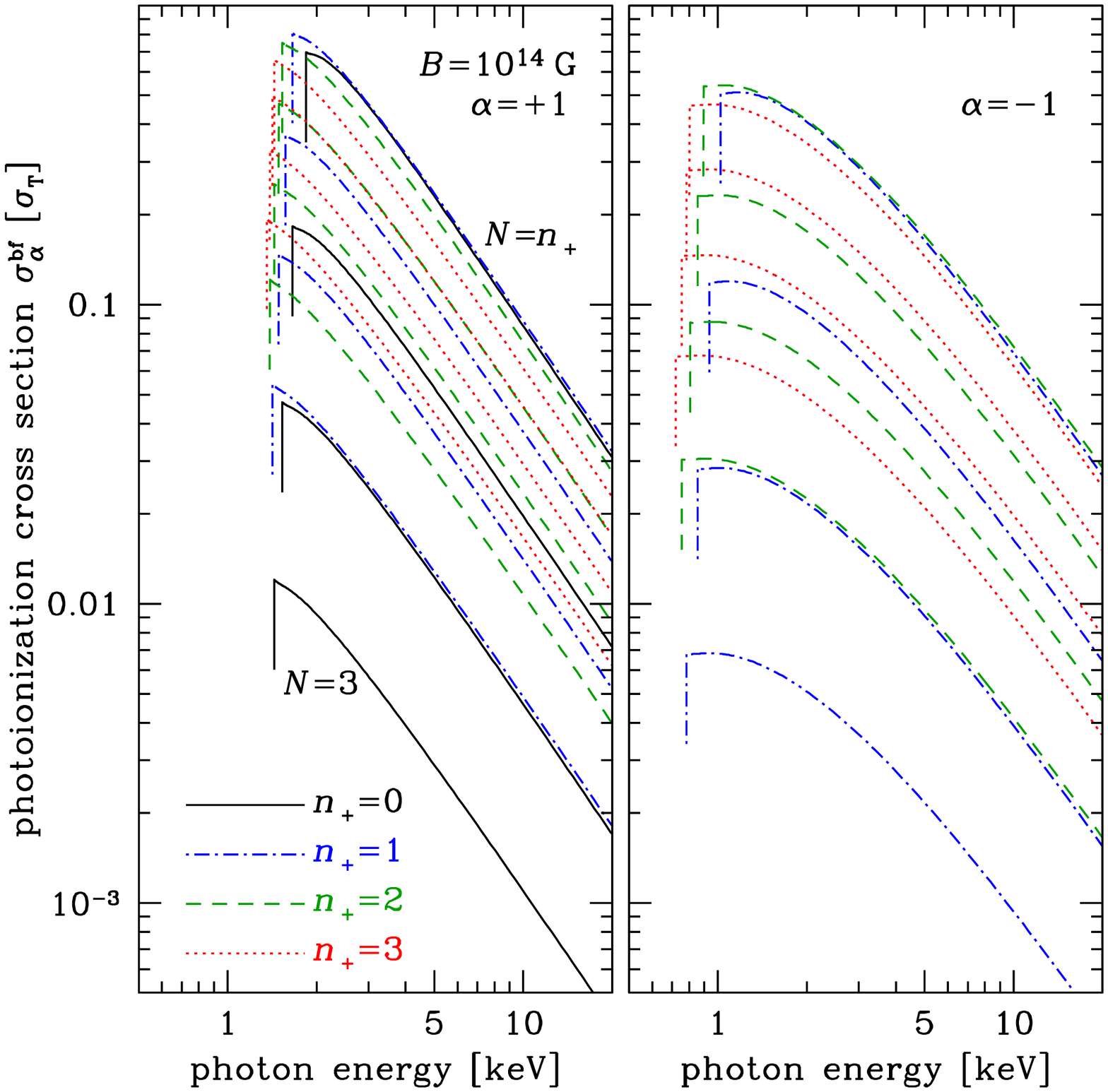}
\caption{The same as in Fig.~\ref{fig:bfap12}, but for $B=10^{14}$~G.
}
\label{fig:bfap14}
\end{figure}

\begin{figure}
\centering
\includegraphics[width=\columnwidth]{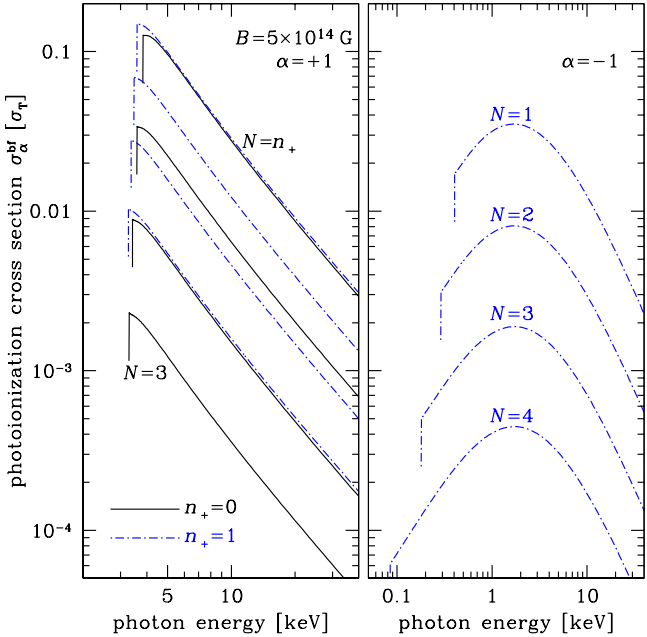}
\caption{The same as in Fig.~\ref{fig:bfap12}, but for
$B=5\times10^{14}$~G and only for $n_+=0$ and 1.
}
\label{fig:bfap147}
\end{figure}

Examples of the photoionization cross sections, given by
\req{sigma_bf_ad} for the circular polarizations $\alpha=\pm1$, are
presented in Figs.~\ref{fig:bfap12}\,--\,\ref{fig:bfap147}.
Figures~\ref{fig:bfap12}, \ref{fig:bfap13}, and \ref{fig:bfap14}
correspond to the field strengths $B=10^{12}$, $10^{13}$, and
$10^{14}$~G, respectively. For $\alpha=+1$, cross sections for the four
smallest values of $n_+$ and the four smallest possible values of
$\mathcal{N}$ at each $n_+$ are shown. For $\alpha=-1$, there are no
lines with $n_+=0$, because absorption of photons with this polarization
by such states is forbidden in the adiabatic dipole approximation. In
Fig.~\ref{fig:bfap147} for $B=5\times10^{14}$~G, only $n_+=0$ and
$n_+=1$ are considered, because the states with $n_+>1$ have positive
energies $E_i$ at such strong field. Although they can be treated as
bound states in the adiabatic approximation, they actually belong to the
continuum and can autoionize due to admixtures of the Landau orbitals
with smaller $n_+$.

Each cross section in Figs.~\ref{fig:bfap12}\,--\,\ref{fig:bfap147}
decreases with increasing photon energy $\hbar\omega$ above the
photoionization threshold $\hbar\omega_\mathrm{thr}= -E_i = |E_i^\|| +
\alpha\hbar\omc{+}$, where the longitudinal energies $E_i^\|$ are
calculated according to the approximation (\ref{tightly}). The cross
sections for the circular polarization become smaller with increasing
magnetic field strength $B$, in agreement with the decrease of the
geometric transverse cross section of the ion, which is proportional
to $\am^2\propto B^{-1}$ (cf.~Sect.~\ref{sect:adiabgeom}). The
photoionization cross sections also become smaller with increasing
$\mathcal{N}$ at fixed $n_+$ and $\omega$.

\section{Conclusions}
\label{sect:concl}

We performed a systematic derivation of practical equations for
computing the basic characteristics of a one-electron ion in different
quantum states in a strong magnetic field: its binding energies,
geometric sizes, oscillator strengths of bound-bound transitions, and
photoionization cross sections. These quantities are necessary
ingredients for construction of models of atmospheres of neutron stars
with strong magnetic fields under the conditions where one-electron ions
can contribute substantially into the atmospheric opacities. We did not
assume that the atomic nucleus is infinitely massive or fixed in space,
but considered the full quantum-mechanical two-body problem. This is
especially important  in sufficiently warm atmospheres with sufficiently
strong magnetic fields, where the thermal motion of the ions cannot be
decoupled from their internal quantum-mechanical structure and the
Rabi-Landau quantization of both the electron and the nucleus must be
taken into account. The obtained results generally confirm, somewhat
correct and extend the previously published quantum-mechanical studies
of an one-electron ion, which moves in a quantizing magnetic field.

In addition, we performed an approximate analytic treatment of the
problem in the adiabatic approximation and derived explicit asymptotic
expressions for the binding energies, transverse geometric sizes, and
cross sections of absorption of radiation, polarized transversely to the
magnetic field. We expect that these analytic expressions can be useful
in the case of superstrong magnetic fields, typical for magnetars.

\begin{acknowledgments}

The work of A.P.{} was partially supported by the Russian Foundation for
Basic Research and Deutsche Forschungsgemeinschaft according to the
research project 19-52-12013.

\end{acknowledgments}

\begin{widetext}
\section*{Appendices}
\renewcommand{\thesection}{}
\renewcommand{\thesubsection}{\Alph{subsection}}
\renewcommand{\theequation}{\Alph{subsection}\arabic{equation}}
\nopagebreak
\subsection{Supplementary relations for $C_{{k}}^{(\mathcal{N},n_+)}$}
\label{sect:suppl}
\setcounter{equation}{0}

Let us consider the operator 
\beq
  \mathcal{B} \equiv \frac{1}{\sqrt{Z}}\left(
  \sqrt{Z-1}\,\hat{b}+\hat{\tilde{a}}
  \right),
\eeq
where $\hat{b}$ and $\hat{\tilde{a}}$ are defined by
Eqs.~(\ref{aa'bb'def}) and (\ref{a+a+'b+b+'})).
It follows from Eqs.~(\ref{aa'bb'}) and
  (\ref{aa'+bb'+}) that
\beq
  [\mathcal{B},\mathcal{B}^\dag]=1.
\label{commutB}
\eeq
 Besides, from
  Eqs.~(\ref{aa'bb'def} and (\ref{a+a+'b+b+'}) we
  see that 
\beq
   \mathcal{B}^\dag \mathcal{B} = \frac{\am^2}{2Z\hbar^2}
   (\bm{\Pi}_\perp - \bm{k}_\perp)^2 - \frac12.
\eeq
Therefore, the eigenvalues of the operator $\mathcal{B}^\dag
\mathcal{B}$ on the  transverse basis states [Sect.~\ref{sect:states}]
equal $n_+$:
\beq
   \mathcal{B}^\dag \mathcal{B} \Psi_{\tilde{N},n,L}'
   = n_+\, \Psi_{\tilde{N},n,L}'.
\eeq
Taking the commutation relation (\ref{commutB}) into account, we
conclude that $\mathcal{B}^\dag$ raises $n_+$ by one, while 
$\mathcal{B}$ decreases $n_+$ by one (it can be proved explicitly using
Eqs.~[\ref{pi+cyclic}]).
On the other hand, using the explicit form of the basis wave functions
(\ref{Psi-sum}) and the definition of $\mathcal{B}$, we obtain
\beq
   \mathcal{B}\,|\tilde{N}',L,n_-,n_+\rangle =
   \sum_{{k}=0}^\mathcal{N} 
\,\mathcal{F}_{\mathcal{N}-{k}-1,\tilde{N}}^{(Z-1)}(\bm{r}_{+,\perp})
\,\mathcal{F}_{n_-,{k}}^{(-1)}(\bm{r}_{-,\perp}-\bm{r}_{+,\perp})
   \left(
   \sqrt{\frac{Z-1}{Z}}\sqrt{\mathcal{N}-k}C_{{k}}^{(\mathcal{N},n_+)}
   +
   \frac{\sqrt{k+1}}{\sqrt{Z}}C_{{k}+1}^{(\mathcal{N},n_+)}
   \right).
\eeq
Therefore
\beq
   \sqrt{n_+}\,C_{{k}}^{(\mathcal{N}-1,n_+ -1)} =
   \sqrt{\frac{Z-1}{Z}}\sqrt{\mathcal{N}-k}\,C_{{k}}^{(\mathcal{N},n_+)}
   +
   \sqrt{\frac{k+1}{Z}}\,C_{{k}+1}^{(\mathcal{N},n_+)}
\label{suprecur1}
\eeq

In the same way, by considering
 $\mathcal{B}^\dag\,|\tilde{N}',L,n_-,n_+\rangle$,
 we find that 
\beq
   \sqrt{n_+ +1}\,C_{{k}}^{(\mathcal{N}+1,n_+ +1)} =
   \sqrt{\frac{Z-1}{Z}}\sqrt{\mathcal{N}+1-k}
   \,C_{{k}}^{(\mathcal{N},n_+)}
   +
   \sqrt{\frac{k}{Z}}\,C_{{k}-1}^{(\mathcal{N},n_+)}.
\label{suprecur2}
\eeq

Furthermore, let us consider operator 
\beq
  \tilde{\mathcal{B}} \equiv \frac{1}{\sqrt{Z}}\left(
  \hat{b} - \sqrt{Z-1}\,\hat{\tilde{a}}
  \right).
\eeq
It is also easy to see that
$[\tilde{\mathcal{B}},\tilde{\mathcal{B}}^\dag]=1$ and
\beq
   \tilde{\mathcal{B}}^\dag \tilde{\mathcal{B}} =
    \hat{b}^\dag\hat{b} + \hat{\tilde{a}}^\dag\hat{\tilde{a}} 
    -\frac{\amZ{Z}^2}{2\hbar^2}\, (\bm{\Pi}_\perp - \bm{k}_\perp)^2 
    +\frac12.
\eeq
Therefore, the eigenvalues of the operator $\mathcal{B}^\dag
\mathcal{B}$ on the  transverse basis states
equal $N$. Taking into account the commutation relations, we obtain that
$\tilde{\mathcal{B}}^\dag$ and $\tilde{\mathcal{B}}$ are 
the creation and annihilation operators with respect to the quantum
number $N+\tilde{n}-n_+ = \mathcal{N} - n_+$. By analogy with the case
of operator $\mathcal{B}$ above, we obtain the following recurrent
relations:
\bea
   \sqrt{\mathcal{N}-n_+}\,C_{{k}}^{(\mathcal{N}-1,n_+)}
   &=&
   \sqrt{\frac{\mathcal{N}-k}{Z}}\,C_{{k}}^{(\mathcal{N},n_+)}
   -
   \sqrt{\frac{Z-1}{Z}}\sqrt{{k}+1}\,C_{{k}+1}^{(\mathcal{N},n_+)},
\label{suprecur3}
\\
   \sqrt{\mathcal{N}-n_++1}\,C_{{k}}^{(\mathcal{N}+1,n_+)}
   &=&
   \sqrt{\frac{\mathcal{N}+1-k}{Z}}\,C_{{k}}^{(\mathcal{N},n_+)}
   -
   \sqrt{\frac{Z-1}{Z}}\sqrt{k}\,C_{{k}-1}^{(\mathcal{N},n_+)}.
\label{suprecur4}
\eea

In the particular case $n_+=0$, relation (\ref{suprecur1}) gives
$\sqrt{\mathcal{N}-{k}}\,\sqrt{Z-1}\,C_{{k}}^{(\mathcal{N},0)} =
-\sqrt{{k}+1}\,C_{{k}+1}^{(\mathcal{N},0)}$, which after ${k}$
iterations yields
\beq
  C_{{k}}^{(\mathcal{N},0)} = (-1)^{k} (Z-1)^{k/2}
    \sqrt{
      \frac{1}{{k}!} \prod_{p=1}^{k} (\mathcal{N}-{k}+p)
      }
      \,C_0^{(\mathcal{N},0)}.
\eeq
Here, $C_0^{(\mathcal{N},0)}$ can be found using \req{suprecur4} at
${k}=n_+=0$, $C_0^{(\mathcal{N}+1,0)}=C_0^{(\mathcal{N},0)}/\sqrt{Z}$,
which yields $C_0^{(\mathcal{N},0)}=C_0^{(0,0)}/Z^{\mathcal{N}/2} = 
1/Z^{\mathcal{N}/2}$. The result can be written in the form
\beq
   C_{{k}}^{(\mathcal{N},0)} =(-1)^{k}
    \frac{(Z-1)^{k/2}}{Z^{\mathcal{N}/2}}
   \sqrt{\frac{\mathcal{N}!}{k!\,(\mathcal{N}-k)!}}
   \equiv(-1)^{k}
    \frac{(Z-1)^{k/2}}{Z^{\mathcal{N}/2}}
   \sqrt{\left( \begin{array}{c} \mathcal{N} \\ {k} \end{array}
   \right)}.
\label{CN0}
\eeq
Now, using \req{suprecur2}, we can find $C_{{k}}^{(\mathcal{N},n_+)}$
with different $n_+$.

\subsection{Proof of Equation (\ref{eta_int})}
\label{sect:eta_int}
\setcounter{equation}{0}

The substitution of \req{v0} into
\req{eta_int1} gives
\beq
 \eta_\mathrm{int}(z) \approx
    -\frac{2}{\aBred}\int_0^{z/\am\sqrt2} \dd\zeta
    \sum_{k=0}^{\mathcal{N}} \frac{1}{k!}
    \left(C_k^{(\mathcal{N},n_+)}\right)^2
   \int_0^\infty\dd\rho\,
    \frac{\rho^k\erm{-\rho}}{\sqrt{\rho+\zeta^2}}.
\eeq
Interchanging the summation and integration orders, we obtain
\bea
  \eta_\mathrm{int}(z) &\approx&
   -\frac{2}{\aBred} \sum_{k=0}^{\mathcal{N}} \frac{1}{k!}
    \left(C_k^{(\mathcal{N},n_+)}\right)^2
   \int_0^\infty\rho^k\erm{-\rho}
   \left[\ln\left(\zeta+\sqrt{\rho+\zeta^2}\right)
    - \frac12\ln\rho \right]\dd\rho
\nonumber\\
  & = & -\frac{2}{\aBred} \sum_{k=0}^{\mathcal{N}}
    \left(C_k^{(\mathcal{N},n_+)}\right)^2
   \left[ \ln\zeta +  \frac{1}{k!}\int_0^\infty \rho^k\erm{-\rho}
     \ln\left(1+\sqrt{1+\rho/\zeta^2}\right)\,\dd\rho
      -  \frac{1}{2k!} \int_0^\infty
       \rho^k\erm{-\rho}\ln\rho\,\dd\rho \right].
\eea
The first integral in the square brackets can be evaluated at
$\zeta\gg1$ as $k!\ln2+O(\zeta^{-2})$.
The last integral equals (\cite{GR}, 4.352)
\beq
   \int_0^\infty\rho^k\erm{-\rho}\ln\rho\,\dd\rho
= -k!\,\psi(k) = k! \,(H_k - \gamma_\mathrm{E}),
\label{eta3}
\eeq
where $\psi(k)$ is the digamma function (\ref{digamma})
and
$
  H_k=\sum_{n=1}^k n^{-1}
$
is the $k$th harmonic number.
Taking into account the normalization of $C_k^{(\mathcal{N},n_+)}$
[\req{orthoC}] and setting $z=z_0$, we obtain \req{eta_int}.

\subsection{Estimate of $J_{\mathcal{N}0}$ at large $\mathcal{N}$}
\label{sect:Jlarge}
\setcounter{equation}{0}

Let us consider \req{JNn} at $n_+=0$ and $\mathcal{N}\gg1$:
\beq
   J_{\mathcal{N}0} = \frac{1}{Z^\mathcal{N}}
   \sum_{k=1}^{\mathcal{N}} H_k\,(Z-1)^k
   \left( \begin{array}{c} \mathcal{N} \\ {k} \end{array}
   \right),
\label{JN0}
\eeq
where we have used \req{CN0} for $\left(C_k^{(\mathcal{N},0)}\right)^2$.
According to the Stirling's approximation for factorials,
\beq
   \left( \begin{array}{c} \mathcal{N} \\ {k} \end{array}
   \right) \sim \left(
     \frac{1}{(k/\mathcal{N})^{k/\mathcal{N}}\,
     (1-k/\mathcal{N})^{1-k/\mathcal{N}}} + o(1)
     \right)^\mathcal{N}.
\eeq
This function is strongly peaked at $k\approx\mathcal{N}/2$.
Therefore, we can take out $H_k$ at $k\approx\mathcal{N}/2$ 
from under the
sum sign and obtain (for $\mathcal{N}\gg1$)
\beq
    J_{\mathcal{N}0} \sim \frac{H_{[\mathcal{N}/2]}}{Z^\mathcal{N}}
   \sum_{k=1}^{\mathcal{N}}
   \left( \begin{array}{c} \mathcal{N} \\ {k} \end{array}
   \right) (Z-1)^k \approx \frac{H_{[\mathcal{N}/2]}}{Z^\mathcal{N}}
   \sum_{k=0}^{\mathcal{N}}
   \left( \begin{array}{c} \mathcal{N} \\ {k} \end{array}
   \right) (Z-1)^k
   = H_{[\mathcal{N}/2]}
   \frac{(1+(Z-1))^\mathcal{N}}{Z^\mathcal{N}}
   = H_{[\mathcal{N}/2]}.
\eeq
Now from the double inequality \cite{Young91}
\beq
   \frac{1}{2(k+1)} < H_k - \ln k - \gamma_\mathrm{E} < \frac{1}{2k}
\eeq
we have $H_{[\mathcal{N}/2]} = \ln\mathcal{N}-\ln2 +
\gamma_\mathrm{E} + O(1/\mathcal{N})$ and
\beq
   J_{\mathcal{N}0}\sim\ln\mathcal{N}+O(1).
\eeq

\subsection{Calculation of $\mbox{Re}\,\Theta(\ii\efqm)$}
\label{sect:ReTheta}
\setcounter{equation}{0}

The function $\mbox{Re}\,\Theta(\ii\efqm)$, which is given by
\req{ReTheta1}, can be represented as (e.g., \cite{AS}, 6.3.17)
\beq
\mbox{Re}\,\Theta(\ii\efqm) =
\ln\efqm +\gamma_\mathrm{E} - \sum_{k=1}^{\infty}
       \frac{(\efqm)^2}{k\left[k^2+(\efqm)^2\right]}
       =
       \ln\efqm +\gamma_\mathrm{E} - \sum_{k=1}^{\infty}
       \frac{1}{k\,(1+\epsilon k^2)},
\label{ReTheta}
\eeq
where $\epsilon$ is the dimensionless longitudinal energy
defined by \req{efqunum}. At large $\epsilon$ (small $\efqm$) the series
on the right-hand side converges well, but with decreasing $\epsilon$
the convergence becomes progressively slower, which can be easily
understood from the fact that the series diverges logarithmically at
$\epsilon\to0$. At this limit, one can use the formula
(e.g., \cite{AS}, 6.3.19)
\beq
\mbox{Re}\,\psi(1+\ii\efqm) = \ln\efqm
    + \sum_{n=1}^\infty \frac{(-1)^{n-1}B_{2n}}{2n\efqm^{2n}},
\label{ReThetas}
\eeq
where $B_{2n}$ are the Bernoulli numbers.
However, due to the asymptotic nature of the latter formula, its
accuracy rapidly worsens with
increasing $\epsilon$ at any fixed number of terms. 

In this appendix we propose a method to calculate
$\mbox{Re}\,\Theta(\ii\efqm)$ with keeping the number of terms of the
sum in \req{ReTheta} reasonably small at intermediate $\epsilon$. Let us
consider the integral
\beq
    \int_{K}^{\infty} f_\epsilon(x)\,\dd x =
    \frac12\,\ln\left(1+\frac{1}{\epsilon K^2}\right),
\label{intK}
\quad\mbox{where~~}
    f_\epsilon(x) \equiv \frac{1}{x\,(1+\epsilon x^2)}.
\eeq
According to the first mean value theorem for integrals,
\beq
   \exists\,\xi_k : k < \xi_k < k+1,
\quad
    \int_{k}^{k+1} f_\epsilon(x)\,\dd x =
     \frac{f_\epsilon(k+1)-f_\epsilon(k)}{2}
         - \frac{1}{12}\,f_\epsilon''(\xi_k).
\eeq
Assuming that $K\in\mathbb{N}$, we can write
\beq
   \int_{K}^{\infty} f_\epsilon(x)\,\dd x = \sum_{k=K}^\infty
      \int_{k}^{k+1} f_\epsilon(x)\,\dd x = 
      \frac{f_\epsilon(K)}{2}+ \sum_{k=K+1}^\infty f_\epsilon(k) - R_K,
\quad\mbox{where~~}
        R_K = \frac{1}{12}\sum_{k=K}^\infty
        f_\epsilon''(\xi_k).
\label{sumK}
\eeq
Using explicit $f_\epsilon(x)$ in \req{intK}, one can show that
\beq
   \frac{2}{\xi^2}\,f_\epsilon(\xi) < f_\epsilon''(\xi)
   < \frac{12}{\xi^2}\,f_\epsilon(\xi)
\label{f''}
\eeq
 for any $\xi>0$ and $\epsilon>0$. Therefore,
\beq
 0 <  R_K <  \sum_{k=K}^\infty \frac{1}{\xi_k^3\,(1+\epsilon\,\xi_k^2)}
   < \sum_{k=K}^\infty \frac{1}{k^3\,(1+\epsilon k^2)}
   < \sum_{k=K}^\infty \frac{1}{k^3\,(1+\epsilon K^2)}
   = \frac{1}{2\,K^2\,(1+\epsilon\,K^2)}.
\label{R_K}
\eeq
Equations (\ref{intK}) and (\ref{sumK}) allow us to rewrite the sum on
the right-hand side of \req{ReTheta} as
\beq
    \sum_{k=1}^{\infty}
       \frac{1}{k\,(1+\epsilon k^2)}
       =
        \sum_{k=1}^{K-1}
       \frac{1}{k\,(1+\epsilon k^2)}
       +
       \frac{1}{2K(1+\epsilon K^2)}
       +
       \frac12\ln\left(1+\frac{1}{\epsilon K^2}\right)
       + R_K.
\label{sumK2}
\eeq
Recalling that $\efqm=1/\sqrt{\epsilon}$ and using
Eqs.~(\ref{sumK2}) and (\ref{R_K}), we transform
\req{ReTheta} into
\beq
   \mbox{Re}\,\Theta(\ii\efqm) =
   \gamma_\mathrm{E} - \sum_{k=1}^{K}
       \frac{1}{k+\epsilon k^3}
       +
       \frac{1}{2K(1+\epsilon K^2)}
       -
       \frac12\ln\left(\epsilon + \frac{1}{K^2}\right)
       - \frac{a}{K^2\,(1+\epsilon K^2)},
\label{ReTheta2}
\eeq
where $0 < a < 1/2$.
This transformation allows us to greatly
reduce the number $K$ of terms in the sum, that are needed to attain a
required accuracy. For example, to reproduce four digits of
$\mbox{Re}\,\Theta(\ii\efqm)=-0.09465$ at $\efqm=1$, we must retain
more than 300 terms
in the original formula (\ref{ReTheta}), while $K=15$ suffices in
\req{ReTheta2} with $a=0$ and only $K=7$ with $a=0.25$.

In practice, we calculate $\mbox{Re}\,\Theta(\ii\efqm)$ using
\req{ReTheta2} with $K=3+[11\efqm]$ and $a=0.25$ at $\efqm<3$
($\epsilon>1/9$). At $\efqm\geq3$,
we substitute \req{ReThetas}
retaining eight terms into \req{ReTheta1}, which gives
\beq
   \mbox{Re}\,\Theta(\ii\efqm) \approx
     - \frac{\epsilon}{12} - \frac{\epsilon^2}{120}
      - \frac{\epsilon^3}{252} - \frac{\epsilon^4}{240}
      - \frac{\epsilon^5}{132} - \frac{691\,\epsilon^6}{32760}
      - \frac{\epsilon^7}{12} - \frac{3617}{8160}\,\epsilon^8.
\eeq
This recipe ensures that both absolute and fractional errors of
calculated $\mbox{Re}\,\Theta(\ii\efqm)$ are 
less than $10^{-6}$ for any~$\efqm$.

\subsection{Proof of Equation (\ref{kCk})}
\label{sect:d}
\setcounter{equation}{0}

Let us rewrite equations (\ref{suprecur1}) and (\ref{suprecur3}),
respectively, as
\bea
   \sqrt{k+1}\,C_{{k}+1}^{(\mathcal{N},n_+)}
   & = &
   \sqrt{Z\,n_+}\,C_{{k}}^{(\mathcal{N}-1,n_+ -1)}
   -
   \sqrt{(Z-1)(\mathcal{N}-k)}\,C_{{k}}^{(\mathcal{N},n_+)},
\\
  (Z-1) \sqrt{k+1}\,C_{{k}+1}^{(\mathcal{N},n_+)}
   & = &
   \sqrt{(Z-1)(\mathcal{N}-k)}\,C_{{k}}^{(\mathcal{N},n_+)}
   -
   \sqrt{Z(Z-1)(\mathcal{N}-n_+)}\,C_{{k}}^{(\mathcal{N}-1,n_+)}.
\eea
Taking the sum of the left and the right parts of these equations, 
we exclude the term
$\sqrt{(Z-1)(\mathcal{N}-k)}\,C_{{k}}^{(\mathcal{N},n_+)}$
and, having divided both parts by $Z$, obtain
\beq
    \sqrt{k+1}\,C_{{k}+1}^{(\mathcal{N},n_+)}
    =
    \sqrt{\frac{n_+}{Z}}\,C_{{k}}^{(\mathcal{N}-1,n_+ -1)}
    -
    \sqrt{\frac{Z-1}{Z}}\,\sqrt{\mathcal{N}-n_+}\,
    \,C_{{k}}^{(\mathcal{N}-1,n_+)}.
\label{kCk1}
\eeq
Dropping the  term with $k=0$ (which equals zero) from the sum in the
left-hand side of \req{kCk} and shifting the summation index $k$
to $k+1$, we can write
\beq
   \sum_{k=0}^{\mathcal{N}} k
     \left(C_k^{(\mathcal{N},n_+)}\right)^2
     = \sum_{k=0}^{\mathcal{N}-1} (k+1)
     \left(C_{k+1}^{(\mathcal{N},n_+)}\right)^2.
\eeq
The substitution of \req{kCk1} gives
\bea
   \sum_{k=0}^{\mathcal{N}} k
     \left(C_k^{(\mathcal{N},n_+)}\right)^2
   & = &  
     \sum_{k=0}^{\mathcal{N}-1}
     \left(
        \sqrt{\frac{n_+}{Z}}\,C_{{k}}^{(\mathcal{N}-1,n_+ -1)}
    -
    \sqrt{\frac{Z-1}{Z}}\,\sqrt{\mathcal{N}-n_+}\,
    \,C_{{k}}^{(\mathcal{N}-1,n_+)}
    \right)^2
\nonumber\\
   & = &  
     \frac{n_+}{Z} \sum_{k=0}^{\mathcal{N}-1} 
     \left(C_k^{(\mathcal{N}-1,n_+ - 1)}\right)^2
     -
     \frac{2}{Z}\sqrt{(Z-1)\,n_+\,(\mathcal{N}-n_+)}
     \sum_{k=0}^{\mathcal{N}-1} 
     C_{{k}}^{(\mathcal{N}-1,n_+ -1)} C_{{k}}^{(\mathcal{N}-1,n_+)}
\nonumber\\&&
     +
     \frac{Z-1}{Z}\,(\mathcal{N} - n_+)
     \sum_{k=0}^{\mathcal{N}-1} 
     \left(C_k^{(\mathcal{N}-1,n_+)}\right)^2.
\eea
According to the orthonormality relation (\ref{orthoC}),
the first and third sums on the right-hand side equal one,
and the second sum equals zero. Thus we are left with \req{kCk}.

\newpage

\end{widetext}

\newcommand{\artref}[4]{\textit{#1} \textbf{#3}, #4 (#2).}
\newcommand{\AandA}[3]{\artref{Astron.\ Astrophys.}{#1}{#2}{#3}}
\newcommand{\ApJ}[3]{\artref{Astrophys.\ J.}{#1}{#2}{#3}}
\newcommand{\ApSS}[3]{\artref{Astrophys.\ Space Sci.}{#1}{#2}{#3}}
\newcommand{\JPB}[3]{\artref{J.\ Phys. B: At.\ Mol.\ Opt.\ Phys.}{#1}{#2}{#3}}
\newcommand{\PR}[4]{\artref{Phys.\ Rev. #1}{#2}{#3}{#4}}
\newcommand{\RPP}[3]{\artref{Rep.\ Prog.\ Phys.}{#1}{#2}{#3}}

\end{document}